\documentclass[a4paper,12pt]{article}
\pdfoutput=1
\usepackage[DIV=14,BCOR=0mm,headinclude=true,footinclude=false]{typearea}
\usepackage[utf8]{inputenc}
\usepackage[T1]{fontenc}

\usepackage{style}

\usepackage{subcaption}
\usepackage{tipa}
\usepackage{graphicx}
\usepackage[bottom]{footmisc}
\usepackage[titletoc]{appendix}
\usepackage{hyperref}
\usepackage{listings}
\usepackage{float}
\usepackage{amsmath}
\usepackage{mathrsfs}
\usepackage{amssymb}
\usepackage{mathtools}
\usepackage{amsfonts}
\usepackage{braket}
\usepackage{slashed}
\usepackage{dashbox}
\usepackage[colorinlistoftodos]{todonotes}
\usepackage{tikz-feynman}   
\usetikzlibrary{snakes}     
\usepackage[normalem]{ulem}

\usepackage{scalerel}
\usepackage{multirow}  
\usepackage{makecell}  

\newcommand\reallywidehat[1]{\arraycolsep=0pt\relax%
	\begin{array}{c}
		\stretchto{
			\scaleto{
				\scalerel*[\widthof{\ensuremath{#1}}]{\kern-.5pt\bigwedge\kern-.5pt}
				{\rule[-\textheight/2]{1ex}{\textheight}} 
			}{\textheight} %
		}{0.5ex}\\           
		#1\\                 
		\rule{-1ex}{0ex}
	\end{array}
}

\newcommand{\vev}[1]{\ensuremath{\left\langle #1 \right\rangle}}
\def\be{\begin{equation}}
\def\ee{\end{equation}}
\def\ba{\begin{aligned}}
\def\ea{\end{aligned}}

\newcommand{\SL}{\text{SL}\left(2,\mathbb{R}\right)}

\numberwithin{equation}{section}
\numberwithin{table}{section}

\title{Love numbers and Love symmetries for $p$-form and gravitational perturbations of higher-dimensional spherically symmetric black holes}

\author[a,b]{Panagiotis Charalambous\footnote{\texttt{pcharala@sissa.it}}}

\affiliation[a]{International School for Advanced Studies (SISSA), \\
Via Bonomea 265, 34136 Trieste, Italy}
\affiliation[b]{Center for Cosmology and Particle Physics, Department of Physics,
New York University, \\
New York, NY 10003, USA}
\date{}

\abstract{
	The static Love numbers of four-dimensional asymptotically flat, isolated, general-relativistic black holes are known to be identically vanishing. The Love symmetry proposal suggests that such vanishings are addressed by selection rules following from the emergence of an enhanced $\SL$ (``Love'') symmetry in the near-zone region; more specifically, it is the fact that the black hole perturbations belong to a highest-weight representation of this near-zone $\SL$ symmetry, rather than the existence of the Love symmetry itself, that outputs the vanishings of the corresponding Love numbers. In higher spacetime dimensions, some towers of magic zeroes with regards to the black hole response problem have also been reported for scalar, electromagnetic and gravitational perturbations of the Schwarzschild-Tangherlini black hole. Here, we extend these results by supplementing with $p$-form perturbations of the Schwarzschild-Tangherlini black hole. We furthermore analytically extract the static Love numbers and the leading order dissipation numbers associated with spin-$0$ scalar and spin-$2$ tensor-type tidal perturbations of the higher-dimensional Reissner-Nordstr\"om black hole. We find that Love symmetries exist and that the vanishings of the static Love numbers are captured by representation theory arguments even for these higher spin perturbations of the higher-dimensional spherically symmetric black holes of General Relativity. Interestingly, these near-zone $\SL$ structures acquire extensions to Witt algebras. Our setup allows to also study the $p$-form response problem of a static spherically symmetric black hole in a generic theory of gravity. We perform explicit computations for some black holes in the presence of string-theoretic corrections and investigate under what geometric conditions Love symmetries emerge in the near-zone.
}

\begin{document}

\maketitle

\section{Introduction}
\label{sec:Intro}

Ever since the first ever confirmed observation of transient gravitational waves emitted during the final stages of the coalescence of a binary system of black holes~\cite{LIGOScientific:2016aoc}, the number of gravitational wave detections has been increasingly growing. The current state-of-the-art third Gravitational-Wave Transient Catalog (GWTC-3)~\cite{KAGRA:2021vkt} of the recently formed LIGO-VIRGO-KAGRA collaboration enumerates a total of $90$ candidate compact binary coalescences and will continue to improve in sensitivity in the future~\cite{Saleem:2021iwi,LIGOScientific:2016wof,Punturo:2010zza,2017arXiv170200786A,Reitze:2019iox}.

More notably, the space-based LISA~\cite{2017arXiv170200786A}, planned to lunch in 2037 and operating in the low frequency range $10^{-4}-1\,\text{Hz}$, compared to LIGO's sensitivity in the $10-10^3\,\text{Hz}$ frequency range, will allow to observe gravitational waves from compact binary systems at much wider orbits and, hence, better study the early stages of the inspiraling phase of the binary system. This regime is particularly relevant for studying tidal effects. More generally, the early stages of the inspiraling phase are accurately described by a Post-Newtonian (PN) description of the dynamics of the system, and observing a larger window of this stage will allow to probe higher PN orders. More importantly, this will allow to probe Love numbers~\cite{Love:1912}, conservative Green's functions associated with the response problem of compact bodies~\cite{PoissonWill2014}, and, hence, probe the internal structure of the involved relativistic configurations, e.g. the elusive nuclear equation of state of Neutron Stars~\cite{LIGOScientific:2017vwq,Raithel:2018ncd,Chatziioannou:2020pqz,Pacilio:2021jmq}. For gravitational interactions, the leading order conservative tidal effects enter at $5$PN order in four spacetime dimensions, and are encoded in the quadrupolar static tidal Love number of each of the bodies involved in the binary system~\cite{Flanagan:2007ix}. This ``effacement'' makes the measurement of Love numbers challenging, requiring high signal-to-noise ratio signals covering a large window of the coalescence~\cite{Chatziioannou:2020pqz,Pacilio:2021jmq,Piovano:2022ojl}.

Besides directly probing the internal structure of the involved bodies~\cite{LIGOScientific:2017vwq,Raithel:2018ncd}, Love numbers have also found applications in proposals for testing strong-field gravity~\cite{Cardoso:2017cfl,Franzin:2017mtq,Cardoso:2018ptl,Katagiri:2023yzm}, as well as for lifting a degeneracy in measuring the luminosity distance and inclination plane~\cite{Usman:2018imj} through ``I-Love-Q'' relations~\cite{Xie:2022brn,Yagi:2013bca,Yagi:2013awa,Yagi:2015pkc,Yagi:2016qmr}.

Their predictable imprint on gravitational wave signatures is a direct consequence of the employment of the worldline Effective Field Theory (EFT) as a toolkit for constructing gravitational waveform templates~\cite{Goldberger:2004jt,Porto:2005ac,Porto:2016pyg,Levi:2015msa,Levi:2018nxp,Goldberger:2022ebt}. Within the worldline EFT, a compact body is treated as an effective point-particle propagating along a worldline, dressed with multipole moments that couple to curvature tensors and capture finite-size effects. In this framework, Love numbers enter as Wilson coefficients for operators quadratic in the curvature reducing their computation to a matching condition.

Applying this algorithm to the case of isolated, asymptotically flat, general relativistic black holes in four spacetime dimensions, one arrives at a theoretically intriguing property: the static Love numbers for scalar, electromagnetic and gravitational perturbations of rotating black holes vanish identically~\cite{Fang:2005qq,Damour:2009vw,Binnington:2009bb,Gurlebeck:2015xpa,Bicak:1977,Bicak:1976,Poisson:2014gka,Landry:2015zfa,Pani:2015hfa,Pani:2015nua,LeTiec:2020spy,LeTiec:2020bos,Chia:2020yla,Charalambous:2021mea,Ivanov:2022qqt,Ivanov:2022hlo}. This property of general-relativistic black holes makes Love numbers relevant for probing new physics. For instance, non-zero Love numbers for compact bodies with masses above the Tolman-Oppenheimer-Volkoff (TOV) limit~\cite{Tolman:1939jz,Oppenheimer:1939ne,Kalogera:1996ci} can be used to probe the environments of black holes~\cite{DeLuca:2021ite}; in fact, small Love numbers themselves can be magnified by supermassive compact objects~\cite{Pani:2019cyc} making this prospect even more promising. Furthermore, the surprisingly rigid ``I-Love-Q'' relations one encounters for general-relativistic compact bodies become non-universal once one departs from General Relativity~\cite{Pani:2015tga,Uchikata:2016qku}. 
On the more theoretical side, the vanishings of the black hole Love numbers provide with an example of ``magic zeroes'' from the worldline EFT point of view, raising naturalness concerns and calling upon the existence of enhanced symmetry structures that are expected to output appropriate selection rules~\cite{tHooft:1979rat,Porto:2016zng}.

Related to this, there have been various works indicating a persisting hidden conformal structure of asymptotically flat black holes. This has been utilized to propose holographic correspondences of black holes with thermal states in a dual $\text{CFT}_2$, for instance, the extremal~\cite{Bardeen:1999px,Kunduri:2007vf,Amsel:2009et,Guica:2008mu,Lu:2008jk} and the non-extremal~\cite{Castro:2010fd,Krishnan:2010pv,Chen:2010zwa,Lowe:2011aa} Kerr/CFT conjectures, which propose that the temperatures of the left-movers and right-movers and the associated central charges in this dual CFT description are directly related to geometric properties of the associated black hole. More recently, the resemblance of the equations of motion governing perturbations of asymptotically flat black holes were contrasted to the BPZ equation satisfied by Liouville correlators involving the insertion of a particular degenerate state, and were used to set up gauge-gravity dictionaries between CFTs and black hole perturbations~\cite{Aminov:2020yma,Bonelli:2021uvf,Consoli:2022eey,Bautista:2023sdf}. Furthermore, the astrophysically relevant photon rings have also been shown to be equipped with conformal structures, providing with proposals for their detectable implications on polarimetric observations~\cite{Johnson:2019ljv,Himwich:2020msm,Raffaelli:2021gzh,Hadar:2022xag,Chen:2022fpl}.

Another conformal structure that lies in the spirit of the Kerr/CFT conjecture and that proves relevant in addressing the vanishing of the static Love numbers is the ``Love symmetry''~\cite{Charalambous:2021kcz,Charalambous:2022rre,Charalambous:2023jgq}. According to this proposal, an enhanced, globally defined, $\SL$ (``Love'') symmetry manifests in the near-zone region; the regime relevant for defining the response problem. The selection rules outputting the vanishing of the static Love numbers are then identified from the fact that the corresponding black hole perturbations belong to highest-weight representations of the Love $\SL$ symmetry. Other symmetry attempts of identifying selection rules relevant for the vanishing of the static Love numbers that act directly at the IR level have also been proposed. These include the ``ladder symmetries'' proposal \cite{Hui:2021vcv,Hui:2022vbh,Berens:2022ebl,Katagiri:2022vyz,Sharma:2024hlz}, whose origins can be traced back to the notion of ``mass ladder operators'' for spacetimes admitting closed conformal Killing vectors~\cite{Cardoso:2017qmj}, and the manifestation of a Schr\"{o}dinger symmetry at the level of the exact static response problem~\cite{BenAchour:2022uqo,BenAchour:2023dgj}. More interestingly, the Love symmetry appears to be closely related to the enhanced $\SL$ isometry subgroup of the near-horizon throat of extremal black holes~\cite{Bardeen:1999px,Kunduri:2007vf,Charalambous:2021kcz,Charalambous:2022rre,Charalambous:2023jgq}.

Similar results turn out to exist in higher spacetime dimensions as well. More specifically, the static Love numbers associated with spin-$0$ (massless scalar), spin-$1$ (electromagnetic) and spin-$2$ (gravitational) perturbations of the $d$-dimensional Schwarzschild-Tangherlini black hole have already been computed in Refs.~\cite{Kol:2011vg,Hui:2020xxx}. While the static Love numbers have a much richer structure and are in general in agreement with Wilsonian naturalness arguments, there exist towers of resonant conditions that depend on the multipolar order $\ell$ of the perturbation for which they vanish, again hinting at the existence of an enhanced symmetry explanation. For the case of spin-$0$ perturbations, for instance, the static Love numbers turn out vanish whenever $\ell/\left(d-3\right)$ is an integer. Nevertheless, Love symmetry has been shown to still exist for any multipolar order and spacetime dimensionality for the case of spin-$0$ perturbations of the Schwarzschild-Tangherlini black hole and is in perfect agreement with these results~\cite{Charalambous:2021kcz,Charalambous:2022rre}.

Here, we extend this analysis to higher spin perturbations of higher-dimensional asymptotically flat, static and spherically symmetric black holes. In particular, we supplement with the computation of $p$-form Love numbers and provide with a Love symmetry explanation beyond scalar, notably, electromagnetic and gravitational perturbations, plus $p$-form perturbations with $p>1$. The structure of this paper is then as follows. In Section~\ref{sec:BHs_SphericallySymmetricDd}, we present the background geometry of a static spherically symmetric black hole in a generic theory of gravity and set up the framework for covariantly studying perturbations via a $2+\left(d-2\right)$ decomposition on the sphere, in the spirit of Refs.~\cite{Ishibashi:2003ap,Martel:2005ir}.

In Section~\ref{sec:EOM_NZ}, we analyze various types of perturbations of such black holes to identify the relevant master variables, working directly at the level of the action~\cite{Hui:2020xxx}. More specifically, we generalize the analysis of Ref.~\cite{Hui:2020xxx} to the scenario where the background geometry is a generic spherically symmetric black hole, possibly non-general-relativistic, for the cases of spin-$0$ massless scalar and spin-$1$ electromagnetic perturbations. We then further extend this construction along the lines of Ref.~\cite{Yoshida:2019tvk} to incorporate the case of $p$-form perturbations of a generic spherically symmetric black hole. For completeness, we also review the analysis of Ref.~\cite{Hui:2020xxx} around spin-$2$ gravitational perturbations of the Schwarzschild-Tangherlini black hole.

In Section~\ref{sec:LoveNumbers}, we present the definition of Love numbers within the worldline EFT and introduce the notion of the near-zone expansion as a necessary tool for performing matching computations. Using this, we employ particular near-zone splittings of the equations of motion obeyed by the master variables, and find the Love numbers associated to $p$-form and gravitational perturbations of Schwarzschild-Tangherlini black holes at leading order in the near-zone expansions. We find that the Love numbers can be collectively written in terms of two parameters: the orbital number $\ell$, and an index $j$ capturing the $SO\left(d-1\right)$ sector that the perturbation belongs to. After categorizing the results in three classes depending on the values of the index $j$, we then analyze the behavior of the static Love numbers in each class, commenting on their running and the existence of towers of resonant conditions for which they vanish. We also consider spin-$0$ scalar and spin-$2$ tensor-type tidal perturbations of the higher-dimensional Reissner-Nordstr\"{o}m black hole. By matching onto the worldline EFT, we are then able to analytically confirm the conjectured expressions of the relevant static Love numbers presented in Ref.~\cite{Pereniguez:2021xcj}, while we also extract the leading order dissipative viscosity numbers.

Resuming the investigation on the response properties of general-relativistic black holes, we reveal in Section~\ref{sec:LoveSymmetryDd} the manifestation of enhanced Love $\SL$ symmetries within the near-zone region, which turn out to be completely independent of the orbital number of the perturbation. Using representation theory arguments, we then identify the existence of selection rules that are in $1$-to-$1$ correspondence with all the resonant conditions for which the static Love numbers vanish, hence restoring the naturalness of General Relativity with respect to the black hole response problem. Interestingly, we also recognize that the Love $\SL$ symmetries have unique extensions to centerless Virasoro (Witt) algebras~\cite{Ortin:2012mt} and comment on their extended representations.

In Section~\ref{sec:LoveSymmetryModGR}, we consider black holes in modified theories of gravity beyond General Relativity. We perform explicit calculations of the static Love numbers for higher-dimensional black holes in the presence of string theoretic corrections, namely, the Callan-Myers-Perry black hole of bosonic/heterotic string theory~\cite{Callan:1988hs,Myers:1998gt} and the $\alpha^{\prime3}$-corrected black hole solution of type-II superstring theory~\cite{Myers:1987qx}. We find that the static Love numbers for these black holes do not exhibit any resonant conditions of vanishings, hence not requiring the existence of enhanced Love symmetries. We attempt to generalize this statement by extracting necessary geometric constraints for the existence of near-zone $\SL$ symmetries and confirm that these constraints are in accordance with the current known results around the static response problems for black holes in various theories of gravity. 

We conclude with a discussion of the results of the current work in Section~\ref{sec:Discussion}. For convenience, we also include Appendix~\ref{app:2F1Gamma}, containing information about the $\Gamma$-function and Euler's hypergeometric function, that are involved in solving the near-zone equations of motion and extracting the black hole response coefficients.

\textit{Notation and conventions}: We will be working in geometrized units with $c=1$, adopting the mostly-positive metric Lorentzian signature, $\left(\eta_{\mu\nu}\right) = \text{diag}\left(-1,+1,+1,\dots\right)$. Small Greek letters will denote spacetime indices running from $0$ to $d-1$, for a $d$-dimensional spacetime, with $x^0$ the temporal coordinate, and repeated indices will be summed over. In performing the $2+\left(d-2\right)$ decomposition of the perturbations around the static spherically symmetric black hole background, capital Latin letters from the beginning of the alphabet, $A$, $B$, $\dots$, will denote spherical indices, running from $1$ to $d-2$ and labeling the $d-2$ spherical coordinates $\theta^{A}$, $A=1,\dots,d-2$ charting the unit $\left(d-2\right)$ sphere, while Small Latin letters from the beginning of the alphabet, $a$, $b$, $\dots$, will denote the remaining directions of the manifold, running from $0$ to $1$ and labeling the temporal and radial coordinates, e.g., in Schwarzschild coordinates, $x^{a}=\left(t,r\right)$. Capital bold symbols and hatted capital bold symbols will be used to refer to differential forms on the full spacetime and differential forms on $\mathbb{S}^{d-2}$ respectively; curly hatted capital symbols will be used for co-exact differential forms on $\mathbb{S}^{d-2}$.

To avoid a large number of indices notations, in Sections~\ref{sec:WorldlineEFT}-\ref{sec:LoveNumbersMatching}, spatial indices running from $1$ to $d-1$ will also be labeled by small Latin letters from the beginning of the alphabet. We will also adapt the multi-index notation $a_1\dots a_{\ell}\equiv L$, within which $x^{a_1}\dots x^{a_{\ell}}\equiv x^{L}$ and $\partial_{a_1}\dots\partial_{a_{\ell}}\equiv\partial_{L}$. The symmetric trace-free part of a tensor with respect to a set of indices will be denoted by enclosing the indices within angular brackets, e.g. $T_{\left\langle a_1a_2 \dots \right\rangle b_1b_2\dots}$ is the symmetric trace-free part of the tensor $T_{a_1a_2\dots b_1b_2\dots}$ with respect to the indices $\left\{a_1,a_2,\dots\right\}$.

\section{Spherically symmetric black holes in higher spacetime dimensions}
\label{sec:BHs_SphericallySymmetricDd}

We start with a few key properties of the background geometry. We will be dealing with a general asymptotically flat, static, spherically symmetric and non-extremal black hole, which can always be brought to the form
\be\label{eq:SphericallySymemtricBHDd}
	ds^2 = -f_{t}\left(r\right)dt^2 + \frac{dr^2}{f_{r}\left(r\right)} + r^2d\Omega_{d-2}^2 \,,
\ee
where $d\Omega_{d-2}^2 = \Omega_{AB}\left(\theta\right)d\theta^{A}d\theta^{B}$ is the metric on $\mathbb{S}^{d-2}$, with angular coordinates labeled by capital Latin indices, $A,B=1,\dots,d-2$, and the argument ``$\theta$'' in $\Omega_{AB}\left(\theta\right)$ collectively indicating all the angular coordinates. In the above parameterization of the geometry, the radial coordinate is an areal radius. The asymptotic flatness and non-extremality conditions are imposed by the requirements
\be
	\begin{gathered}
		\lim_{r\rightarrow\infty}f_{t,r}\left(r\right)=1 \,, \\
		f_{t}\left(r_{\text{h}}\right)=f_{r}\left(r_{\text{h}}\right)=0 \quad \text{and} \quad f_{t}^{\prime}\left(r_{\text{h}}\right),f_{r}^{\prime}\left(r_{\text{h}}\right)\ne0 \,,
	\end{gathered}
\ee
with $r=r_{\text{h}}$ the location of the event horizon. Similar to the $4$-dimensional Schwarzschild black hole, the event horizon is a Killing horizon with respect to the Killing vector $K=\partial_{t}$.

To analyze the behavior of the perturbations near the event horizon, it is necessary to employ the tortoise coordinate,
\be
	dr_{\ast} = \frac{dr}{\sqrt{f_{t}\left(r\right)f_{r}\left(r\right)}} \,,
\ee
in terms of which the advanced ($+$) and retarded ($-$) null coordinates $\left(t_{\pm},r,\theta^{A}\right)$ are defined as
\be\label{eq:AdvancedRetarded_SchwazrschildDd}
	dt_{\pm} = dt \pm dr_{\ast} \,.
\ee
In particular, the near-horizon behavior of the tortoise coordinate can be extracted explicitly to be
\be\label{eq:Tortoise_SchwazrschildDd}
	r_{\ast} \sim \frac{\beta}{2}\ln\left|\frac{r-r_{\text{h}}}{r_{\text{h}}}\right| \quad \text{as $r\rightarrow r_{\text{h}}$} \,,
\ee
with $\beta$ the inverse surface gravity,
\be\label{eq:betaDd}
	\beta = \kappa^{-1} = \frac{2}{\sqrt{f_{t}^{\prime}\left(r_{\text{h}}\right)f_{r}^{\prime}\left(r_{\text{h}}\right)}} \,.
\ee
Then, monochromatic waves of frequency $\omega$ that are ingoing at the future ($+$)/past ($-$) event horizon behave as
\be\label{eq:NHbehavior_SphericalSymmetric}
	e^{-i\omega t_{\pm}} \sim e^{-i\omega t}\left(\frac{r-r_{\text{h}}}{r_{\text{h}}}\right)^{\mp i\beta\omega/2} \,.
\ee

For General Relativity, the most general spherically symmetric and asymptotically flat black hole geometry is the higher-dimensional Reissner-Nordstr\"{o}m(-Tangherlini) solution~\cite{Tangherlini:1963bw,Hollands:2012xy},
\be\ba\label{eq:RNDdGeometry}
	f_{t}\left(r\right) = f_{r}\left(r\right) &= 1 - \left(\frac{r_{s}}{r}\right)^{d-3} + \left(\frac{r_{Q}}{r}\right)^{2\left(d-3\right)} \\
	&=  \left[1-\left(\frac{r_{+}}{r}\right)^{d-3}\right]\left[1-\left(\frac{r_{-}}{r}\right)^{d-3}\right] \,.
\ea\ee
where the Schwarzschild radius $r_{s}$ and the charge parameter $r_{Q}$ are related to the ADM mass $M$ and the electric charge $Q$ (in CGS units) of the black hole according to
\be
	r_{s}^{d-3} = \frac{16\pi GM}{\left(d-2\right)\Omega_{d-2}} \,,\quad r_{Q}^{2\left(d-3\right)} = \frac{32\pi^2 GQ^2}{\left(d-2\right)\left(d-3\right)\Omega_{d-2}^2} \,,
\ee
while the inner and outer horizons are expressed in terms of $r_{s}$ and $r_{Q}$ as
\be
	r_{\pm}^{d-3} = \frac{1}{2}\left[r_{s}^{d-3} \pm \sqrt{r_{s}^{2\left(d-3\right)}-4r_{Q}^{2\left(d-3\right)}}\right] \,.
\ee
In the above expressions, $\Omega_{d-2}=2\pi^{\left(d-1\right)/2}/\Gamma\left(\frac{d-1}{2}\right)$ is the surface area of the unit $\left(d-2\right)$-sphere. The essential singularity at $r\rightarrow0$ is hidden behind an event horizon as long as the magnitude of the electric charge is bounded from above from the mass of the black hole,
\be
	Q^2 \le 2\frac{d-3}{d-2}GM^2 \,,
\ee
with the saturation of the inequality indicating the extremality condition.

\subsection{$2+\left(d-2\right)$ decomposition}
\label{sec:2pDm2Decomposition}

Let us now review the key elements of covariantly studying the perturbations of the above generic spherically symmetric black hole. A higher-dimensional version of the Newman-Penrose formalism is possible~\cite{Pravda:2004ka,Coley:2004jv,Pravdova:2008gp,Durkee:2010xq} and the separability of perturbations around an algebraically special background geometry relevant for black holes has been shown explicitly for the class of the so-called Kundt spacetimes~\cite{Durkee:2010qu,Godazgar:2011sn}. Even though this class includes the higher-dimensional Schwarzschild-Tangherlini black hole, we choose here to work in a less involved formalism that is also more reminiscent of the early days of studying the stability of the four-dimensional Schwarzschild black hole by Regge and Wheeler~\cite{Regge:1957td} and Zerilli~\cite{Zerilli:1970se}.

The background geometry is of the form of a generic $\mathcal{M}^{\left(2\right)}\times\mathbb{S}^{d-2}$ manifold equipped with a time-like Killing vector $t^{a}$,
\be
	\begin{gathered}
		ds^2 = g_{ab}\left(x\right)dx^{a}dx^{b} + r^2\left(x\right)\Omega_{AB}\left(\theta\right)d\theta^{A}d\theta^{B} \,, \\
		\mathcal{L}_{t}g_{ab} = 0 \,,\quad t^{a}\nabla_{a}\Omega_{AB} = 0 \,,\quad t^{a}\nabla_{a}r = 0 \,,
	\end{gathered}
\ee
where small Latin indices run over $\mathcal{M}^{\left(2\right)}$, $a,b=0,1$, and capital Latin indices run over the spherical coordinates $\theta^{A}$ of $\mathbb{S}^{d-2}$, $A=1,2,\dots,d-2$. With respect to $\mathcal{M}^{\left(2\right)}$, $r\left(x\right)$ is a scalar. In order to perform a covariant $2+\left(d-2\right)$ decomposition, we follow Refs.~\cite{Ishibashi:2003ap,Martel:2005ir} (see also Refs.~\cite{Kodama:2003jz,Kodama:2003kk,Ishibashi:2011ws}) and introduce the $\mathcal{M}^{\left(2\right)}$ co-vector normal to surfaces of constant $r\left(x\right)$,
\be
	r_{a} \equiv \nabla_{a}r \,.
\ee
In Schwarzschild coordinates, $r_{a}=\left(0,1\right)$. This allows to covariantly define $f_{t}\left(r\right)$ and $f_{r}\left(r\right)$ as
\be
	f_{t}\left(r\right) = -t_{a}t^{a} \,,\quad f_{r}\left(r\right) = r_{a}r^{a} \,.
\ee
The time-like Killing vector $t^{a}$ and the $2$-vector $r^{a}$ are orthogonal to each other, $t_{a}r^{a}=0$, and serve as a basis for $\mathcal{M}^{\left(2\right)}$. For example, the metric tensor $g_{ab}$ and the Levi-Civita tensor $\varepsilon_{ab}$ on $\mathcal{M}^{\left(2\right)}$ can be written as
\be
	g_{ab} =  -\frac{1}{f_{t}}t_{a}t_{b} + \frac{1}{f_{r}}r_{a}r_{b} \,,\quad \varepsilon_{ab} = -\frac{1}{\sqrt{f_{t}f_{r}}}\left(t_{a}r_{b}-r_{a}t_{b}\right) \,.
\ee
A zweibein for $\mathcal{M}^{\left(2\right)}$ would then be $\ell^{a} = t^{a}/\sqrt{f_{t}}$ and $n^{a} = r^{a}/\sqrt{f_{r}}$, such that $g_{ab} = -\ell_{a}\ell_{b} + n_{a}n_{b}$.

Let us now see how to decompose covariant derivatives. First of all, for spacetime scalar functions,
\be
	\nabla_{a}\phi = D_{a}\phi = -\frac{1}{f_{t}}t^{b}D_{b}\phi\,t_{a} + \frac{1}{f_{r}}r^{b}D_{b}\phi\,r_{a} \,,\quad \nabla_{A}\phi = D_{A}\phi \,,
\ee
where $D_{a}$ and $D_{A}$ are the covariant derivatives compatible with $g_{ab}$ and the unit-sphere metric $\Omega_{AB}$ respectively. For higher-spin fields, we need the $2+\left(d-2\right)$ decomposition of Christoffel symbols,
\be\ba
	\Gamma^{a}_{bA} &= 0 \,,\quad \Gamma^{a}_{AB} = -rr^{a}\Omega_{AB} \,, \\
	\Gamma^{A}_{ab} &= 0 \,,\quad \Gamma^{A}_{aB} = \frac{1}{r}r_{a}\delta^{A}_{B} \,.
\ea\ee
Then, we can see that, for a dual vector field $V_{\mu}$,
\be
	\begin{gathered}
		\nabla_{b}V_{a} = D_{b}V_{a} \,,\quad \nabla_{A}V_{a} = D_{A}V_{a} - \frac{1}{r}r_{a}V_{A} \\
		\nabla_{a}V_{A} = rD_{a}\left(\frac{V_{A}}{r}\right) \,,\quad \nabla_{B}V_{A} = D_{B}V_{A} + rr^{a}V_{a}\Omega_{AB} \,,
	\end{gathered}
\ee
while, for a rank-$2$ co-tensor $T_{\mu\nu}$,
\be
	\begin{gathered}
		\nabla_{c}T_{ab} = D_{c}T_{ab} \,,\quad \nabla_{A}T_{ab} = D_{A}T_{ab} - \frac{1}{r}\left(r_{a}T_{Ab}+r_{b}T_{aA}\right) \,, \\
		\nabla_{b}T_{aA} = rD_{b}\left(\frac{T_{aA}}{r}\right) \,,\quad \nabla_{B}T_{aA} = D_{B}T_{aA} - \frac{1}{r}r_{a}T_{BA} + rr^{b}T_{ab}\Omega_{AB} \,, \\
		\nabla_{a}T_{AB} = r^2 D_{a}\left(\frac{T_{AB}}{r^2}\right) \,,\quad  \nabla_{C}T_{AB} = D_{C}T_{AB} + rr^{a}\left(T_{aB}\Omega_{AC}+T_{Aa}\Omega_{BC}\right) \,.
	\end{gathered}
\ee

Furthermore, it will be useful to have explicit formulas for the covariant derivatives of the vectors $t^{a}$ and $r^{a}$. It is straightforward to show that
\be
	D_{a}t_{b} = -\frac{f_{t}^{\prime}}{2f_{t}}\left(t_{a}r_{b}-r_{a}t_{b}\right) \,,\quad D_{a}r_{b} = f_{r}\left[-\frac{f_{t}^{\prime}}{2f_{t}^2}\,t_{a}t_{b} + \frac{f_{r}^{\prime}}{2f_{r}^2}\,r_{a}r_{b}\right] \,,
\ee
where primes denote derivatives with respect to $r$ and we used that $r^{a}D_{a}F\left(r\right)=F^{\prime}\left(r\right)r_{a}r^{a} = F^{\prime}\left(r\right)f_{r}\left(r\right)$.
\section{Equations of motion and master variables}
\label{sec:EOM_NZ}

We now begin extracting the equations of motion governing perturbations of the background spherical symmetric black hole geometry. We will employ the aforementioned formalism of covariantly performing a $2+\left(d-2\right)$ decomposition but follow the prescription of Ref.~\cite{Hui:2020xxx} of working directly at the level of the action.

\subsection{Spin-$0$ perturbations}
\label{sec:EOM_Seq0}

We start from the action for a free scalar field minimally coupled to gravity,
\be
	S^{\left(0\right)} = \int d^{d}x\sqrt{-g}\left[-\frac{1}{2}\left(\nabla\Phi\right)^2 - \frac{1}{2}m^2\Phi^2\right] \,.
\ee
The scalar field is $\left(2+\left(d-2\right)\right)$-decomposed in spherical harmonic modes according to
\be
	\Phi\left(x\right) = \sum_{\ell,\mathbf{m}}\frac{\Psi^{\left(0\right)}_{\ell,\mathbf{m}}\left(t,r\right)}{r^{\left(d-2\right)/2}}Y_{\ell,\mathbf{m}}\left(\theta\right) \,.
\ee
The resulting reduced action then describes a scalar field minimally coupled to $2$-d gravity,
\be
	\begin{gathered}
		S^{\left(0\right)} = \sum_{\ell,\mathbf{m}}S^{\left(0\right)}_{\ell,\mathbf{m}} \,, \\
		S^{\left(0\right)}_{\ell,\mathbf{m}} = \int d^2x\sqrt{-g^{\left(2\right)}}\,\left[-\frac{1}{2}D_{a}\bar{\Psi}^{\left(0\right)}_{\ell,\mathbf{m}}D^{a}\Psi^{\left(0\right)}_{\ell,\mathbf{m}}-\frac{1}{2}V^{\left(0\right)}_{\ell}\left(r\right)\left|\Psi^{\left(0\right)}_{\ell,\mathbf{m}}\right|^2\right] \,,
	\end{gathered}
\ee
with the potential given by
\be\ba\label{eq:V0S}
	V^{\left(0\right)}_{\ell}\left(r\right) &= \frac{\ell\left(\ell+d-3\right)}{r^2} + \frac{\left(d-2\right)\left(d-4\right)}{4r^2}r_{a}r^{a} + \frac{d-2}{2r}D_{a}r^{a} + m^2 \\
	&=\frac{\ell\left(\ell+d-3\right)}{r^2} + \frac{\left(d-2\right)\left(d-4\right)}{4r^2}f_{r} + \frac{d-2}{2r}\frac{\left(f_{t}f_{r}\right)^{\prime}}{2f_{t}} + m^2 \,.
\ea\ee

Working with the tortoise coordinate, this reduces to the action for a scalar field propagating in $2$-d flat spacetime under the influence of a potential,
\be
	S^{\left(0\right)}_{\ell,\mathbf{m}} = \frac{1}{2}\int dtdr_{\ast} \left[\frac{1}{2}\left|\partial_{t}\Psi^{\left(0\right)}_{\ell,\mathbf{m}}\right|^2-\frac{1}{2}\left|\partial_{r_{\ast}}\Psi^{\left(0\right)}_{\ell,\mathbf{m}}\right|^2 - \frac{1}{2}f_{t}\left(r\right)V^{\left(0\right)}_{\ell}\left(r\right)\left|\Psi^{\left(0\right)}_{\ell,\mathbf{m}}\right|^2\right] \,,
\ee
and the equation of motion for scalar field perturbations reduces to a Shr\"{o}dinger-like equation,
\be
	\left[\partial_{r_{\ast}}^2-\partial_{t}^2-f_{t}\left(r\right)V^{\left(0\right)}_{\ell}\left(r\right)\right]\Psi^{\left(0\right)}_{\ell,\mathbf{m}} = 0 \,.
\ee

\subsection{Spin-$1$ perturbations}
\label{sec:EOM_Seq1}

Next, for electromagnetic perturbations, we focus to an electrically neutral black hole background such that there is no background electric field\footnote{This simply ensures that we will not need to deal with coupled equations of motion involving gravitational perturbation modes. It also eliminates the need to refer to a particular gravitational theory and allows to treat the spin-$1$ response problem of spherically symmetric black holes in a generic theory of gravity.}. To treat the Maxwell action,
\be
	S^{\left(1\right)} = \int d^{d}x\sqrt{-g}\left[-\frac{1}{4}F_{\mu\nu}F^{\mu\nu}\right] \,,\quad F_{\mu\nu} = \partial_{\mu}A_{\nu}-\partial_{\nu}A_{\mu} \,,
\ee
the $2+\left(d-2\right)$ decomposition involves first decomposing the components of the gauge field into irreducible representations of $SO\left(d-1\right)$,
\be
	A_{\mu}\left(x\right) =
	\begin{pmatrix}
		A_{a}\left(x\right) \\
		D_{A}A^{\left(\text{L}\right)}\left(x\right) + A_{A}^{\left(\text{T}\right)}\left(x\right)
	\end{pmatrix} \,.
\ee
With respect to $SO\left(d-1\right)$ transformations, $A_{a}$ and $A^{\left(\text{L}\right)}$ are scalars, while $A_{A}^{\left(\text{T}\right)}$ is a transverse co-vector, $D^{A}A_{A}^{\left(\text{T}\right)}=0$. Under gauge transformations $\delta_{\Lambda}A_{\mu}=\partial_{\mu}\Lambda$, the $SO\left(d-1\right)$-decomposed components transform according to
\be
	\delta_{\Lambda}A_{a} = \partial_{a}\Lambda \,,\quad \delta_{\Lambda}A^{\left(\text{L}\right)} = \Lambda \,,\quad \delta_{\Lambda}A_{A}^{\left(\text{T}\right)} = 0 \,.
\ee
Notably, the transverse vectors are gauge invariant, while we also see how the longitudinal modes $A^{\left(\text{L}\right)}$ are redundant degrees of freedom; they are pure gauge. A second gauge invariant quantity can then be constructed as
\be
	\mathcal{A}_{a} = A_{a} - D_{a}A^{\left(\text{L}\right)}
\ee
and the $\left(2+\left(d-2\right)\right)$-decomposed field strength tensor reads
\be\ba
	F_{ab} &= D_{a}\mathcal{A}_{b} - D_{b}\mathcal{A}_{a} \\
	F_{aA} &= D_{a}A^{\left(\text{T}\right)}_{A} - D_{A}\mathcal{A}_{a} \\
	F_{AB} &= D_{A}A^{\left(\text{T}\right)}_{B} - D_{B}A^{\left(\text{T}\right)}_{A} \,.
\ea\ee

The next step is to expand the scalars into scalar spherical harmonic modes $Y_{\ell,\mathbf{m}}\left(\theta\right)$ and the transverse vector into transverse vector spherical harmonic modes\footnote{For more information on the scalar, vector and tensor spherical harmonics in higher dimensions, we refer to Refs.~\cite{Hui:2020xxx,Chodos:1983zi,Higuchi:1986wu}.} $Y^{\left(\text{T}\right)A}_{\ell,\mathbf{m}}\left(\theta\right)$,
\be\ba
	\mathcal{A}^{a}\left(x\right) &= \sum_{\ell,\mathbf{m}}\mathcal{A}^{a}_{l,\mathbf{m}}\left(t,r\right)Y_{\ell,\mathbf{m}}\left(\theta\right) \,, \\
	A^{\left(\text{T}\right)}_{A}\left(x\right) &= \sum_{\ell,\mathbf{m}}A^{\left(\text{V}\right)}_{l,\mathbf{m}}\left(t,r\right)Y^{\left(\text{T}\right)}_{A;\ell,\mathbf{m}}\left(\theta\right) \,.
\ea\ee
Spherical symmetry of the background ensures that the scalar modes $\mathcal{A}^{a}_{\ell,\mathbf{m}}$ and the vector modes $A^{\left(\text{V}\right)}_{\ell,\mathbf{m}}$ will completely decouple from each other. Indeed, the Maxwell action after this expansion reads\footnote{The common sum over the ``azimuthal'' multi-index $\mathbf{m}$ is a bit misleading. In contrast to the $2$-sphere, the scalar and vector, and also tensor and $p$-form, spherical harmonics have a different degeneracy depending on their rank~\cite{Hui:2020xxx,Chodos:1983zi,Higuchi:1986wu} and, thus, each $SO\left(d-1\right)$ sector will have its own sum over $\mathbf{m}$. For simplicity, however, we will keep writing a common sum over $\mathbf{m}$ for all $SO\left(d-1\right)$ sectors as a book-keeping prescription; after all, this subtlety will be completely irrelevant for the study of the black hole Love numbers, which are independent of $\mathbf{m}$ by virtue of the spherical symmetry of the background.}
\be
	\begin{gathered}
		S^{\left(1\right)} = \sum_{\ell,\mathbf{m}}\left(S^{\left(\text{V}\right)}_{\ell,\mathbf{m}} + S^{\left(\text{S}\right)}_{\ell,\mathbf{m}}\right) \,, \\
		\ba
			S^{\left(\text{V}\right)}_{\ell,\mathbf{m}} &= \int d^2x\sqrt{-g^{\left(2\right)}}\,r^{d-4} \left[-\frac{1}{2}D_{a}\bar{A}^{\left(\text{V}\right)}_{\ell,\mathbf{m}}D^{a}A^{\left(\text{V}\right)}_{\ell,\mathbf{m}} - \frac{1}{2}\frac{\left(\ell+1\right)\left(\ell+d-4\right)}{r^2}\left|A^{\left(\text{V}\right)}_{\ell,\mathbf{m}}\right|^2 \right] \,, \\
			S^{\left(\text{S}\right)}_{\ell,\mathbf{m}} &= \int d^2x\sqrt{-g^{\left(2\right)}}\,r^{d-2} \left[-\frac{1}{4}\bar{\mathcal{F}}_{ab;\ell,\mathbf{m}}\mathcal{F}^{ab}_{\ell,\mathbf{m}} - \frac{1}{2}\frac{\ell\left(\ell+d-3\right)}{r^2}\bar{\mathcal{A}}_{a;\ell,\mathbf{m}}\mathcal{A}^{a}_{\ell,\mathbf{m}}\right] \,,
		\ea
	\end{gathered}
\ee
where
\be
	\mathcal{F}^{ab}_{\ell,\mathbf{m}} \equiv D^{a}\mathcal{A}^{b}_{\ell,\mathbf{m}} - D^{b}\mathcal{A}^{a}_{\ell,\mathbf{m}} \,.
\ee

\subsubsection{Vector modes}
We begin by studying the decoupled vector modes. Similar to the scalar field case, these will be governed by the action for a scalar field minimally coupled to $2$-d gravity propagating under the influence of a potential. More explicitly, writing
\be
	A^{\left(\text{V}\right)}_{\ell,\mathbf{m}}\left(t,r\right) = \frac{\Psi^{\left(\text{V}\right)}_{\ell,\mathbf{m}}\left(t,r\right)}{r^{\left(d-4\right)/2}} \,,
\ee
the reduced action for the vector modes reads
\be
	S_{\ell,\mathbf{m}}^{\left(\text{V}\right)} = \int d^2x\sqrt{-g^{\left(2\right)}}\left[-\frac{1}{2}D_{a}\bar{\Psi}^{\left(\text{V}\right)}_{\ell,\mathbf{m}}D^{a}\Psi^{\left(\text{V}\right)}_{\ell,\mathbf{m}} - \frac{1}{2}V^{\left(\text{V}\right)}_{\ell}\left(r\right)\left|\Psi^{\left(\text{V}\right)}_{\ell,\mathbf{m}}\right|^2\right]
\ee
with the potential given by
\be\ba\label{eq:V1V}
	V^{\left(\text{V}\right)}_{\ell}\left(r\right) &= \frac{\left(\ell+1\right)\left(\ell+d-4\right)}{r^2} + \frac{\left(d-4\right)\left(d-6\right)}{4r^2}r_{a}r^{a} + \frac{d-4}{2r}D_{a}r^{a} \\
	&= \frac{\left(\ell+1\right)\left(\ell+d-4\right)}{r^2} + \frac{\left(d-4\right)\left(d-6\right)}{4r^2}f_{r} + \frac{d-4}{2r}\frac{\left(f_{t}f_{r}\right)^{\prime}}{2f_{t}} \,.
\ea\ee
Working with the tortoise coordinate, this becomes the action for a scalar field in $2$-d flat spacetime,
\be
	S^{\left(\text{V}\right)}_{\ell,\mathbf{m}} = \frac{1}{2}\int dtdr_{\ast} \left[\frac{1}{2}\left|\partial_{t}\Psi^{\left(\text{V}\right)}_{\ell,\mathbf{m}}\right|^2-\frac{1}{2}\left|\partial_{r_{\ast}}\Psi^{\left(\text{V}\right)}_{\ell,\mathbf{m}}\right|^2 - \frac{1}{2}f_{t}\left(r\right)V^{\left(\text{V}\right)}_{\ell}\left(r\right)\left|\Psi^{\left(\text{V}\right)}_{\ell,\mathbf{m}}\right|^2\right] \,,
\ee
with Schr\"{o}dinger-like equations of motion,
\be
	\left[\partial_{r_{\ast}}^2-\partial_{t}^2-f_{t}\left(r\right)V^{\left(\text{V}\right)}_{\ell}\left(r\right)\right]\Psi^{\left(\text{V}\right)}_{\ell,\mathbf{m}} = 0 \,.
\ee

\subsubsection{Scalar modes}
We next analyze the action for the scalar modes. For the sake of this, inspired by Ref.~\cite{Hui:2020xxx}, we introduce an auxiliary $2$-d scalar field $\Psi^{\left(\text{S}\right)}_{\ell,\mathbf{m}}\left(t,r\right)$ and consider the action
\be\ba
	\tilde{S}^{\left(\text{S}\right)}_{\ell,\mathbf{m}} = \int d^2x\sqrt{-g^{\left(2\right)}} &\bigg[\frac{1}{2}\sqrt{\ell\left(\ell+d-3\right)}\,r^{\left(d-4\right)/2}\text{Re}\left\{\bar{\Psi}^{\left(\text{S}\right)}_{\ell,\mathbf{m}}\varepsilon_{ab}\mathcal{F}^{ab}_{\ell,\mathbf{m}}\right\} \\
	&- \frac{1}{2}\frac{\ell\left(\ell+d-3\right)}{r^2}\left(\left|\Psi^{\left(\text{S}\right)}_{\ell,\mathbf{m}}\right|^2 + r^{d-2}\bar{\mathcal{A}}_{a;\ell,\mathbf{m}}\mathcal{A}^{a}_{\ell,\mathbf{m}}\right)\bigg] \,.
\ea\ee
Classically, this is equivalent to the original action $S^{\left(\text{S}\right)}_{\ell,\mathbf{m}}$ for the scalar modes as can be seen by putting the auxiliary field on-shell,
\be
	\Psi^{\left(\text{S}\right)}_{\ell,\mathbf{m}} = \frac{1}{2}\frac{r^{d/2}}{\sqrt{\ell\left(\ell+d-3\right)}}\varepsilon_{ab}\mathcal{F}^{ab}_{\ell,\mathbf{m}} \,.
\ee

The upshot of this alternative action is that it can be recast in a form similar to the scalar field modes and the gauge field vector modes. This is achieved by integrating out $\mathcal{A}^{a}_{\ell,\mathbf{m}}$ in $\tilde{S}^{\left(\text{S}\right)}_{\ell,\mathbf{m}}$,
\be
	\mathcal{A}^{a}_{\ell,\mathbf{m}} = \frac{r^{-\left(d-4\right)/2}}{\sqrt{\ell\left(\ell+d-3\right)}}\left[\varepsilon^{ab}D_{b} - t^{a}\frac{d-4}{2r}\sqrt{\frac{f_{r}}{f_{t}}}\right]\Psi^{\left(\text{S}\right)}_{\ell,\mathbf{m}} \,.
\ee
As a result,
\be\ba
	\tilde{S}^{\left(\text{S}\right)}_{\ell,\mathbf{m}} &= \int d^2x\sqrt{-g^{\left(2\right)}}\left[-\frac{1}{2}D_{a}\bar{\Psi}^{\left(\text{S}\right)}_{\ell,\mathbf{m}}D^{a}\Psi^{\left(\text{S}\right)}_{\ell,\mathbf{m}} - \frac{1}{2}V^{\left(\text{S}\right)}_{\ell}\left(r\right)\left|\Psi^{\left(\text{S}\right)}_{\ell,\mathbf{m}}\right|^2\right] \\
	&= \int dtdr_{\ast}\left[\frac{1}{2}\left|\partial_{t}\Psi^{\left(\text{S}\right)}_{\ell,\mathbf{m}}\right|^2 -\frac{1}{2}\left|\partial_{r_{\ast}}\Psi^{\left(\text{S}\right)}_{\ell,\mathbf{m}}\right|^2 - \frac{1}{2}f_{t}\left(r\right)V^{\left(\text{S}\right)}_{\ell}\left(r\right)\left|\Psi^{\left(\text{S}\right)}_{\ell,\mathbf{m}}\right|^2\right] \,,
\ea\ee
with the potential given by
\be\ba\label{eq:V1S}
	V^{\left(\text{S}\right)}_{\ell}\left(r\right) &= \frac{\ell\left(\ell+d-3\right)}{r^2} + \frac{\left(d-2\right)\left(d-4\right)}{4r^2}r_{a}r^{a} - \frac{d-4}{2r}D_{a}r^{a} \\
	&= \frac{\ell\left(\ell+d-3\right)}{r^2} + \frac{\left(d-2\right)\left(d-4\right)}{4r^2}f_{r} - \frac{d-4}{2r}\frac{\left(f_{t}f_{r}\right)^{\prime}}{2f_{t}} \,.
\ea\ee

\subsection{$p$-form perturbations}

In higher spacetime dimensions, it is also possible to have $p$-form perturbations, generated by a completely antisymmetric gauge field tensor $A_{\mu_1\mu_2\dots\mu_{p}}$ of rank $p\le d-3$ or, in $p$-form notation, by the object
\be
	\mathbf{A}^{\left(p\right)} = \frac{1}{p!}A_{\mu_1\mu_2\dots\mu_{p}}\,\text{d}x^{\mu_1}\wedge\text{d}x^{\mu_2}\wedge\dots\wedge\text{d}x^{\mu_{p}} \,.
\ee
Focusing to spherically symmetric black hole backgrounds that are not charged under the $p$-form, the task now is to study the $p$-form extension of the Maxwell action
\be
	S^{\left(p\right)} = -\frac{1}{2}\int \mathbf{F}^{\left(p+1\right)}\wedge \star\mathbf{F}^{\left(p+1\right)} \,,
\ee
where $\mathbf{F}^{\left(p+1\right)}=\text{d}\mathbf{A}^{\left(p\right)}$ is the $\left(p+1\right)$-form field strength tensor, and $\star$ is the Hodge dual operation. In the traditional index notation,
\be
	F_{\mu_1\mu_2\dots\mu_{p+1}} = \left(p+1\right)\partial_{[\mu_1}A_{\mu_2\dots\mu_{p+1}]}
\ee
and the $p$-form action reads
\be
	S^{\left(p\right)} = -\frac{1}{2\left(p+1\right)!}\int d^{d}x\sqrt{-g}\,F_{\mu_1\mu_2\dots\mu_{p+1}}F^{\mu_1\mu_2\dots\mu_{p+1}} \,.
\ee

The $2+\left(d-2\right)$ decomposition of the $p$-form gauge field into irreducible representations of $SO\left(d-1\right)$ is now achieved via the Hodge decomposition on $\mathbb{S}^{d-2}$. Such a perturbation analysis has been developed in Ref.~\cite{Yoshida:2019tvk} which we extend here to the more general spherically symmetric background geometry with $f_{t}\ne f_{r}$. We will adopt the notation of Ref.~\cite{Yoshida:2019tvk} and distinguish a $p$-form on $\mathbb{S}^{d-2}$ from a $p$-form on the full manifold by hatting it. For instance,
\be
	\hat{\mathbf{A}}^{\left(p\right)} \equiv \frac{1}{p!}A_{A_1 A_2\dots A_{p}}\,\text{d}\theta^{A_1}\wedge\text{d}\theta^{A_2}\wedge\dots\wedge\text{d}\theta^{A_{p}} \,.
\ee
The spacetime $p$-form gauge field $\mathbf{A}^{\left(p\right)}$ can then at a first step be tensorially decomposed as~\cite{Yoshida:2019tvk}
\be
	\mathbf{A}^{\left(p\right)} = \frac{1}{2}\text{d}x^{a}\wedge\text{d}x^{b}\wedge\hat{\mathbf{T}}_{ab}^{\left(p-2\right)} + \text{d}x^{a}\wedge\hat{\mathbf{V}}_{a}^{\left(p-1\right)} + \hat{\mathbf{X}}^{\left(p\right)} \,,
\ee
where the components of the forms $\hat{\mathbf{T}}_{ab}^{\left(p-2\right)}$, $\hat{\mathbf{V}}_{a}^{\left(p-1\right)}$ and $\hat{\mathbf{X}}^{\left(p\right)}$ on the sphere have been identified with the relevant components of the spacetime $p$-form gauge field, that is,
\be
	\left(T_{ab}\right)_{A_1\dots A_{p-2}} \equiv A_{abA_1\dots A_{p-2}} \,,\quad \left(V_{a}\right)_{A_1\dots A_{p-1}} \equiv A_{aA_1\dots A_{p-1}} \quad\text{and}\quad X_{A_1\dots A_{p}} \equiv A_{A_1\dots A_{p}} \,.
\ee
The Hodge decomposition on $\mathbb{S}^{d-2}$ now consists of decomposing a general $p$-form $\hat{\mathbf{A}}^{\left(p\right)}$ on the sphere into a ``longitudinal'' $\left(p-1\right)$-form $\hat{\mathbf{A}}^{\left(p-1\right)}$ and a ``transverse'' $p$-form $\hat{\mathcal{A}}^{\left(p\right)}$. More explicitly,
\be
	\hat{\mathbf{A}}^{\left(p\right)} = \hat{\text{d}}\hat{\mathbf{A}}^{\left(p-1\right)} + \hat{\mathcal{A}}^{\left(p\right)} \,,
\ee
with $\hat{\mathcal{A}}^{\left(p\right)}$ a co-exact $p$-form on $\mathbb{S}^{d-2}$, that is\footnote{The coderivative operator of a $p$-form on a $d$-dimensional spacetime is defined as $\delta\equiv\left(-1\right)^{d\left(p+1\right)+1}\star\text{d}\star$ and, like the exterior derivative, it is also nilpotent, i.e. $\delta^2=0$.} $\hat{\delta}\hat{\mathcal{A}}^{\left(p\right)} = 0$. In components form, this is indeed the tranversality condition,
\be
	\hat{\delta}\hat{\mathcal{A}}^{\left(p\right)} = 0 \Leftrightarrow D^{A_1}\mathcal{A}_{A_1 A_2\dots A_{p+1}} = 0 \,.
\ee
The longitudinal mode can be further Hodge decomposed as $\hat{\mathbf{A}}^{\left(p-1\right)} = \hat{\text{d}}\hat{\mathbf{A}}^{\left(p-2\right)} + \hat{\mathcal{A}}^{\left(p-1\right)}$, with $\hat{\mathcal{A}}^{\left(p-1\right)}$ a co-exact $\left(p-1\right)$-form on the sphere. The nilpotency of the exterior derivative then implies that a general form on $\mathbb{S}^{d-2}$ is expressible entirely by co-exact form fields. The $2+\left(d-2\right)$ decomposition of the $p$-form gauge field into irreducible representations of $SO\left(d-1\right)$ is therefore the following
\be\ba
	\mathbf{A}^{\left(p\right)} &= \frac{1}{2}\text{d}x^{a}\wedge\text{d}x^{b}\wedge\left(\hat{\text{d}}\hat{\mathcal{T}}_{ab}^{\left(p-3\right)}+\hat{\mathcal{T}}_{ab}^{\left(p-2\right)}\right) \\
	&\quad+ \text{d}x^{a}\wedge\left(\hat{\text{d}}\hat{\mathcal{V}}_{a}^{\left(p-2\right)}+\hat{\mathcal{V}}_{a}^{\left(p-1\right)}\right) \\
	&\quad+ \hat{\text{d}}\hat{\mathcal{X}}^{\left(p-1\right)} + \hat{\mathcal{X}}^{\left(p\right)} \,,
\ea\ee
where all the forms that appear are now co-exact on $\mathbb{S}^{d-2}$.

The $p$-form action is invariant under the gauge transformations $\delta_{\Lambda}\mathbf{A}^{\left(p\right)} = \text{d}\Lambda^{\left(p-1\right)}$. After decomposing the gauge parameter $\left(p-1\right)$-form into co-exact forms on the sphere, one can work out that
\be\ba
	\text{d}\Lambda^{\left(p-1\right)} &= \frac{1}{2}\text{d}x^{a}\wedge\text{d}x^{b}\wedge\left(2\,\hat{\text{d}}\hat{\varLambda}_{ab}^{\left(p-3\right)}+D_{a}\hat{\varLambda}_{b}^{\left(p-2\right)}\right) \\
	&\quad+ \text{d}x^{a}\wedge\left(-\hat{\text{d}}\hat{\varLambda}_{a}^{\left(p-2\right)} + D_{a}\hat{\varLambda}^{\left(p-1\right)}\right) + \hat{\text{d}}\hat{\varLambda}^{\left(p-1\right)} \,.
\ea\ee
As a result, the $SO\left(d-1\right)$-decomposed components of the $p$-form gauge field transform according to
\be
	\begin{gathered}
		\delta_{\Lambda}\hat{\mathcal{T}}_{ab}^{\left(p-3\right)} = \hat{\varLambda}_{ab}^{\left(p-3\right)} \,,\quad \delta_{\Lambda}\hat{\mathcal{V}}_{a}^{\left(p-2\right)} = -\hat{\varLambda}_a^{\left(p-2\right)} \,,\quad \delta_{\Lambda}\hat{\mathcal{X}}^{\left(p-1\right)} = \hat{\varLambda}^{\left(p-1\right)} \,, \\
		\delta_{\Lambda}\hat{\mathcal{T}}_{ab}^{\left(p-2\right)} = 2D_{[a}\hat{\varLambda}_{b]}^{\left(p-2\right)} \,,\quad \delta_{\Lambda}\hat{\mathcal{V}}_{a}^{\left(p-1\right)} = D_{a}\hat{\varLambda}^{\left(p-1\right)} \,,\quad \delta_{\Lambda}\hat{\mathcal{X}}^{\left(p\right)} = 0 \,.
	\end{gathered}
\ee
The first line shows that the longitudinal modes are pure gauge. Instead of fixing the gauge, however, we will directly work with gauge invariant combinations. In particular, we can rearrange the independent degrees of freedom into the gauge invariant co-exact $p$-form on the sphere, $\hat{\mathcal{X}}^{\left(p\right)}$, plus the following gauge invariant combinations
\be
	\hat{\mathcal{H}}_{ab}^{\left(p-2\right)} = \hat{\mathcal{T}}_{ab}^{\left(p-2\right)} + 2D_{[a}\hat{\mathcal{V}}_{b]}^{\left(p-2\right)} \,,\quad \hat{\mathcal{A}}_{a}^{\left(p-1\right)} = \hat{\mathcal{V}}_{a}^{\left(p-1\right)} - D_{a}\hat{\mathcal{X}}^{\left(p-1\right)} \,.
\ee
In terms of these, the field strength $\left(p+1\right)$-form is written as
\be\ba
	\mathbf{F}^{\left(p+1\right)} &= \frac{1}{2}\text{d}x^{a}\wedge\text{d}x^{b}\wedge\left(2D_{a}\hat{\mathcal{A}}_{b}^{\left(p-1\right)} + \left(p-2\right)!\,\hat{\text{d}}\hat{\mathcal{H}}_{ab}^{\left(p-2\right)}\right) \\
	&\quad+ \text{d}x^{a}\wedge\left(D_{a}\hat{\mathcal{X}}^{\left(p\right)} - \hat{\text{d}}\hat{\mathcal{A}}_{a}^{\left(p-1\right)}\right) + \hat{\text{d}}\hat{\mathcal{X}}^{\left(p\right)} \,.
\ea\ee

We can now expand into co-exact $p$-form spherical harmonics $Y^{\left(\text{T}\right)A_1 \dots A_{p}}_{\ell,\mathbf{m}}\left(\theta\right)$ on the sphere,
\be\ba
	\left(\mathcal{H}^{ab}\right)^{A_1\dots A_{p-2}}\left(x\right) &= \sum_{\ell,\mathbf{m}}\mathcal{H}^{ab}_{\ell,\mathbf{m}}\left(t,r\right)Y^{\left(\text{T}\right)A_1 \dots A_{p-2}}_{\ell,\mathbf{m}}\left(\theta\right) \,, \\
	\left(\mathcal{A}^{a}\right)^{A_1\dots A_{p-1}}\left(x\right) &= \sum_{\ell,\mathbf{m}}\mathcal{A}^{a}_{\ell,\mathbf{m}}\left(t,r\right)Y^{\left(\text{T}\right)A_1 \dots A_{p-1}}_{\ell,\mathbf{m}}\left(\theta\right) \,, \\
	\mathcal{X}^{A_1\dots A_{p}}\left(x\right) &= \sum_{\ell,\mathbf{m}}\mathcal{X}_{\ell,\mathbf{m}}\left(t,r\right)Y^{\left(\text{T}\right)A_1 \dots A_{p}}_{\ell,\mathbf{m}}\left(\theta\right) \,.
\ea\ee
The important property of the co-exact $p$-form spherical harmonics besides their transversality, $D_{A_1}Y^{\left(\text{T}\right)A_1 \dots A_{p}}_{\ell,\mathbf{m}}=0$, is that they satisfy the eigenvalue problem\footnote{More information on the co-exact $p$-form spherical harmonics can be found in~\cite{Yoshida:2019tvk,Camporesi1994}.}
\be
	D_{B}D^{B}Y^{\left(\text{T}\right)A_1 \dots A_{p}}_{\ell,\mathbf{m}} = -\left[\ell\left(\ell+d-3\right)-p\right]Y^{\left(\text{T}\right)A_1 \dots A_{p}}_{\ell,\mathbf{m}} \,.
\ee

The $p$-form action after this expansion then reduces to
\be
	\begin{gathered}
		S^{\left(p\right)} = \sum_{\ell,\mathbf{m}}\left(S^{\left(p\right)}_{\ell,\mathbf{m}} + S^{\left(p-1\right)}_{\ell,\mathbf{m}} + S^{\left(p-2\right)}_{\ell,\mathbf{m}}\right) \,, \\
		\ba
			S^{\left(p\right)}_{\ell,\mathbf{m}} &= \int d^2x\sqrt{-g^{\left(2\right)}}\,r^{d-2p-2} \left[-\frac{1}{2p!}D_{a}\bar{\mathcal{X}}_{\ell,\mathbf{m}}D^{a}\mathcal{X}_{\ell,\mathbf{m}} - \frac{1}{2p!}\frac{\left(\ell+p\right)\left(\ell+d-p-3\right)}{r^2}\left|\mathcal{X}_{\ell,\mathbf{m}}\right|^2 \right] \,, \\
			S^{\left(p-1\right)}_{\ell,\mathbf{m}} &= \int d^2x\sqrt{-g^{\left(2\right)}}\,r^{d-2p} \bigg[-\frac{1}{4\left(p-1\right)!}\bar{\mathcal{F}}_{ab;\ell,\mathbf{m}}\mathcal{F}^{ab}_{\ell,\mathbf{m}} \\
			&\qquad\qquad\qquad\qquad\qquad\quad- \frac{1}{2\left(p-1\right)!}\frac{\left(\ell+p-1\right)\left(\ell+d-p-2\right)}{r^2}\bar{\mathcal{A}}_{a;\ell,\mathbf{m}}\mathcal{A}^{a}_{\ell,\mathbf{m}}\bigg] \,, \\
			S^{\left(p-2\right)}_{\ell,\mathbf{m}} &= \int d^2x\sqrt{-g^{\left(2\right)}}\,r^{d-2p+2} \left[- \frac{\left(p-2\right)!}{4}\frac{\left(\ell+p-2\right)\left(\ell+d-p-1\right)}{r^2}\bar{\mathcal{H}}_{ab;\ell,\mathbf{m}}\mathcal{H}^{ab}_{\ell,\mathbf{m}}\right] \,,
		\ea
	\end{gathered}
\ee
where we have defined
\be
	\mathcal{F}^{ab}_{\ell,\mathbf{m}} \equiv D^{a}\mathcal{A}^{b}_{\ell,\mathbf{m}} - D^{b}\mathcal{A}^{a}_{\ell,\mathbf{m}} \,.
\ee
The first thing to observe is that the $\left(p-2\right)$-form sector generated by the spherical harmonic modes $\mathcal{H}^{ab}_{\ell,\mathbf{m}}$ is trivial,
\be
	\mathcal{H}^{ab}_{\ell,\mathbf{m}} = 0 \,.
\ee

\subsubsection{$p$-form modes}
Similar to the case of spin-$1$ perturbations, we start with the simplest, $p$-form, sector. Performing the field redefinition
\be
	\mathcal{X}_{\ell,\mathbf{m}}\left(t,r\right) = \sqrt{p!}\frac{\Psi^{\left(p\right)}_{\ell,\mathbf{m}}\left(t,r\right)}{r^{\left(d-2p-2\right)/2}} \,,
\ee
the reduced action for $p$-form modes takes the canonical form
\be
	S_{\ell,\mathbf{m}}^{\left(p\right)} = \int d^2x\sqrt{-g^{\left(2\right)}}\left[-\frac{1}{2}D_{a}\bar{\Psi}^{\left(p\right)}_{\ell,\mathbf{m}}D^{a}\Psi^{\left(p\right)}_{\ell,\mathbf{m}} - \frac{1}{2}V^{\left(p\right)}_{\ell}\left(r\right)\left|\Psi^{\left(p\right)}_{\ell,\mathbf{m}}\right|^2\right]
\ee
with potential
\be\ba\label{eq:Vpp}
	V^{\left(p\right)}_{\ell}\left(r\right) &= \frac{\left(\ell+p\right)\left(\ell+d-p-3\right)}{r^2} + \frac{\left(d-2p-2\right)\left(d-2p-4\right)}{4r^2}r_{a}r^{a} + \frac{d-2p-2}{2r}D_{a}r^{a} \\
	&= \frac{\left(\ell+p\right)\left(\ell+d-p-3\right)}{r^2} + \frac{\left(d-2p-2\right)\left(d-2p-4\right)}{4r^2}f_{r} + \frac{d-2p-2}{2r}\frac{\left(f_{t}f_{r}\right)^{\prime}}{2f_{t}} \,.
\ea\ee

\subsubsection{$\left(p-1\right)$-form modes}
For the analysis of the $\left(p-1\right)$-form sector, we introduce an auxiliary $2$-d scalar field $\Psi^{\left(\tilde{p}\right)}_{\ell,\mathbf{m}}\left(t,r\right)$ and consider the action
\be\ba
	\tilde{S}^{\left(p-1\right)}_{\ell,\mathbf{m}} = \int d^2x&\sqrt{-g^{\left(2\right)}} \bigg[\frac{1}{2}\sqrt{\frac{\left(\ell+p-1\right)\left(\ell+d-p-2\right)}{\left(p-1\right)!}}\,r^{\left(d-2p-2\right)/2}\text{Re}\left\{\bar{\Psi}^{\left(\tilde{p}\right)}_{\ell,\mathbf{m}}\varepsilon_{ab}\mathcal{F}^{ab}_{\ell,\mathbf{m}}\right\} \\
	&\quad- \frac{1}{2}\frac{\left(\ell+p-1\right)\left(\ell+d-p-2\right)}{r^2}\left(\left|\Psi^{\left(\tilde{p}\right)}_{\ell,\mathbf{m}}\right|^2 + \frac{r^{d-2p}}{\left(p-1\right)!}\bar{\mathcal{A}}_{a;\ell,\mathbf{m}}\mathcal{A}^{a}_{\ell,\mathbf{m}}\right)\bigg] \,,
\ea\ee
which is classically equivalent to the original action $S^{\left(p-1\right)}_{\ell,\mathbf{m}}$; a fact that becomes evident after putting the auxiliary field on-shell,
\be
	\Psi^{\left(\tilde{p}\right)}_{\ell,\mathbf{m}} = \frac{1}{2}\frac{r^{\left(d-2p+2\right)/2}}{\sqrt{\left(p-1\right)!\left(\ell+p-1\right)\left(\ell+d-p-2\right)}}\varepsilon_{ab}\mathcal{F}^{ab}_{\ell,\mathbf{m}} \,.
\ee

Integrating out $\mathcal{A}^{a}_{\ell,\mathbf{m}}$ in $\tilde{S}^{\left(p-1\right)}_{\ell,\mathbf{m}}$ instead,
\be
	\mathcal{A}^{a}_{\ell,\mathbf{m}} = \sqrt{\frac{\left(p-1\right)!}{\left(\ell+p-1\right)\left(\ell+d-p-2\right)}}r^{-\left(d-2p-2\right)/2}\left[\varepsilon^{ab}D_{b} - t^{a}\frac{d-2p-2}{2r}\sqrt{\frac{f_{r}}{f_{t}}}\right]\Psi^{\left(\tilde{p}\right)}_{\ell,\mathbf{m}} \,,
\ee
results to the canonically normalized reduced action
\be
	\tilde{S}^{\left(p-1\right)}_{\ell,\mathbf{m}} = \int d^2x\sqrt{-g^{\left(2\right)}}\left[-\frac{1}{2}D_{a}\bar{\Psi}^{\left(\tilde{p}\right)}_{\ell,\mathbf{m}}D^{a}\Psi^{\left(\tilde{p}\right)}_{\ell,\mathbf{m}} - \frac{1}{2}V^{\left(\tilde{p}\right)}_{\ell}\left(r\right)\left|\Psi^{\left(\tilde{p}\right)}_{\ell,\mathbf{m}}\right|^2\right] \,,
\ee
with the potential given by
\be\ba\label{eq:Vpph}
	V^{\left(\tilde{p}\right)}_{\ell}\left(r\right) &= \frac{\left(\ell+p-1\right)\left(\ell+d-p-2\right)}{r^2} + \frac{\left(d-2p\right)\left(d-2p-2\right)}{4r^2}r_{a}r^{a} - \frac{d-2p-2}{2r}D_{a}r^{a} \\
	&= \frac{\left(\ell+p-1\right)\left(\ell+d-p-2\right)}{r^2} + \frac{\left(d-2p\right)\left(d-2p-2\right)}{4r^2}f_{r} - \frac{d-2p-2}{2r}\frac{\left(f_{t}f_{r}\right)^{\prime}}{2f_{t}} \,.
\ea\ee
In fact, this is the same as the potential in Eq.~\eqref{eq:Vpp} for the $p$-form modes after replacing the rank of the $p$-form with its dual on the sphere,
\be
	p \rightarrow \tilde{p} = d-p-2 \,.
\ee
More collectively, the $p$-form perturbation potential can be rewritten as
\be\ba\label{eq:Vpj}
	V^{\left(j\right)}_{\ell}\left(r\right) &= \frac{\left(\ell+j\right)\left(\ell+d-j-3\right)}{r^2} + \frac{\left(d-2j-2\right)\left(d-2j-4\right)}{4r^2}r_{a}r^{a} + \frac{d-2j-2}{2r}D_{a}r^{a} \\
	&= \frac{\left(\ell+j\right)\left(\ell+d-j-3\right)}{r^2} + \frac{\left(d-2j-2\right)\left(d-2j-4\right)}{4r^2}f_{r} + \frac{d-2j-2}{2r}\frac{\left(f_{t}f_{r}\right)^{\prime}}{2f_{t}} \,,
\ea\ee
where the index $j$ is either equal to $p$, for the $p$-form perturbation modes, or equal to $\tilde{p}=d-p-2$, for the $\left(p-1\right)$-form perturbation modes. This nicely also captures the spin-$1$ vector ($j=1$) and scalar ($j=d-3$) sectors of Eq.~\eqref{eq:V1V} and Eq.~\eqref{eq:V1S} respectively, as well as the spin-$0$ scalar ($j=0$) sector of Eq.~\eqref{eq:V0S}.

\subsection{Spin-$2$ perturbations}
\label{sec:EOM_Seq2}

Before writing down the relevant action for gravitational (spin-$2$) perturbations, let us first study the decomposition of the metric perturbations $h_{\mu\nu}$ into irreducible $SO\left(d-1\right)$ representation and the construction of gauge invariants. The $\frac{d\left(d+1\right)}{2}$ components of $h_{\mu\nu}$ are rearranged according to
\be\ba
	h_{ab}\left(x\right) &= H_{ab}\left(x\right) \,, \\
	h_{aA}\left(x\right) &= D_{A}H^{\left(\text{S}\right)}_{a}\left(x\right) + h^{\left(\text{V}\right)}_{aA}\left(x\right) \,, \\
	h_{AB}\left(x\right) &= r^2\left(K\left(x\right)\Omega_{AB} + D_{\langle A}D_{B\rangle}G\left(x\right) + D_{(A}h^{\left(\text{V}\right)}_{B)}\left(x\right) + h_{AB}^{\left(\text{TT}\right)}\left(x\right) \right)
\ea\ee
into seven $SO\left(d-1\right)$ scalars
\be
	H_{ab}\left(x\right) \,,\quad H^{\left(\text{L}\right)}_{a}\left(x\right) \,,\quad K\left(x\right) \,\quad\text{and} \,\quad G\left(x\right) \,,
\ee
three $SO\left(d-1\right)$ transverse vectors carrying $d-3$ degrees of freedom each,
\be
	\begin{gathered}
		h^{\left(\text{V}\right)}_{aA}\left(x\right) \,\quad\text{and}\,\quad h^{\left(\text{V}\right)}_{A}\left(x\right) \,, \\
		D^{A}h_{aA}^{\left(\text{V}\right)}\left(x\right) = 0 \,,\quad  D^{A}h_{A}^{\left(\text{V}\right)}\left(x\right) = 0 \,,
	\end{gathered}
\ee
and one $SO\left(d-1\right)$ transverse symmetric tracefree tensor carrying $\frac{\left(d-1\right)\left(d-4\right)}{2}$ degrees of freedom,
\be
	h^{\left(\text{TT}\right)}_{AB}\left(x\right) \,,\quad D^{A}h^{\left(\text{TT}\right)}_{AB}\left(x\right) = 0 \,,\quad \Omega^{AB}h^{\left(\text{TT}\right)}_{AB}\left(x\right) = 0 \,.
\ee
In four-spacetime dimensions, there is no analogue of $h^{\left(\text{TT}\right)}_{AB}\left(x\right)$, which vanishes identically.

Under infinitesimal diffeomorphisms $x^{\mu}\rightarrow x^{\mu}+\xi^{\mu}\left(x\right)$, the metric perturbations transform according to
\be
	\delta_{\xi}h_{\mu\nu} = \nabla_{\mu}\xi_{\nu} + \nabla_{\nu}\xi_{\mu} \,.
\ee
Decomposing the $d$ gauge parameters $\xi_{\mu}$ into $SO\left(d-1\right)$ irreducible representations to three scalars, $\xi_{a}$ and $\xi^{\left(\text{S}\right)}$, and one transverse vector, $\xi^{\left(\text{V}\right)}_{A}$, $D^{A}\xi^{\left(\text{V}\right)}_{A}=0$,
\be
	\xi_{a}\left(x\right) \,,\quad \xi_{A}\left(x\right) = D_{A}\xi^{\left(\text{S}\right)}\left(x\right) + \xi^{\left(\text{V}\right)}\left(x\right) \,,
\ee
the gauge transformation properties of the various $SO\left(d-1\right)$-decomposed components of $h_{\mu\nu}$ can be read to be
\be\ba
	\delta_{\xi}H_{ab} &= D_{a}\xi_{b} + D_{b}\xi_{a} \,,\quad \delta_{\xi}H^{\left(\text{S}\right)}_{a} = \xi_{a} + D_{a}\xi^{\left(\text{S}\right)} - \frac{2}{r}r_{a}\xi^{\left(\text{S}\right)} \,, \\
	\delta_{\xi}K &= \frac{2}{r}r^{a}\xi_{a} + \frac{2}{d-2}\frac{1}{r^2}D_{A}D^{A}\xi^{\left(\text{S}\right)} \,,\quad \delta_{\xi}G = \frac{2}{r^2}\xi^{\left(\text{S}\right)} \,, \\
	\delta_{\xi}h^{\left(\text{V}\right)}_{aA} &= D_{a}\xi^{\left(\text{V}\right)}_{A} - \frac{2}{r}r_{a}\xi^{\left(\text{V}\right)}_{A} \,,\quad \delta_{\xi}h^{\left(\text{V}\right)}_{A} = \frac{2}{r^2}\xi^{\left(\text{V}\right)}_{A} \,,\quad \delta_{\xi}h^{\left(\text{TT}\right)}_{AB} = 0 \,.
\ea\ee
One sees, in particular, that the transverse symmetric tracefree tensor is gauge invariant, while $H_{a}^{\left(\text{S}\right)}$, $G$ and $h_{a}^{\left(\text{V}\right)}$ are redundant degrees of freedom. Instead of fixing the gauge, let us work with gauge invariant quantities. For tensor modes, this is just the transverse symmetric tracefree tensor $h^{\left(\text{TT}\right)}_{AB}$. For vector modes, the gauge invariant combination is
\be
	\mathcal{H}^{\left(\text{V}\right)}_{aA} = h^{\left(\text{V}\right)}_{aA} - \frac{1}{2}r^2D_{a}h^{\left(\text{V}\right)}_{A} \,.
\ee
Last, for scalar modes, there are two sets of gauge invariant combinations,
\be\ba
	\mathcal{H}_{ab} &= H_{ab} - 2D_{(a}H^{\left(\text{S}\right)}_{b)} + D_{(a}\left(r^2D_{b)}G\right) \,, \\
	\mathcal{K} &= K - \frac{1}{d-2}D_{A}D^{A}G + rr^{a}D_{a}G - \frac{2}{r}r^{a}H^{\left(\text{S}\right)}_{a} \,.
\ea\ee

In performing the $2+\left(d-2\right)$ decomposition of the field, we expand in scalar, transverse vector and transverse symmetric tracefree tensor spherical harmonics~\cite{Hui:2020xxx,Chodos:1983zi,Higuchi:1986wu},
\be\ba
	\mathcal{H}_{ab}\left(x\right) &= \sum_{\ell,\mathbf{m}}\mathcal{H}_{ab;\ell,\mathbf{m}}\left(t,r\right)Y_{\ell,\mathbf{m}}\left(\theta\right) \,, \\
	\mathcal{K}\left(x\right) &= \sum_{\ell,\mathbf{m}}\mathcal{K}_{\ell,\mathbf{m}}\left(t,r\right)Y_{\ell,\mathbf{m}}\left(\theta\right) \,, \\
	\mathcal{H}^{\left(\text{V}\right)}_{aA}\left(x\right) &= \sum_{\ell,\mathbf{m}}\mathcal{H}_{a;\ell,\mathbf{m}}\left(t,r\right)Y^{\left(\text{T}\right)}_{A;\ell,\mathbf{m}}\left(\theta\right) \,, \\
	h^{\left(\text{TT}\right)}_{AB}\left(x\right) &= \sum_{\ell,\mathbf{m}}h^{\left(\text{T}\right)}_{\ell,\mathbf{m}}\left(t,r\right)Y^{\left(\text{TT}\right)}_{AB;\ell,\mathbf{m}}\left(\theta\right) \,.
\ea\ee

We now look at an explicit action. Solely on the premises of working with equations of motion that are at most second-order in the derivatives, the most general such local theory of gravity is Lovelock gravity~\cite{Lovelock:1971yv}. Treating General Relativity as a low-energy effective field theory, one can write down an infinite number of higher-order curvature corrections in the gravity action~\cite{Donoghue:2017pgk}. As an elementary analysis though, we will focus here to General Relativity, described by the Einstein-Hilbert action. Perturbations around an asymptotically flat vacuum background will then be described by the massless Fierz-Pauli action,
\be
	S^{\left(\text{gr}\right)} = \int d^{d}x\sqrt{-g}\left[-\frac{1}{2}\nabla_{\rho}h_{\mu\nu}\nabla^{\rho}h^{\mu\nu} + \nabla_{\rho}h_{\mu\nu}\nabla^{\nu}h^{\mu\rho} - \nabla_{\mu}h\nabla_{\nu}h^{\mu\nu} + \frac{1}{2}\nabla_{\mu}h\nabla^{\mu}h\right] \,,
\ee
where we are using canonical variables, i.e. the perturbed metric around a background $g_{\mu\nu}$ is $g_{\mu\nu}^{\text{full}} = g_{\mu\nu} + \sqrt{32\pi G}h_{\mu\nu}$. In the presence of matter and other radiation fields, e.g. for a charged black hole, one should furthermore add the corresponding perturbations in the above action, which will also involve coupling of background stress energy-momentum tensor to gravitational perturbations.

Let us ignore for the moment other fields in the system and focus to this pure gravity quadratic action. These other fields would ultimately modify the potentials we will present below by additive pieces and also result in sources in the equations of motion. Inserting the spherical harmonic expansions of the metric perturbations as described above and after a few manipulations, we find the following decoupling of the tensor (``$\left(\text{T}\right)$''), vector (``$\left(\text{RW}\right)$'') and scalar (``$\left(\text{Z}\right)$'') modes
\be
	\begin{gathered}
		S^{\left(\text{gr}\right)} = \sum_{\ell,\mathbf{m}}\left(S^{\left(\text{T}\right)}_{\ell,\mathbf{m}} + S^{\left(\text{RW}\right)}_{\ell,\mathbf{m}} + S^{\left(\text{Z}\right)}_{\ell,\mathbf{m}}\right) \\
		\ba
			S^{\left(\text{T}\right)}_{\ell,\mathbf{m}} &= \int d^2x\sqrt{-g^{\left(2\right)}}\,r^{d-2} \bigg[-\frac{1}{2}D_{a}\bar{h}^{\left(\text{T}\right)}_{\ell,\mathbf{m}}D^{a}h^{\left(\text{T}\right)}_{\ell,\mathbf{m}} - -\frac{1}{r}r^{a}D_{a}\left|h^{\left(\text{T}\right)}_{\ell,\mathbf{m}}\right|^2 \\
			&\quad\quad\quad\quad\quad\quad\quad\quad\quad\quad- \frac{1}{2}\frac{\ell\left(\ell+d-3\right)+2\left(d-3\right)}{r^2}\left|h^{\left(\text{T}\right)}_{\ell,\mathbf{m}}\right|^2\bigg] \,, \\
			S^{\left(\text{RW}\right)}_{\ell,\mathbf{m}} &= \int d^2x\sqrt{-g^{\left(2\right)}}\,2r^{d-4} \bigg[ -\frac{1}{4}\mathcal{F}_{ab;\ell,\mathbf{m}}\mathcal{F}^{ab}_{\ell,\mathbf{m}}-\frac{2}{r}r_{a}\text{Re}\left\{\bar{\mathcal{H}}^{b}_{\ell,\mathbf{m}}D_{b}\mathcal{H}^{a}_{\ell,\mathbf{m}}\right\} \\
			&\quad\quad\quad\quad-\frac{1}{2}\left(\frac{\left(\ell+1\right)\left(\ell+d-4\right)}{r^2}\bar{\mathcal{H}}_{a;\ell,\mathbf{m}}\mathcal{H}^{a}_{\ell,\mathbf{m}}-4\left|r_{a}\mathcal{H}^{a}_{\ell,\mathbf{m}}\right|^2\right) \bigg] \,,
		\ea
	\end{gathered}
\ee
\be\ba
	{}&S^{\left(\text{Z}\right)}_{\ell,\mathbf{m}} = \int d^2x\sqrt{-g^{\left(2\right)}}\,r^{d-2} \bigg[ -\frac{1}{2}D_{c}\bar{\mathcal{H}}_{ab;\ell,\mathbf{m}}D^{c}\mathcal{H}^{ab}_{\ell,\mathbf{m}} + D_{c}\bar{\mathcal{H}}_{ab;\ell,\mathbf{m}}D^{b}\mathcal{H}^{ac}_{\ell,\mathbf{m}} - \text{Re}\left\{D_{a}\bar{\mathcal{H}}_{\ell,\mathbf{m}}D_{b}\mathcal{H}^{ab}_{\ell,\mathbf{m}}\right\} \\
	&+ \frac{1}{2}D_{a}\bar{\mathcal{H}}_{\ell,\mathbf{m}}D^{a}\mathcal{H}_{\ell,\mathbf{m}} + \frac{\left(d-2\right)\left(d-3\right)}{2}D_{a}\bar{\mathcal{K}}_{\ell,\mathbf{m}}D^{a}\mathcal{K}_{\ell,\mathbf{m}} - \left(d-2\right)\text{Re}\left\{D_{a}\bar{\mathcal{K}}_{\ell,\mathbf{m}}\left(D_{b}\mathcal{H}^{ab}_{\ell,\mathbf{m}}-D^{a}\mathcal{H}_{\ell,\mathbf{m}}\right)\right\} \\
	&-\frac{d-2}{r}\text{Re}\left\{\left(D_{a}\bar{\mathcal{H}}_{\ell,\mathbf{m}}+\left(d-4\right)D_{a}\bar{\mathcal{K}}_{\ell,\mathbf{m}}\right)\left(r_{b}\mathcal{H}^{ab}_{\ell,\mathbf{m}}-r^{a}\mathcal{K}_{\ell,\mathbf{m}}\right)\right\} \\
	&-\frac{1}{2}\frac{\ell\left(\ell+d-3\right)}{r^2}\left( \bar{\mathcal{H}}_{ab;\ell,\mathbf{m}}\mathcal{H}^{ab}_{\ell,\mathbf{m}} - \left|\mathcal{H}_{\ell,\mathbf{m}}\right|^2 - \left(d-3\right)\left(d-4\right)\left|\mathcal{K}_{\ell,\mathbf{m}}\right|^2 - 2\left(d-3\right)\text{Re}\left\{\bar{\mathcal{H}}_{\ell,\mathbf{m}}\mathcal{K}_{\ell,\mathbf{m}}\right\} \right) \bigg] \,.
\ea\ee
In the above expressions, we have introduced the notation
\be
	\mathcal{F}^{ab}_{\ell,\mathbf{m}} \equiv D^{a}\mathcal{H}^{b}_{\ell,\mathbf{m}}-D^{b}\mathcal{H}^{a}_{\ell,\mathbf{m}} \,,\quad \mathcal{H}_{\ell,\mathbf{m}}\equiv g_{ab}\mathcal{H}^{ab}_{\ell,\mathbf{m}} \,.
\ee

\subsubsection{Tensor modes}
We begin with the easier case of the tensor modes and perform the field redefinition
\be
	h^{\left(\text{T}\right)}_{\ell,\mathbf{m}} = \frac{\Psi^{\left(\text{T}\right)}_{\ell,\mathbf{m}}}{r^{\left(d-2\right)/2}} \,.
\ee
The resulting action after integration by parts takes the canonical form
\be
	S^{\left(\text{T}\right)}_{\ell,\mathbf{m}} = \int d^2x\sqrt{-g^{\left(2\right)}}\left[ -\frac{1}{2}D_{a}\bar{\Psi}^{\left(\text{T}\right)}_{\ell,\mathbf{m}}D^{a}\Psi^{\left(\text{T}\right)}_{\ell,\mathbf{m}} - \frac{1}{2}V^{\left(\text{T}\right)}_{\ell}\left(r\right)\left|\Psi^{\left(\text{T}\right)}_{\ell,\mathbf{m}}\right|^2 \right] \,,
\ee
with the tensor modes potential given by
\be\ba\label{eq:V2T}
	V^{\left(\text{T}\right)}_{\ell}\left(r\right) &= \frac{\ell\left(\ell+d-3\right)+2\left(d-3\right)}{r^2}+\frac{d^2-14d+32}{4r^2}r_{a}r^{a}+\frac{d-6}{2r}D_{a}r^{a} \\
	&= \frac{\ell\left(\ell+d-3\right)+2\left(d-3\right)}{r^2}+\frac{d^2-14d+32}{4r^2}f_{r}+\frac{d-6}{2r}\frac{\left(f_{t}f_{r}\right)^{\prime}}{2f_{t}} \,.
\ea\ee

\subsubsection{Vector (Regge-Wheeler) modes}
Next, for the vector modes we follow a procedure similar to the scalar modes for the spin-$1$ perturbations. We introduce an auxiliary Regge-Wheeler variable $\Psi^{\left(\text{RW}\right)}_{\ell,\mathbf{m}}$ and consider the following action~\cite{Hui:2020xxx}
\be\ba
	\tilde{S}^{\left(\text{RW}\right)}_{\ell,\mathbf{m}} = \int d^2x\sqrt{-g^{\left(2\right)}}&\bigg[ \sqrt{\frac{F_{\ell}\left(r\right)}{2}}r^{\left(d-6\right)/2}\text{Re}\left\{\bar{\Psi}^{\left(\text{RW}\right)}_{\ell,\mathbf{m}}\left(\varepsilon_{ab}\mathcal{F}^{ab}_{\ell,\mathbf{m}}-\frac{4}{r}\sqrt{\frac{f_{r}}{f_{t}}}\,t_{a}\mathcal{H}^{a}_{\ell,\mathbf{m}}\right)\right\} \\
	&-\frac{1}{2}\frac{F_{\ell}\left(r\right)}{r^2}\left(\left|\Psi^{\left(\text{RW}\right)}_{\ell,\mathbf{m}}\right|^2 + 2r^{d-4}\bar{\mathcal{H}}_{a;\ell,\mathbf{m}}\mathcal{H}^{a}_{\ell,\mathbf{m}}\right) \bigg] \,,
\ea\ee
with
\be
	F_{\ell}\left(r\right) \equiv \left(\ell+1\right)\left(\ell+d-4\right) - 2\left(d-3\right)r_{a}r^{a}-2r D_{a}r^{a} \,.
\ee
For Schwarzschild-Tangherlini black holes, $F_{\ell}\left(r\right) = \left(\ell-1\right)\left(\ell+d-2\right)$ becomes a constant.

This alternative action retrieves the original action $S^{\left(\text{RW}\right)}_{\ell,\mathbf{m}}$ for the vector modes after integrating out the auxiliary field,
\be
	\Psi^{\left(\text{RW}\right)}_{\ell,\mathbf{m}} = \frac{r^{\left(d-2\right)/2}}{\sqrt{2F_{\ell}\left(r\right)}}\left[\varepsilon_{ab}\mathcal{F}^{ab}_{\ell,\mathbf{m}}-\frac{4}{r}\sqrt{\frac{f_{r}}{f_{t}}}\,t_{a}\mathcal{H}^{a}_{\ell,\mathbf{m}}\right] \,.
\ee
By integrating out the fields $\mathcal{H}^{a}_{\ell,\mathbf{m}}$ instead,
\be
	\mathcal{H}^{a}_{\ell,\mathbf{m}} = \frac{r^{-\left(d-6\right)/2}}{\sqrt{2F_{\ell}\left(r\right)}}\left[\varepsilon^{ab}D_{b} - t^{a}\left(\frac{d-2}{2r}+\frac{F_{\ell}^{\prime}\left(r\right)}{2F_{\ell}\left(r\right)}\right)\sqrt{\frac{f_{r}}{f_{t}}}\right]\Psi^{\left(\text{RW}\right)}_{\ell,\mathbf{m}} \,,
\ee
we end up with a canonically normalized action for the field $\Psi^{\left(\text{RW}\right)}_{\ell,\mathbf{m}}$,
\be
	\tilde{S}^{\left(\text{RW}\right)}_{\ell,\mathbf{m}} = \int d^2x\sqrt{-g^{\left(2\right)}}\left[-\frac{1}{2}D_{a}\bar{\Psi}^{\left(\text{RW}\right)}_{\ell,\mathbf{m}}D^{a}\Psi^{\left(\text{RW}\right)}_{\ell,\mathbf{m}} - \frac{1}{2}V^{\left(\text{RW}\right)}_{\ell}\left(r\right)\left|\Psi^{\left(\text{RW}\right)}_{\ell,\mathbf{m}}\right|^2\right] \,,
\ee
with the Regge-Wheeler potential given by
\be\ba\label{eq:V2RW}
	V^{\left(\text{RW}\right)}_{\ell}\left(r\right) &= \frac{\left(\ell+1\right)\left(\ell+d-4\right)}{r^2} + \frac{\left(d-4\right)\left(d-6\right)}{4r^2}r_{a}r^{a} - \frac{d+2}{2r}D_{a}r^{a} \\
	&+\left[\frac{F_{\ell}^{\prime}}{2F_{\ell}}\left(\frac{d-4}{r}+\frac{F_{\ell}^{\prime}}{2F_{\ell}}\right)-\left(\frac{F_{\ell}^{\prime}}{2F_{\ell}}\right)^{\prime}\right]r_{a}r^{a} - \frac{F_{\ell}^{\prime}}{2F_{\ell}}D_{a}r^{a} \,.
\ea\ee
For Schwarzschild-Tangherlini black holes, the terms in the second line are zero.

\subsubsection{Scalar (Zerilli) modes}
The first observation to find the master variable relevant for the gravitoelectric response of the black hole is that the scalar modes $\mathcal{H}^{ab}_{\ell,\mathbf{m}}$ are, in fact, auxiliary fields. To see this, we break down the three independent components of $\mathcal{H}^{ab}_{\ell,\mathbf{m}}$ into the trace $\mathcal{H}_{\ell,\mathbf{m}}=g_{ab}\mathcal{H}^{ab}_{\ell,\mathbf{m}}$ and two fields $A_{\ell,\mathbf{m}}$ and $B_{\ell,\mathbf{m}}$ that compose the traceless part as
\be
	\mathcal{H}^{ab}_{\ell,\mathbf{m}} = A_{\ell,\mathbf{m}}I_{\left(1\right)}^{\braket{ab}}+B_{\ell,\mathbf{m}}I_{\left(2\right)}^{\braket{ab}} + \frac{1}{2}g^{ab}\mathcal{H}_{\ell,\mathbf{m}} \,,
\ee
with
\be
	I_{\left(1\right)}^{\braket{ab}} = \frac{1}{f_{t}}t^{a}t^{b}+\frac{1}{f_{r}}r^{a}r^{b} \,,\quad I_{\left(2\right)}^{\braket{ab}} = \frac{2}{\sqrt{f_{t}f_{r}}}t^{(a}r^{b)}
\ee
the two independent STF rank-2 tensors in $2$-d, satisfying
\be
	\begin{gathered}
		I_{\left(1\right)\braket{ab}}I_{\left(1\right)}^{\braket{ac}} = \delta_{b}^{c} \,,\quad I_{\left(1\right)\braket{ab}}I_{\left(2\right)}^{\braket{ac}} = \varepsilon_{b}^{\,\,\,c} \,,\quad I_{\left(2\right)\braket{ab}}I_{\left(2\right)}^{\braket{ac}} = -\delta_{b}^{c} \,, \\
		D^{c}I_{\left(1\right)}^{\braket{ab}} = -\frac{f_{t}^{\prime}}{f_{t}}\sqrt{\frac{f_{r}}{f_{t}}}\,t^{c}I^{\braket{ab}}_{\left(2\right)} \,,\quad D^{c}I_{\left(2\right)}^{\braket{ab}} = -\frac{f_{t}^{\prime}}{f_{t}}\sqrt{\frac{f_{r}}{f_{t}}}\,t^{c}I^{\braket{ab}}_{\left(1\right)} \,.
	\end{gathered}
\ee
Then the kinetic term of the scalar modes $\mathcal{H}^{ab}_{\ell,\mathbf{m}}$ turns out to include no time-derivatives,
\be\ba
	-\frac{1}{2}D_{c}\bar{\mathcal{H}}_{ab;\ell,\mathbf{m}}D^{c}\mathcal{H}^{ab}_{\ell,\mathbf{m}} + D_{c}\bar{\mathcal{H}}_{ab;\ell,\mathbf{m}}D^{b}\mathcal{H}^{ac}_{\ell,\mathbf{m}} & \\
	- \text{Re}\left\{D_{a}\bar{\mathcal{H}}_{\ell,\mathbf{m}}D_{b}\mathcal{H}^{ab}_{\ell,\mathbf{m}}\right\} + \frac{1}{2}D_{a}\bar{\mathcal{H}}_{\ell,\mathbf{m}}D^{a}\mathcal{H}_{\ell,\mathbf{m}} &\supset -2\varepsilon_{ab}\text{Re}\left\{D^{a}\bar{A}_{\ell,\mathbf{m}}D^{b}B_{\ell,\mathbf{m}}\right\} \,,
\ea\ee
deeming all components of $\mathcal{H}^{ab}_{\ell,\mathbf{m}}$ auxiliary. More explicitly, the full Zerilli action for the scalar modes reads, after some integrations by parts,
\be
	S^{\left(\text{Z}\right)}_{\ell,\mathbf{m}} = \int d^2x\sqrt{-g^{\left(2\right)}}\,r^{d-2}\left(\mathcal{L}^{\left(AB\right)}_{\ell,\mathbf{m}} + \mathcal{L}^{\left(KK\right)}_{\ell,\mathbf{m}} + \mathcal{L}^{\left(HH\right)}_{\ell,\mathbf{m}} + \mathcal{L}^{\left(ABK\right)}_{\ell,\mathbf{m}} + \mathcal{L}^{\left(ABH\right)}_{\ell,\mathbf{m}} + \mathcal{L}^{\left(KH\right)}_{\ell,\mathbf{m}} \right) \,,
\ee
\be\ba
	\mathcal{L}^{\left(AB\right)}_{\ell,\mathbf{m}} &= -2\varepsilon^{ab}\text{Re}\left\{D_{a}\bar{A}_{\ell,\mathbf{m}}D_{b}B_{\ell,\mathbf{m}}\right\} - \frac{M_{\ell}\left(r\right)}{r^2}\left(\left|A_{\ell,\mathbf{m}}\right|^2 - \left|B_{\ell,\mathbf{m}}\right|^2 \right) \,, \\
	\mathcal{L}^{\left(KK\right)}_{\ell,\mathbf{m}} &= \frac{\left(d-2\right)\left(d-3\right)}{2}D_{a}\bar{\mathcal{K}}_{\ell,\mathbf{m}}D^{a}\mathcal{K}_{\ell,\mathbf{m}} + \frac{L_{\ell}\left(r\right)}{2r^2}\left|K_{\ell,\mathbf{m}}\right|^2 \,, \\
	\mathcal{L}^{\left(HH\right)}_{\ell,\mathbf{m}} &= \frac{G_{\ell}\left(r\right)}{4r^2}\left|\mathcal{H}_{\ell,\mathbf{m}}\right|^2 \,, \\
	\mathcal{L}^{\left(ABK\right)}_{\ell,\mathbf{m}} &= \left(d-2\right)\text{Re}\left\{\left(\bar{A}_{\ell,\mathbf{m}}I_{\left(1\right)}^{\braket{ab}}+\bar{B}_{\ell,\mathbf{m}}I_{\left(2\right)}^{\braket{ab}}\right)\,D_{a}D_{b}\mathcal{K}_{\ell,\mathbf{m}} \right\} \\
	&\quad+\frac{2\left(d-2\right)}{r}\text{Re}\left\{\left(\bar{A}_{\ell,\mathbf{m}}r^{a}+\sqrt{\frac{f_{r}}{f_{t}}}\,\bar{B}_{\ell,\mathbf{m}}t^{a}\right)D_{a}\mathcal{K}_{\ell,\mathbf{m}}\right\} \\
	\mathcal{L}^{\left(ABH\right)} &= -\frac{d-2}{r}\text{Re}\left\{\left(\bar{A}_{\ell,\mathbf{m}}r^{a}+\sqrt{\frac{f_{r}}{f_{t}}}\,\bar{B}_{\ell,\mathbf{m}}t^{a}\right)D_{a}\mathcal{H}_{\ell,\mathbf{m}}\right\} \,, \\
	\mathcal{L}^{\left(KH\right)}_{\ell,\mathbf{m}} &= \frac{d-2}{2}\text{Re}\left\{\bar{\mathcal{H}}_{\ell,\mathbf{m}}\left[-D_{a}D^{a}\mathcal{K}_{\ell,\mathbf{m}} - \frac{2d-5}{r}r^{a}D_{a}\mathcal{K}_{\ell,\mathbf{m}} + \frac{N_{\ell}\left(r\right)}{r^2}\mathcal{K}_{\ell,\mathbf{m}}\right] \right\} \,,
\ea\ee
where we have defined
\be\ba
	M_{\ell}\left(r\right) &\equiv \ell\left(\ell+d-3\right) - \left[\left(d-2\right)\frac{rf_{t}^{\prime}}{f_{t}} + r^2\left(\frac{f_{t}^{\prime}}{f_{t}}\right)^{\prime}\right]r_{a}r^{a} - \frac{rf_{t}^{\prime}}{f_{t}}rD_{a}r^{a} \,, \\
	G_{\ell}\left(r\right) &\equiv \ell\left(\ell+d-3\right) + \left(d-2\right)\left(d-3\right)r_{a}r^{a} + \left(d-2\right)rD_{a}r^{a} \,, \\
	L_{\ell}\left(r\right) &\equiv \left(d-4\right)\left[\left(d-2\right)\ell\left(\ell+d-3\right) - G_{\ell}\left(r\right)\right] \,, \\
	N_{\ell}\left(r\right) &\equiv \frac{1}{d-2}\left[\left(2d-5\right)\ell\left(\ell+d-3\right) - G_{\ell}\left(r\right)\right] \,.
\ea\ee
By integrating out the auxiliary variables $A_{\ell,\mathbf{m}}$, $B_{\ell,\mathbf{m}}$ and $\mathcal{H}_{\ell,\mathbf{m}}$, it is then possible to write down a Schr\"{o}dinger-like equation of motion for a master variable built from $\mathcal{K}_{\ell,\mathbf{m}}$ and its derivatives. The procedure is quite cumbersome and not enlightening for generic $f_{t}$ and $f_{r}$ so we will just write down the results for the Schwarzschild-Tangherlini black hole, for which $f_{t} = f_{r} = 1 -\left(r_{s}/r\right)^{d-3} \equiv f\left(r\right)$ and the above functions reduce to constants,
\be
	\begin{gathered}
		M_{\ell} = \ell\left(\ell+d-3\right) \,,\quad L_{\ell} =\left(d-3\right)\left(d-4\right)\left[\ell\left(\ell+d-3\right)-\left(d-2\right)\right] \,, \\
		G_{\ell} = \ell\left(\ell+d-3\right) + \left(d-2\right)\left(d-3\right) \,,\quad N_{\ell} = 2\ell\left(\ell+d-3\right) - \left(d-3\right) \,.
	\end{gathered}
\ee
After all, this is the only relevant case according to our current setup for the gravitoelectric response of an asymptotically flat and electrically neutral general-relativistic black hole in vacuum.

The Zerilli master variable $\Psi_{\ell,\mathbf{m}}^{\left(\text{Z}\right)}$ is constructed as~\cite{Hui:2020xxx,Ishibashi:2003ap,Kodama:2003jz,Kodama:2003kk,Ishibashi:2011ws}
\be
	\Psi^{\left(\text{Z}\right)}_{\ell,\mathbf{m}} = \frac{4fr^{\frac{d-4}{2}}}{H_{\ell}\left(r\right)}\sqrt{\left(d-2\right)\left(d-3\right)\lambda_{\ell}\ell\left(\ell+d-3\right)}\,\mathcal{V}_{\ell,\mathbf{m}} \,,
\ee
\be
	\mathcal{V}_{\ell,\mathbf{m}} = -\frac{r\mathcal{K}_{\ell,\mathbf{m}}}{2\sqrt{f_{t}f_{r}}}-\frac{\left(d-2\right)r}{2M_{\ell}}\sqrt{\frac{f_{r}}{f_{t}}}\left[A_{\ell,\mathbf{m}}+\frac{1}{2}\mathcal{H}_{\ell,\mathbf{m}}-\frac{1}{f_{r}}r^{a}D_{a}\left(r\mathcal{K}_{\ell,\mathbf{m}}\right)+\frac{rf_{t}^{\prime}}{2f_{t}}\mathcal{K}_{\ell,\mathbf{m}}\right] \,,
\ee
where we have defined
\be
	\lambda_{\ell} \equiv \left(\ell-1\right)\left(\ell+d-2\right)
\ee
and
\be\ba
	H_{\ell}\left(r\right) &= 2\ell\left(\ell+d-3\right) - 2\left(d-2\right)f\left(r\right) + \left(d-2\right)rf^{\prime}\left(r\right) \\
	&= 2\lambda_{\ell} + \left(d-1\right)\left(d-2\right)\left(\frac{r_{s}}{r}\right)^{d-3} \,.
\ea\ee
It satisfies a Schr\"{o}dinger-like equation,
\be
	\left[\partial_{r_{\ast}}^2-\partial_{t}^2-f\left(r\right)V_{\ell}^{\left(\text{Z}\right)}\right]\Psi_{\ell,\mathbf{m}}^{\left(\text{Z}\right)} = 0 \,,
\ee
with the Zerilli potential given by~\cite{Hui:2020xxx,Ishibashi:2003ap,Kodama:2003jz,Kodama:2003kk,Ishibashi:2011ws}
\be
	V_{\ell}^{\left(\text{Z}\right)}\left(r\right) = V_{\ell}^{\left(0\right)}\left(r\right) - \frac{2f^{\prime}\left(r\right)}{r}\frac{\left[2\lambda_{\ell}+\left(d-1\right)\left(d-2\right)\right]\left[H_{\ell}\left(r\right)+2\left(d-3\right)\lambda_{\ell}\right]}{H_{\ell}^2\left(r\right)} \,,
\ee
where $V^{\left(0\right)}\left(r\right)$ is the scalar field potential in Eq.~\ref{eq:V0S}.
\section{Love numbers of spherically symmetric black holes in General Relativity}
\label{sec:LoveNumbers}

In this section, we will explicitly compute the static Love numbers associated with perturbations of general-relativistic spherically symmetric black holes, for which
\be
	f_{t}\left(r\right) = f_{r}\left(r\right) \equiv f\left(r\right) \,.
\ee
We will begin by presenting the definition of scalar, $p$-form and tidal Love numbers through the worldline EFT. We will proceed to study $p$-form and gravitational perturbations of the higher-dimensional Schwarzschild-Tangherlini black hole. The matching onto the worldline EFT definition of the Love numbers will be achieved by employing a near-zone expansion of the relevant equations of motion, a procedure that will give rise to a closed expression for the dynamical response coefficients for each type of perturbations to leading order in the near-zone expansion. In the static limit, we will find that static Love numbers have a very rich structure. In particular, the behavior of the static Love numbers of the higher-dimensional Schwarzschild-Tangherlini black hole will be the expected one, as dictated by power counting arguments within the worldline EFT, except for a discrete tower of resonant conditions associated with the orbital number of the perturbation, for which the static Love numbers will turn out to be exactly zero. Although not necessary for the strictly static responses, the near-zone expansion of the equations of motion will be crucial in revealing the emergence of enhanced symmetries that precisely addresses these examples of ``magic zeroes'' in the black hole response problem. These computations have already be done in Ref.~\cite{Hui:2020xxx} for the cases of scalar, electromagnetic and gravitational static perturbations of the higher-dimensional Schwarzschild-Tangherlini black hole. The new element of this section, besides the study of these response problems within the near-zone expansion, is the study of the $p$-form Love numbers, with $2\le p \le d-3$, which to our best knowledge have so far not been studied in the literature.

We will also study the spin-$0$ scalar and spin-$2$ tensor modes of the higher-dimensional Reissner-Nordstr\"{o}m black hole. A similar study was conducted in Ref.~\cite{Pereniguez:2021xcj}; we will here prove their conjectured expressions for the tensor-type static tidal Love numbers of the electrically charged black hole by explicit analytical calculations.

\subsection{Definition of Love numbers for relativistic compact bodies}
\label{sec:WorldlineEFT}

The worldline EFT description of a compact body is based on its universal behavior of appearing as a point-particle when viewed from very large distances. One can then describe any compact body as an effective point-particle propagating along a worldline $x_{\text{cm}}^{\mu}\left(\lambda\right)$ that passes through the center of mass of the body and is parameterized by an affine parameter $\lambda$. This effective point-particle is then dressed with multipole moments accounting for finite-size effects, that is, couplings of the worldline with curvature tensors accounting for deviations from geodesic motion.

The worldline effective action for a spherically symmetric and non-rotating compact body can then be written down as~\cite{Goldberger:2004jt,Porto:2005ac,Porto:2016pyg,Levi:2015msa,Levi:2018nxp,Goldberger:2022ebt}
\be
	S_{\text{EFT}}\left[x_{\text{cm}},\phi\right] = -M\int d\tau + S_{\text{bulk}}\left[\phi\right] + S_{\text{finite-size}}\left[x_{\text{cm}},\phi\right] \,.
\ee
The first term is just the minimal point-particle action for a non-spinning body, with the affine parameter chosen to be the proper time $\tau$. The minimal part of the effective action also contains the bulk action, $S_{\text{bulk}}\left[\phi\right]$, which captures the dynamics of the long-distance interaction fields, here collectively denoted by ``$\phi$''. For systems interacting via general-relativistic gravitational ($g_{\mu\nu}$) forces or scalar ($\Phi$) or $p$-form ($A_{\mu_1\dots\mu_{p}}$) forces minimally coupled to gravity, for instance, the bulk action would be
\be
	S_{\text{bulk}}\left[g,\Phi,A\right] = \int d^{d}x\sqrt{-g}\left[\frac{1}{16\pi G}R -\frac{1}{2}\left(\nabla\Phi\right)^2 - \sum_{p=1}^{d-3}\frac{1}{2\left(p+1\right)!}F_{\mu_1\dots\mu_{p+1}}F^{\mu_1\dots\mu_{p+1}}\right] \,,
\ee
with $R$ the Ricci scalar and $F_{\mu_1\mu_2\dots\mu_{p+1}} = \left(p+1\right)\partial_{[\mu_1}A_{\mu_2\dots\mu_{p+1}]}$ the $\left(p+1\right)$-form field strength. In the presence of background interaction fields, this term is expanded around the background to give rise to bulk interaction vertices. For example, for asymptotically flat spacetimes, one writes $g_{\mu\nu} = \eta_{\mu\nu} + \sqrt{32\pi G}\,h_{\mu\nu}$ and performs an expansion in the graviton field $h_{\mu\nu}$. We also remark here that we are omitting gauge-fixing terms that should be included before performing EFT calculations.

The last term, $S_{\text{finite-size}}\left[x_{\text{cm}},\phi\right]$, contains non-minimal coupling of the worldline to curvature tensors. In particular, the leading finite-size effects come from quadratic couplings of symmetric trace-free derivatives of curvature tensors whose Wilson coefficients define the static Love numbers of each type of perturbation,
\be
	\begin{gathered}
		S_{\text{finite-size}} \supset S_{\text{Love}}^{\left(0\right)} + S_{\text{Love}}^{\left(\text{gr}\right)} + \sum_{p=1}^{d-3}S_{\text{Love}}^{\left(p\right)} \,, \\
		S_{\text{Love}}^{\left(0\right)} = \sum_{\ell=0}^{\infty}\frac{C_{\ell}^{\left(0\right)}}{2\ell!}\int d\tau\, \mathcal{E}_{L}^{\left(0\right)}\left(x_{\text{cm}}\left(\tau\right)\right)\mathcal{E}^{\left(0\right)L}\left(x_{\text{cm}}\left(\tau\right)\right) \,, \\
		\ba
			S_{\text{Love}}^{\left(\text{gr}\right)} = \sum_{\ell=2}^{\infty} &\bigg[\frac{C_{\ell}^{\mathcal{E},\left(\text{gr}\right)}}{2\ell!}\int d\tau\, \mathcal{E}_{L}^{\left(\text{gr}\right)}\left(x_{\text{cm}}\left(\tau\right)\right)\mathcal{E}^{\left(\text{gr}\right)L}\left(x_{\text{cm}}\left(\tau\right)\right) \\
			&+ \frac{C_{\ell}^{\mathcal{B},\left(\text{gr}\right)}}{2\ell!}\int d\tau\, \mathcal{B}_{L|b}^{\left(\text{gr}\right)}\left(x_{\text{cm}}\left(\tau\right)\right)\mathcal{B}^{\left(\text{gr}\right)L|b}\left(x_{\text{cm}}\left(\tau\right)\right) \\
			&+ \frac{C_{\ell}^{\mathcal{T},\left(\text{gr}\right)}}{2\ell!}\int d\tau\, \mathcal{T}_{L|bc}^{\left(\text{gr}\right)}\left(x_{\text{cm}}\left(\tau\right)\right)\mathcal{T}^{\left(\text{gr}\right)L|bc}\left(x_{\text{cm}}\left(\tau\right)\right)\bigg] \,,
		\ea \\
		\ba
			S_{\text{Love}}^{\left(p\right)} = \sum_{\ell=1}^{\infty} &\bigg[ \frac{C_{\ell}^{\mathcal{E},\left(p\right)}}{2\ell!}\frac{1}{p}\int d\tau\, \mathcal{E}_{L|b_1\dots b_{p-1}}^{\left(p\right)}\left(x_{\text{cm}}\left(\tau\right)\right)\mathcal{E}^{\left(p\right)L|b_1\dots b_{p-1}}\left(x_{\text{cm}}\left(\tau\right)\right) \\
			&+ \frac{C_{\ell}^{\mathcal{B},\left(p\right)}}{2\ell!}\frac{1}{p+1}\int d\tau\, \mathcal{B}_{L|b_1\dots b_{p}}^{\left(p\right)}\left(x_{\text{cm}}\left(\tau\right)\right)\mathcal{B}^{\left(p\right)L|b_1\dots b_{p}}\left(x_{\text{cm}}\left(\tau\right)\right) \bigg] \,.
		\ea
	\end{gathered}
\ee
In the above expressions small Latin indices are spatial indices and $L\equiv a_1\dots a_{\ell}$. The symmetric trace-free tensors appearing are then defined in terms of the $d$-velocity $u^{\mu} = \frac{dx_{\text{cm}}^{\mu}}{d\tau}$ and a set of local vielbein vector $e^{\mu}_{a}$, satisfying $u_{\mu}e^{\mu}_{a}=0$\footnote{For the current case of spherically symmetric and non-rotating bodies, for which $u_{\mu}u^{\mu}=-1$, $e_{a}^{\mu} = \delta_{a}^{\mu} +\delta_{a}^{\nu}u_{\nu}u^{\mu}$.}. More specifically, for spin-$0$ (scalar) perturbations,
\be
	\mathcal{E}_{L}^{\left(0\right)} = e^{\mu_1}_{a_1}\dots e^{\mu_{\ell}}_{a_{\ell}} \nabla_{\langle \mu_1}\dots\nabla_{\mu_{\ell}\rangle}\Phi \,,
\ee
and $C_{\ell}^{\left(0\right)}$ defines the $\ell$'th static scalar Love number. For spin-$2$ (gravitational) perturbations,
\be\ba
	\mathcal{E}_{L}^{\left(\text{gr}\right)} &= e^{\mu_1}_{a_1}\dots e^{\mu_{\ell-2}}_{a_{\ell-2}} \nabla_{\langle \mu_1}\dots\nabla_{\mu_{\ell-2}}E_{a_{\ell-1}a_{\ell}\rangle}^{\left(\text{gr}\right)} \,,\quad E_{ab}^{\left(\text{gr}\right)} = u^{\mu}e^{\nu}_{a}u^{\rho}e^{\sigma}_{b}C_{\mu\nu\rho\sigma} \,, \\
	\mathcal{B}_{L|b}^{\left(\text{gr}\right)} &= e^{\mu_1}_{a_1}\dots e^{\mu_{\ell-2}}_{a_{\ell-2}} \nabla_{\langle \mu_1}\dots\nabla_{\mu_{\ell-2}}B_{a_{\ell-1}a_{\ell}\rangle b}^{\left(\text{gr}\right)} \,,\quad B_{abc}^{\left(\text{gr}\right)} = u^{\mu}e^{\nu}_{a}e^{\rho}_{b}e^{\sigma}_{c}C_{\mu\nu\rho\sigma} \,, \\
	\mathcal{T}_{L|bc}^{\left(\text{gr}\right)} &= e^{\mu_1}_{a_1}\dots e^{\mu_{\ell-2}}_{a_{\ell-2}} \nabla_{\langle \mu_1}\dots\nabla_{\mu_{\ell-2}}T_{a_{\ell-1}|b|a_{\ell}\rangle c}^{\left(\text{gr}\right)} \,,\quad T_{abcd}^{\left(\text{gr}\right)} = e^{\mu}_{a}e^{\nu}_{b}e^{\rho}_{c}e^{\sigma}_{d}C_{\mu\nu\rho\sigma} \,,
\ea\ee
with $C_{\mu\nu\rho\sigma}$ the spacetime Weyl tensor, and $C_{\ell}^{\mathcal{E},\left(\text{gr}\right)}$, $C_{\ell}^{\mathcal{B},\left(\text{gr}\right)}$ and $C_{\ell}^{\mathcal{T},\left(\text{gr}\right)}$ define the $\ell$'th static gravitoelectric, gravitomagnetic and tensor-type tidal Love number respectively. We note here that tensor-type tidal perturbations are non-trivial only in $d>4$. Last, for $p$-form perturbations,
\be\ba
	\mathcal{E}_{L|b_1\dots b_{p-1}}^{\left(p\right)} &= e_{a_1}^{\mu_1}\dots e_{a_{\ell-1}}^{\mu_{\ell-1}}\nabla_{\langle \mu_1}\dots \nabla_{\mu_{\ell-1}}E_{a_{\ell}\rangle b_1\dots b_{p-1}} \,,\quad E_{a_1a_2\dots a_{p}} = u^{\mu_1}e_{a_1}^{\mu_2}\dots e_{a_{p}}^{\mu_{p+1}}F_{\mu_1\mu_2\dots\mu_{p+1}} \,, \\
	\mathcal{B}_{L|b_1\dots b_{p}}^{\left(p\right)} &= e_{a_1}^{\mu_1}\dots e_{a_{\ell-1}}^{\mu_{\ell-1}}\nabla_{\langle \mu_1}\dots \nabla_{\mu_{\ell-1}}B_{a_{\ell}\rangle b_1\dots b_{p}} \,,\quad B_{a_1a_2\dots a_{p+1}} = e_{a_1}^{\mu_1}e_{a_2}^{\mu_2}\dots e_{a_{p+1}}^{\mu_{p+1}}F_{\mu_1\mu_2\dots\mu_{p+1}} \,,
\ea\ee
and $C_{\ell}^{\mathcal{E},\left(p\right)}$ and $C_{\ell}^{\mathcal{B},\left(p\right)}$ define the static electric-type and magnetic-type $p$-form Love numbers respectively\footnote{The ``electric/magnetic'' terminology used here is borrowed from the $p=1$ case, although these are not of electric or magnetic nature for $p>1$.}.

Dynamical Love numbers can also be defined in a similar fashion by operators involving ``time'' derivatives $D=u^{\mu}\nabla_{\mu}$. For instance, the first dynamical Love number is defined by the Wilson coefficient in front of the quadratic coupling of the worldline with the operators of the form $D\mathcal{E}D\mathcal{E}$. The full dynamical scalar Love part of the finite-size non-minimal couplings in the worldline EFT action for spherically symmetric and non-rotating bodies is then
\be\ba
	S_{\text{dynamic Love}}^{\left(0\right)} &= \sum_{\ell=0}^{\infty}\sum_{n=0}^{\infty}\frac{C_{\ell;n}^{\left(0\right)}}{2\ell!}\int d\tau D^{n}\mathcal{E}_{L}^{\left(0\right)}\left(x_{\text{cm}}\left(\tau\right)\right)D^{n}\mathcal{E}^{\left(0\right)L}\left(x_{\text{cm}}\left(\tau\right)\right) \\
	&= \sum_{\ell=0}^{\infty} \int \frac{d\omega}{2\pi}\frac{C_{\ell}^{\left(0\right)}\left(\omega\right)}{2\ell!}\mathcal{E}_{L}^{\left(0\right)}\left(-\omega\right)\mathcal{E}^{\left(0\right)L}\left(\omega\right) \,,
\ea\ee
where in the second line we have switched to frequency space, gathering the dynamical scalar Love numbers $C_{\ell;n}^{\left(0\right)}$ in a frequency-dependent Wilson ``function'' $C_{\ell}^{\left(0\right)}\left(\omega\right)=\sum_{n=0}^{\infty}\left(-1\right)^{n}\omega^{2n}C_{\ell;n}^{\left(0\right)}$; a completely analogous analysis can also be done for the dynamical gravitoelectric, gravitomagnetic and tensor-type tidal Love numbers, as well as for the dynamical electric-type and magnetic-type $p$-form Love numbers.

\subsection{Matching Love numbers}
\label{sec:LoveNumbersMatching}

With this definition of Love numbers at the level of the worldline EFT action, their computation reduces to employing a matching condition onto a microscopic quantity, ``microscopic'' here referring to the full classical computation, for example, within the framework of black hole perturbation analysis of General Relativity. While an on-shell matching onto scattering observables is possible~\cite{Ivanov:2022hlo,Ivanov:2022qqt,Saketh:2023bul}, we will employ here the off-shell ``Newtonian matching''~\cite{Kol:2011vg,Charalambous:2021mea,Charalambous:2023jgq}, due to its applicability at the level of the equations of motion where enhanced symmetries are more directly manifested.

\subsubsection{The EFT side}
From the EFT side, the Newtonian matching condition consists of switching on a background Newtonian source for the type of interaction field under investigation, characterized by a spin-index $s$ and set of $N$ spatial indices $\mathbf{a}\equiv\left\{a_1,\dots, a_{N}\right\}$ (see Table~\ref{tbl:phiSN}),
\be\ba
	\phi_{\mathbf{a}}^{\left(s\right)}\left(\omega,\mathbf{x}\right) = \bar{\phi}_{\mathbf{a}}^{\left(s\right)}\left(\omega,\mathbf{x}\right) + \delta\phi_{\mathbf{a}}^{\left(s\right)}\left(\omega,\mathbf{x}\right) \,,\quad \bar{\phi}_{\mathbf{a}}^{\left(s\right)}\left(\omega,\mathbf{x}\right) = \frac{\left(\ell-s\right)!}{\ell!}\bar{\phi}_{L|\mathbf{a}}x^{L} \,,
\ea\ee
and matching the $1$-point function of the perturbation,
\be
	\vev{\delta\phi_{\mathbf{a}}^{\left(s\right)}\left(\omega,\mathbf{x}\right)} =
	\vcenter{\hbox{\begin{tikzpicture}
			\begin{feynman}
				\vertex (a0);
				\vertex[right=0.6cm of a0] (gblobaux);
				\vertex[left=0.00cm of gblobaux, blob] (gblob){};
				\vertex[below=1cm of a0] (p1);
				\vertex[above=1cm of a0] (p2);
				\vertex[right=1cm of p1] (a1);
				\vertex[right=1cm of p2] (a2);
				\vertex[right=0.69cm of p2] (a22){$\times$};
				\diagram*{
					(p1) -- [double,double distance=0.5ex] (p2),
					(a1) -- (gblob) -- (a2),
				};
			\end{feynman}
	\end{tikzpicture}}} =
	\underbrace{\vcenter{\hbox{\begin{tikzpicture}
				\begin{feynman}
					\vertex[dot] (a0);
					\vertex[below=1cm of a0] (p1);
					\vertex[above=1cm of a0] (p2);
					\vertex[right=0.4cm of a0, blob] (gblob){};							
					\vertex[right=1.5cm of p1] (b1);
					\vertex[right=1.5cm of p2] (b2);
					\vertex[right=1.19cm of p2] (b22){$\times$};
					\vertex[above=0.7cm of a0] (g1);
					\vertex[above=0.4cm of a0] (g2);
					\vertex[right=0.05cm of a0] (gdtos){$\vdots$};
					\vertex[below=0.7cm of a0] (gN);
					\diagram*{
						(p1) -- [double,double distance=0.5ex] (p2),
						(g1) -- [photon] (gblob),
						(g2) -- [photon] (gblob),
						(gN) -- [photon] (gblob),
						(b1) -- (gblob) -- (b2),
					};
				\end{feynman}
	\end{tikzpicture}}}}_{\text{``source''}} + 
	\underbrace{\vcenter{\hbox{\begin{tikzpicture}
				\begin{feynman}
					\vertex[dot] (a0);
					\vertex[left=0.00cm of a0] (lambda){$C_{\ell}\left(\omega\right)$};
					\vertex[below=1.6cm of a0] (p1);
					\vertex[above=0.4cm of a0] (p2);
					\vertex[right=1.5cm of p1] (b1);
					\vertex[right=1.5cm of p2] (b2);
					\vertex[right=1.19cm of p2] (b22){$\times$};
					\vertex[below=0.4cm of a0] (g1);
					\vertex[below=0.3cm of g1] (g2);
					\vertex[below=0.14cm of g2] (gdotsaux);
					\vertex[right=0.00cm of gdotsaux] (gdtos){$\vdots$};
					\vertex[below=0.5cm of gdotsaux] (gN);
					\vertex[below=0.7cm of a0] (gblobaux);
					\vertex[right=0.4cm of gblobaux, blob] (gblob){};
					\diagram*{
						(p1) -- [double,double distance=0.5ex] (p2),
						(b2) -- (a0) -- (gblob) -- (b1),
						(g1) -- [photon] (gblob),
						(g2) -- [photon] (gblob),
						(gN) -- [photon] (gblob),
					};
				\end{feynman}
	\end{tikzpicture}}}}_{\text{``response''}} \,.
\ee
In the above diagrammatic expression, the double line represents the worldline, straight lines indicate propagators of the interaction field $\delta\phi_{\mathbf{a}}^{\left(s\right)}$, a cross (``$\times$'') represents an insertion of the background field $\bar{\phi}_{\mathbf{a}}^{\left(s\right)}$ and wavy lines represent graviton propagators, whose interactions with the worldline come from the minimal point-particle action and capture relativistic corrections; for instance, for asymptotically flat spacetimes, in a body-centered frame where $x_{\text{cm}}^{\mu}=\left(t,\mathbf{0}\right)$ and $u^{\mu}=\left(1,\mathbf{0}\right)$, we have $-M\int d\tau = -M\int dt\,\sqrt{1-\sqrt{32\pi G}\,h_{00}}$, giving rise to an infinite number of graviton-worldline interaction vertices after expanding the square root. Furthermore, in the second equality we have demonstrated how the worldline EFT definition naturally performs a source/response split, unambiguously distinguishing between relativistic corrections in the ``source'' part of the field profile and actual response effects~\cite{Charalambous:2021mea,Ivanov:2022hlo}. This splitting is equivalent to the method of analytically continuing the spacetime dimensionality $d$~\cite{Kol:2011vg} or the multipolar order $\ell$~\cite{LeTiec:2020bos,Charalambous:2021mea,Ivanov:2022hlo,Creci:2021rkz}, as the ``source'' and ``response'' diagrams then have indicial powers $r^{\alpha}$ with $\alpha=\ell$ and $\alpha=-\left(\ell+d-3\right)$ respectively, while the relativistic corrections on each branch have the form $r^{\alpha-n}$ with positive integer $n$.

\begin{table}
	\centering
	\begin{tabular}{|c|c|c|c|}
		\hline
		Type of interaction field & $s$ & $N=\text{card}(\mathbf{a})$ & $\phi_{\mathbf{a}}^{\left(s\right)}$ \\
		\hline\hline
		Scalar & $0$ & $0$ & $\Phi$ \\
		\hline
		Electric-type $p$-form & $1$ & $p-1$ & $\mathcal{A}^{\mathcal{E},\left(p\right)}_{a_1\dots a_{p-1}} = u^{\mu_1}e^{\mu_2}_{a_1}\dots e^{\mu_{p}}_{a_{p-1}}A_{\mu_1\mu_2\dots\mu_{p}}$ \\
		Magnetic-type $p$-form & $1$ & $p$ & $\mathcal{A}^{\mathcal{B},\left(p\right)}_{a_1\dots a_{p}} = e^{\mu_1}_{a_1}e^{\mu_2}_{a_2}\dots e^{\mu_{p}}_{a_{p}}A_{\mu_1\mu_2\dots\mu_{p}}$ \\
		\hline
		Gravitoelectric & $2$ & $0$ & $\mathcal{H}^{\mathcal{E}} = \frac{1}{2}u^{\mu}u^{\nu}h_{\mu\nu}$ \\
		Gravitomagnetic & $2$ & $1$ & $\mathcal{H}^{\mathcal{B}}_{a} = \frac{1}{2}u^{\mu}e_{a}^{\nu}h_{\mu\nu}^{\left(\text{V}\right)}$ \\
		Tensor-type gravitational & $2$ & $2$ & $\mathcal{H}^{\mathcal{T}}_{ab} = \frac{1}{2}e^{\mu}_{\langle a}e^{\nu}_{b\rangle}h_{\mu\nu}^{\left(\text{T}\right)}$ \\
		\hline
	\end{tabular}
	\caption{The various types of interaction fields $\phi_{\mathbf{a}}^{\left(s\right)}$ entering the worldline EFT and the values of the spin-index $s$ and the number $N$ of spatial indices that characterize them. The superscripts ``$\left(\text{V}\right)$'' and ``$\left(\text{T}\right)$'' indicate the usage of the gauge invariant vector and tensor modes of the gravitational field.}
	\label{tbl:phiSN}
\end{table}

In the Newtonian limit, in a gauge where the interaction fields $\delta\phi_{\mathbf{a}}^{\left(s\right)}$ are canonical variables up to an overall normalization constant $N_{\text{prop}}$ in momentum space (see Table~\ref{tbl:Nprop}) and in the body centered frame, the relativistic corrections are suppressed and one ends up with the characteristic bi-monomial form
\be\ba\label{eq:EFT1pointNewtonian}
	{}&\vev{\delta\phi_{\mathbf{a}}^{\left(s\right)}\left(\omega,\mathbf{x}\right)} \rightarrow
	\vcenter{\hbox{\begin{tikzpicture}
			\begin{feynman}
				\vertex[dot] (a0);
				\vertex[below=1cm of a0] (p1);
				\vertex[above=1cm of a0] (p2);						
				\vertex[right=1cm of p1] (b1);
				\vertex[right=1cm of p2] (b2);
				\vertex[right=0.69cm of p2] (b22){$\times$};
				\diagram*{
					(p1) -- [double,double distance=0.5ex] (p2),
					(b1) -- (b2),
				};
			\end{feynman}
	\end{tikzpicture}}} + 
	\vcenter{\hbox{\begin{tikzpicture}
			\begin{feynman}
				\vertex[dot] (a0);
				\vertex[left=0.00cm of a0] (lambda){$C_{\ell}\left(\omega\right)$};
				\vertex[below=1cm of a0] (p1);
				\vertex[above=1cm of a0] (p2);
				\vertex[right=1.2cm of p1] (b1);
				\vertex[right=1.2cm of p2] (b2);
				\vertex[right=0.89cm of p2] (b22){$\times$};
				\diagram*{
					(p1) -- [double,double distance=0.5ex] (p2),
					(b2) -- (a0) -- (b1),
				};
			\end{feynman}
	\end{tikzpicture}}} \\
	&= \frac{\left(\ell-s\right)!}{\ell!} \left[1 + \frac{2^{\ell-2}\Gamma\left(\ell+\frac{d-3}{2}\right)}{\pi^{\left(d-1\right)/2}}N_{\text{prop}}\frac{C_{\ell}\left(\omega\right)}{r^{2\ell+d-3}}\right]\bar{\phi}_{L|\mathbf{a}}\left(\omega\right)x^{L} \,.
\ea\ee
In the above computation, we used the fact that the Love numbers are time-reversal symmetric, $C_{\ell}\left(-\omega\right)=C_{\ell}\left(\omega\right)$, since they appear in front of \textit{local} operator and, hence, only capture conservative dynamics. We remark here that the absence of dissipative effects is implicit in the use of the in-out formalism above, but can be treated through the in-in (Schwinger-Keldysh) formalism~\cite{Schwinger:1960qe,Keldysh:1964ud,Goldberger:2005cd,Goldberger:2019sya,Goldberger:2020fot,Goldberger:2020wbx,Ivanov:2022hlo}.

\begin{table}
	\centering
	\begin{tabular}{|c|c|c|}
		\hline
		$\phi_{\mathbf{a}}^{\left(s\right)}$ & $P^{\mathbf{b}}_{\mathbf{a}}$ & $N_{\text{prop}}$ \\
		\hline\hline
		$\Phi$ & $1$ & $+1$ \\
		\hline
		$\mathcal{A}^{\mathcal{E},\left(p\right)}_{a_1\dots a_{p-1}}$ & $\delta^{b_1}_{[a_1}\dots\delta^{b_{p-1}}_{a_{p-1}]}$ & $-1$ \\
		$\mathcal{A}^{\mathcal{B},\left(p\right)}_{a_1\dots a_{p}}$ & $\delta^{b_1}_{[a_1}\dots\delta^{b_{p}}_{a_{p}]}$ & $+1$ \\
		\hline
		$\mathcal{H}^{\mathcal{E}}$ & $1$ & $+\frac{d-3}{4\left(d-2\right)}$ \\
		$\mathcal{H}^{\mathcal{B}}_{a}$ & $\delta^{b}_{a}$ & $-\frac{1}{8}$ \\
		$\mathcal{H}^{\mathcal{T}}_{ab}$ & $\delta^{c}_{\langle a}\delta^{d}_{b\rangle}$ & $+\frac{1}{4}$ \\
		\hline
	\end{tabular}
	\caption{The normalization of the momentum space propagator $\vev{\phi^{\left(s\right)}_{\mathbf{a}}\phi^{\left(s\right)\mathbf{b}}}\left(p\right) = N_{\text{prop}}P^{\mathbf{b}}_{\mathbf{a}}\,\frac{-i}{p^2}$ of the various interaction fields.}
	\label{tbl:Nprop}
\end{table}

\subsubsection{The microscopic theory side - Near-zone expansion}
From the microscopic theory side, the task is to solve the linearized equations of motion arising from perturbation analysis around a background geometry and in the possible presence of other background interaction fields $\big\{\phi^{\left(s\right),0}_{\mathbf{a}}\big\}$. The setup consists of a compact body which is adiabatically perturbed by another distant, weak and slowly varying configuration of charges and currents, e.g. another compact body, sourcing a perturbing set of fields $\big\{\bar{\phi}^{\left(s\right)}_{\mathbf{a}}\big\}$. The full equations of motion are typically of the form $\Box^{\left(0\right)}\delta\phi^{\left(s\right)}_{\mathbf{a}} = \bar{J}^{\left(s\right)}_{\mathbf{a}}\big[\bar{\phi}^{\left(s\right)}_{\mathbf{a}}\big]$, where $\Box^{\left(0\right)}$ represents a kinetic operator evaluated in the presence of the background fields and $\bar{J}^{\left(s\right)}_{\mathbf{a}}$ is a current sourced by the perturbing configuration.

In order to match onto the worldline EFT $1$-point function, one should work in the appropriate regime where the EFT is accurate. This is the \textit{near-zone region}, defined by the conditions that the wavelength of the perturbation is large compared to the size $\mathcal{R}$ of the unperturbed compact body and the distance from it~\cite{Starobinsky:1973aij,Starobinskil:1974nkd,Maldacena:1997ih,Castro:2010fd,Chia:2020yla,Charalambous:2021kcz},
\be
	\omega\mathcal{R} \ll 1 \quad\text{ and }\quad \omega\left(r-\mathcal{R}\right) \ll 1 \,,
\ee
with $r$ a radial distance whose origin is the center of the body. The first condition follows from the fact that the worldline EFT arises by integrating out the short-scale degrees of freedom associated with the internal structure of the body. The second condition revolves around the fact that the worldline EFT is a one-body EFT, ignoring the dynamics of the second, perturbing, body in the binary setup. Within the near-zone region, one then expands the kinetic operator $\Box^{\left(0\right)}$ in the above phase space variables and sets $\bar{J}^{\left(s\right)}_{\mathbf{a}}=0$, the presence of the source being encoded in the asymptotic boundary conditions. More specifically, the large $r$ behavior within the near-zone regime takes the form, in frequency space~\cite{Love:1912,PoissonWill2014,Damour:2009vw,Binnington:2009bb,Charalambous:2021mea,Ivanov:2022hlo,Charalambous:2023jgq}
\be
	\delta\phi_{\mathbf{a}}^{\left(s\right)}\left(\omega,\mathbf{x}\right) \xrightarrow{r\rightarrow\infty} \sum_{\ell=s}^{\infty}\frac{\left(\ell-s\right)!}{\ell!}\left[1 + k_{\ell}\left(\omega\right)\left(\frac{\mathcal{R}}{r}\right)^{2\ell+d-3}\right]\bar{\phi}_{L|\mathbf{a}}^{\left(s\right)}\left(\omega\right)x^{L} \,,
\ee
where $\bar{\phi}_{L|\mathbf{a}}^{\left(s\right)}$ are the multipole moments of the perturbing source and $k_{\ell}\left(\omega\right)$ are the response coefficients, i.e. the dimensionless Green's functions associated with the response problem. These contain both the conservative, time-reversal even, and the dissipative, time-reversal odd, effects which in the current case of a spherically symmetric and non-rotating unperturbed body can be identified with the real and the imaginary parts of the frequency space response coefficients respectively~\cite{LeTiec:2020bos,Charalambous:2023jgq,Saketh:2023bul},
\be\ba
	k_{\ell}^{\text{cons}}\left(\omega\right) &= \text{Re}\left\{k_{\ell}\left(\omega\right)\right\} = +k_{\ell}^{\text{cons}}\left(-\omega\right) \,, \\
	k_{\ell}^{\text{diss}}\left(\omega\right) &= \text{Im}\left\{k_{\ell}\left(\omega\right)\right\} = -k_{\ell}^{\text{diss}}\left(-\omega\right) \,.
\ea\ee
In particular, matching the worldline EFT $1$-point function Eq.~\eqref{eq:EFT1pointNewtonian} onto this microscopic computation shows that the Love numbers $C_{\ell}\left(\omega\right)$ are precisely equal to the conservative response coefficients up to an overall constant,
\be
	k_{\ell}^{\text{cons}}\left(\omega\right) = \frac{2^{\ell-2}\Gamma\left(\ell+\frac{d-3}{2}\right)}{\pi^{\left(d-1\right)/2}}N_{\text{prop}}\frac{C_{\ell}\left(\omega\right)}{\mathcal{R}^{2\ell+d-3}} \equiv k_{\ell}^{\text{Love}}\left(\omega\right) \,.
\ee

\subsection{$p$-form and tidal Love numbers of Schwarzschild-Tangherlini black holes}
\label{sec:LNs_SchwarzschildDd}

We can now start computing black hole Love numbers as per the above definition. We begin with the case of the higher-dimensional electrically neutral Schwarzschild-Tangherlini black hole for which
\be
	f\left(r\right) = 1 - \left(\frac{r_{s}}{r}\right)^{d-3} \,.
\ee

\subsubsection{$p$-form Love numbers}
We will first consider $p$-form perturbations, which are captured by the master variables $\Psi^{\left(j\right)}$, with $j$ labeling the $SO\left(d-1\right)$ sector of the perturbation. In particular, $j=p$ for co-exact $p$-form modes and $j=\tilde{p}=d-p-2$ for co-exact $\left(p-1\right)$-form modes on the sphere. Equivalently, $j$ is equal to $n$ dualizations of the rank $p$ of the $p$-form gauge field for the co-exact $\left(p-n\right)$-form modes. In this notation, the cases of scalar field and spin-$1$ perturbations can also be incorporated via
\be\ba
	\text{Spin-$0$}&: \Psi^{\left(j=0\right)}_{\ell,\mathbf{m}} = \Psi^{\left(0\right)}_{\ell,\mathbf{m}} \\
	\text{Spin-$1$}&: \Psi^{\left(j=1\right)}_{\ell,\mathbf{m}} = \Psi^{\left(\text{V}\right)}_{\ell,\mathbf{m}} \,,\quad \Psi^{\left(j=d-3\right)}_{\ell,\mathbf{m}} = \Psi^{\left(\text{S}\right)}_{\ell,\mathbf{m}} \,,
\ea\ee
with of course no analogues of co-exact $\left(p-1\right)$-form modes for the $p=0$ scalar field. Performing the field redefinition
\be
	\Phi^{\left(j\right)}_{\ell,\mathbf{m}} = \frac{\Psi^{\left(j\right)}_{\ell,\mathbf{m}}}{r^{\frac{d-2}{2}}} \,,
\ee
introducing the variable $\rho=r^{d-3}$ and defining $\Delta=\rho^2 f = \rho\left(\rho-\rho_{s}\right)$, the radial equations of motion for $p$-form perturbations can be rewritten as
\be\label{eq:FullpEOMSchwarzschild}
	\begin{gathered}
		\mathbb{O}^{\left(j\right)}_{\text{full}}\Phi^{\left(j\right)}_{\ell,\mathbf{m}} = \hat{\ell}(\hat{\ell}+1)\,\Phi^{\left(j\right)}_{\ell,\mathbf{m}} \,, \\
		\mathbb{O}^{\left(j\right)}_{\text{full}} = \partial_{\rho}\,\Delta\,\partial_{\rho} - \frac{r^2\rho^2}{\left(d-3\right)^2\Delta}\partial_{t}^2 + \frac{\rho_{s}}{\rho}\hat{j}^2 \,,
	\end{gathered}
\ee
where we have also introduced the rescaled orbital number and $SO\left(d-1\right)$ sector index
\be
	\hat{\ell} \equiv \frac{\ell}{d-3} \quad \text{and} \quad \hat{j} \equiv \frac{j}{d-3}
\ee
respectively.

Let us now solve these to extract the Schwarzschild-Tangherlini black hole Love numbers at leading order in the near-zone expansion. There are two near-zone splittings that are of particular interest, controlled by a sign $\sigma=\pm1$,
\be\label{eq:Vp_SchwarzschildD}
	\begin{gathered}
		\mathbb{O}^{\left(j\right)}_{\text{full}} = \partial_{\rho}\,\Delta\,\partial_{\rho} + V_0^{\left(\sigma\right)} + \epsilon\,V_1^{\left(\sigma\right)} \,, \\
		\ba
			V_0^{\left(\sigma\right)} &= -\frac{\rho_{s}^2}{4\Delta}\beta^2\partial_{t}^2 + \frac{\rho_{s}}{\rho}\hat{j}\left(\sigma\beta\,\partial_{t} + \hat{j}\right) \,, \\
			V_1^{\left(\sigma\right)} &= -\frac{r^2\rho^2-r_{s}^2\rho_{s}^2}{\left(d-3\right)^2\Delta}\partial_{t}^2 - \sigma\frac{\rho_{s}}{\rho}\hat{j}\beta\,\partial_{t} \,,
		\ea
	\end{gathered}
\ee
where $\beta = \frac{2r_{s}}{d-3}$ is the inverse surface gravity of the $d$-dimensional Schwarzschild-Tangherlini black hole and $\epsilon$ is a formal expansion parameter. Note that we have introduced a $\partial_{t}$ term; even though this was not present in the original equations of motion, it is still subleading in the near-zone expansion since it does not alter the near-horizon behavior of the solution. This might look like making the equations we want to solve more complicated than necessary, but introducing this term actually makes the problem simpler in the sense that we can now analytically solve the leading order near-zone equations of motion in terms of hypergeometric functions.

Indeed, after separating the variables,
\be
	\Phi^{\left(j\right)}_{\omega\ell,\mathbf{m}}\left(t,\rho\right) = e^{-i\omega t}R^{\left(j\right)}_{\omega\ell,\mathbf{m}}\left(\rho\right) \,,
\ee
and introducing the dimensionless radial distance from the event horizon,
\be
	x = \frac{\rho-\rho_{s}}{\rho_{s}} \,,
\ee
the leading order ($\epsilon=0$) near-zone radial equation of motion reads
\be\label{eq:NZRadialScharzschildDd}
	\left[\frac{d}{dx}\,x\left(1+x\right)\frac{d}{dx} + \frac{\beta^2\omega^2}{4x} - \frac{(\beta\omega+2i\sigma\hat{j})^2}{4\left(1+x\right)} \right]R^{\left(j\right)}_{\omega\ell,\mathbf{m}} = \hat{\ell}(\hat{\ell}+1)\,R^{\left(j\right)}_{\omega\ell,\mathbf{m}}
\ee
and the solution satisfying ingoing boundary conditions at the future event horizon (see Eq.~\eqref{eq:NHbehavior_SphericalSymmetric}) can be analytically found to be 
\be\ba\label{eq:RNearZoneJ}
	R^{\left(j\right)}_{\omega\ell,\mathbf{m}} &= \bar{R}^{\left(j\right)\text{in}}_{\ell,\mathbf{m}}\left(\omega\right)\left(\frac{x}{1+x}\right)^{-i\beta\omega/2} \\
	&\quad \times\left(1+x\right)^{-\sigma\hat{j}}{}_2F_1\left(\hat{\ell}+1-\sigma\hat{j},-\hat{\ell}-\sigma\hat{j};1-i\beta\omega;-x\right) \,.
\ea\ee
Expanding around large distances\footnote{Useful formulae involving the hypergeometric function and the $\Gamma$-function can be found in Appendix~\ref{app:2F1Gamma}.} reveals then that the response coefficients at leading order in the near-zone expansion are
\be\label{eq:ResponseCoefficientspSchwarzschildDd}
	k^{\left(j\right)}_{\ell}\left(\omega\right) = \frac{\Gamma(-2\hat{\ell}-1)\Gamma(\hat{\ell}+1-\sigma\hat{j})\Gamma(\hat{\ell}+1+\sigma\hat{j}-i\beta\omega)}{\Gamma(2\hat{\ell}+1)\Gamma(-\hat{\ell}-\sigma\hat{j})\Gamma(-\hat{\ell}+\sigma\hat{j}-i\beta\omega)} \,.
\ee
At this point, let us remark that the matching has been done directly at the level of the master variables $\Phi^{\left(j\right)}_{\ell,\mathbf{m}}$ which are built at most from derivatives of the actual fields in terms of which the response problem is defined. Therefore, analytically continuing the orbital number $\ell$ or the spacetime dimensionality $d$ is sufficient to unambiguously perform the source/response split. The above coefficients in front of the decaying branches of the master variables $\Phi^{\left(j\right)}_{\ell,\mathbf{m}}$ are then equal to the actual response coefficients we are looking for, up to overall non-zero matching normalization constants. For scalar and electromagnetic perturbations, for instance,~\cite{Hui:2020xxx}
\be
	\begin{gathered}
		k^{\left(j=0\right)}_{\ell}\left(\omega\right) = k^{\left(0\right)}_{\ell}\left(\omega\right) \,, \\
		k^{\left(j=1\right)}_{\ell}\left(\omega\right) = k^{\mathcal{B}\left(1\right)}_{\ell}\left(\omega\right) \,,\quad k^{\left(j=d-3\right)}_{\ell}\left(\omega\right) = -\frac{\ell+d-3}{\ell}k^{\mathcal{E}\left(1\right)}_{\ell}\left(\omega\right) \,.
	\end{gathered}
\ee
For the more general case of $p$-form perturbations, there is a co-exact $p$-form sector and a co-exact $\left(p-1\right)$-form sector, serving as the $p>1$ generalizations of the magnetic and electric sectors respectively that one encounters for electromagnetic perturbations. For these,
\be
	\begin{gathered}
		k^{\left(j=p\right)}_{\ell}\left(\omega\right) = k^{\mathcal{B},\left(p\right)}_{\ell}\left(\omega\right) \,,\quad k^{\left(j=d-p-2\right)}_{\ell}\left(\omega\right) = -\frac{\ell+d-p-2}{\ell+p-1}k^{\mathcal{E},\left(p\right)}_{\ell}\left(\omega\right) \,,
	\end{gathered}
\ee
where the superscripts ``$\mathcal{B}$'' and ``$\mathcal{E}$'' here refer to these extensions of the magnetic-type and electric-type perturbations, although this is just a convention of labeling things; these are not actual magnetic or electric in nature. For the sake of simplicity, however, we will keep referring to Eq.~\eqref{eq:ResponseCoefficientspSchwarzschildDd} as the response coefficients associated with each type of perturbation.

In the static limit, the response coefficients in Eq.~\eqref{eq:ResponseCoefficientspSchwarzschildDd} become purely real and correspond to the static Love numbers for $p$-form perturbations,
\be\ba\label{eq:StaticpLNs_SchwarzschildDd}
	k^{\left(j\right)\text{Love}}_{\ell}\left(\omega=0\right) &= \frac{\Gamma^2(\hat{\ell}+1-\hat{j})\Gamma^2(\hat{\ell}+1+\hat{j})}{\pi\,\Gamma(2\hat{\ell}+1)\Gamma(2\hat{\ell}+2)}\frac{\sin\pi(\hat{\ell}-\hat{j})\sin\pi(\hat{\ell}+\hat{j})}{\sin2\pi\hat{\ell}} \\
	&= \frac{\Gamma^2(\hat{\ell}+1-\hat{j})\Gamma^2(\hat{\ell}+1+\hat{j})}{2\pi\,\Gamma(2\hat{\ell}+1)\Gamma(2\hat{\ell}+2)}\left[\tan\pi\hat{\ell}\cos^2\pi\hat{j}-\cot\pi\hat{\ell}\sin^2\pi\hat{j}\right] \,.
\ea\ee
Ignoring at the moment the specific values of $\hat{j}$, the static Love numbers appear to exhibit the expected behavior. Namely, power counting arguments within the worldline EFT for General Relativity as prescribed in Section 3 of Ref.~\cite{Charalambous:2022rre} show that the static Love numbers should be non-zero and non-running for generic $2\hat{\ell}\notin\mathbb{N}$, while they are expected to exhibit a logarithmic running for $2\hat{\ell}\in\mathbb{N}$, seen above by a diverging behavior either as $\tan\pi\hat{\ell}$ (for $\hat{\ell}\in\mathbb{N}$) or as $\cot\pi\hat{\ell}$ (for $\hat{\ell}\in\mathbb{N}+\frac{1}{2}$)~\cite{Kol:2011vg,Hui:2020xxx,Charalambous:2023jgq}. However, taking into consideration the explicit possible values of $\hat{j}$ we see a very rich structure depending on the values of the rank of the $p$-form gauge field. First of all, for the static scalar, static magnetic and static electric susceptibilities
\be\ba\label{eq:Static01LNs_SchwarzschildDd}
	{}&k^{\left(j=0\right)}_{\ell} = \frac{\Gamma^{4}(\hat{\ell}+1)}{2\pi\,\Gamma(2\hat{\ell}+1)\Gamma(2\hat{\ell}+2)}\tan\pi\hat{\ell} \,, \\
	{}&k^{\left(j=1\right)}_{\ell} = \frac{\Gamma^2(\hat{\ell}+1-\frac{1}{d-3})\Gamma^2(\hat{\ell}+1+\frac{1}{d-3})}{\pi\,\Gamma(2\hat{\ell}+1)\Gamma(2\hat{\ell}+2)}\frac{\sin\pi(\hat{\ell}-\frac{1}{d-3})\sin\pi(\hat{\ell}+\frac{1}{d-3})}{\sin2\pi\hat{\ell}} \\
	{}&\text{and} \quad k^{\left(j=d-3\right)}_{\ell} = \frac{\Gamma^{2}(\hat{\ell})\Gamma^{2}(\hat{\ell}+2)}{2\pi\,\Gamma(2\hat{\ell}+1)\Gamma(2\hat{\ell}+2)}\tan\pi\hat{\ell}
\ea\ee
respectively, which agree with ones already obtained in Ref.~\cite{Hui:2020xxx}. Hence, there are still hints of fine-tuning coming from the vanishing of the static scalar susceptibilities ($j=0$) and the static electric susceptibilities ($j=d-3$) whenever $\hat{\ell}\in\mathbb{N}$. We also get the opportunity to see how the electric/magnetic duality is no longer present in $d>4$, namely, the static magnetic susceptibilities ($j=1$) vanish under the \textit{different} resonant conditions $\hat{\ell}\pm\frac{1}{d-3}\in\mathbb{N}$, which are always non-overlapping with the $\hat{\ell}\in\mathbb{N}$ case in $d>4$.

For generic $0< p\le d-3$, we can break down the investigation of the results into three classes, after also noting that $0<\hat{j}\le1$. The first class of $p$-form perturbations is when $\hat{j}$ is an integer, i.e. when $\hat{j}=1$. This corresponds to $p=d-3$ for the co-exact $p$-form $SO\left(d-1\right)$ sector or $p=1$ for the co-exact $\left(p-1\right)$-form $SO\left(d-1\right)$ sector. The latter is simply the electric-type electromagnetic response we saw above. The former is a new category of magnetic-like-type perturbations that emerges in $d>4$ and whose Love numbers are again identical to the static electric susceptibilities. These perturbations are just the Hodge dual version of the electric-type electromagnetic perturbations and they are merely a reflection of the Hodge duality symmetry, $\mathbf{F}^{\left(p+1\right)}\rightarrow \star\mathbf{F}^{\left(p+1\right)}$, of the $p$-form action. The qualitative behavior of static responses under $p$-form perturbations in this class is demonstrated in Table~\ref{tbl:Staticp1LNs_SchwarzschildDd}.

The second class of $p$-form perturbations is when $\hat{j}$ is a half-integer, i.e. when $\hat{j}=\frac{1}{2}$. This occurs only for odd spacetime dimensionalities, $d=5,7,\dots$, and now corresponds to $p=\frac{d-3}{2}$ for the co-exact $p$-form $SO\left(d-1\right)$ sector or $p=\frac{d-1}{2}$ for the co-exact $\left(p-1\right)$-form $SO\left(d-1\right)$ sector, the two types of perturbations again being related by Hodge duality. The static Love numbers for these cases read
\be
	k^{\left(j=\left(d-3\right)/2\right)}_{\ell} = -\frac{\Gamma^2\left(\hat{\ell}+\frac{1}{2}\right)\Gamma^2\left(\hat{\ell}+\frac{3}{2}\right)}{2\pi\,\Gamma(2\hat{\ell}+1)\Gamma(2\hat{\ell}+2)}\cot\pi\hat{\ell} = -\frac{1}{2^{4\hat{\ell}+3}}\frac{1}{k_{\ell}^{\left(j=0\right)}} \,,
\ee
where, in the second equality, we used the Legendre duplication formula for the $\Gamma$-function to compare with the static scalar Love numbers. We therefore see that the behavior of the static Love numbers for this class of $p$-form perturbations is opposite to that of the electric-type Love numbers, namely, they are non-zero and non-running for $2\hat{\ell}\notin\mathbb{N}$, they are logarithmically running for $\hat{\ell}\in\mathbb{N}$ and they are vanishing for $\hat{\ell}\in\mathbb{N}+\frac{1}{2}$, see Table~\ref{tbl:Staticp2LNs_SchwarzschildDd}.

The final, third, class of $p$-form perturbations contains all the other cases, for which $2\hat{j}\notin\mathbb{N}$. From the general expression for the static Love numbers in Eq.~\eqref{eq:StaticpLNs_SchwarzschildDd}, we see that these are non-zero and non-running for generic $\hat{\ell}$, they are logarithmically running for $2\hat{\ell}\in\mathbb{N}$ and are vanishing for $\hat{\ell}\pm\hat{j}\in\mathbb{N}$, see Table~\ref{tbl:Staticp3LNs_SchwarzschildDd}.

\begin{table}[t]
	\centering
	\begin{tabular}{|c||c|}
		\hline
		\Gape[7pt]{Range of parameters} & Behavior of $k_{\ell}^{\left(j\right)}\left(\omega=0\right)$ \\
		\hline\hline
		$\hat{\ell}\in\mathbb{N}$ & \Gape[8pt]{Vanishing} \\
		\hline
		$\hat{\ell}\in\mathbb{N}+\frac{1}{2}$ & \Gape[9pt]{Running} \\
		\hline
		$2\hat{\ell}\notin\mathbb{N}$ & \Gape[10pt]{Non-vanishing and Non-running} \\
		\hline
	\end{tabular}
	\caption[Behavior of static Love numbers for the first class of $p$-form perturbations of the higher-dimensional Schwarzschild-Tangherlini black hole.]{Behavior of static Love numbers for the first class of $p$-form perturbations of the higher-dimensional Schwarzschild-Tangherlini black hole, for which $\hat{j}=\frac{j}{d-3}$ is an integer. This class contains the static scalar susceptibilities ($j=0$) and the static electric susceptibilities as well as the Hodge dual co-exact $p$-form $SO\left(d-1\right)$ sector of $p$-form perturbations with $p=d-3$ ($j=d-3$). For generic orbital number, the static Love numbers for $p$-form perturbations in this class are non-zero and non-running. They are zero for integer $\hat{\ell}=\frac{\ell}{d-3}$ and they exhibit a classical RG flow for half-integer $\hat{\ell}$. As we will see later, the static electric-type and the static tensor-type tidal Love numbers also behave as prescribed here.}
	\label{tbl:Staticp1LNs_SchwarzschildDd}
\end{table}

\begin{table}[t]
	\centering
	\begin{tabular}{|c||c|}
		\hline
		\Gape[7pt]{Range of parameters} & Behavior of $k_{\ell}^{\left(j\right)}\left(\omega=0\right)$ \\
		\hline\hline
		$\hat{\ell}\in\mathbb{N}$ & \Gape[8pt]{Running} \\
		\hline
		$\hat{\ell}\in\mathbb{N}+\frac{1}{2}$ & \Gape[9pt]{Vanishing} \\
		\hline
		$2\hat{\ell}\notin\mathbb{N}$ & \Gape[10pt]{Non-vanishing and Non-running} \\
		\hline
	\end{tabular}
	\caption[Behavior of static Love numbers for the second class of $p$-form perturbations of the higher-dimensional Schwarzschild-Tangherlini black hole.]{Behavior of static Love numbers for the second class of $p$-form perturbations of the higher-dimensional Schwarzschild-Tangherlini black hole, for which $j=\frac{d-3}{2}$. This class exists only for odd spacetime dimensionalities and contains the static magnetic susceptibilities in $d=5$ ($j=1$). For generic orbital number, the static Love numbers for $p$-form perturbations in this class are non-zero and non-running. They are now zero for half-integer $\hat{\ell}=\frac{\ell}{d-3}$ and they exhibit a classical RG flow for integer $\hat{\ell}$.}
	\label{tbl:Staticp2LNs_SchwarzschildDd}
\end{table}

\begin{table}[t]
	\centering
	\begin{tabular}{|c||c|}
		\hline
		\Gape[7pt]{Range of parameters} & Behavior of $k_{\ell}^{\left(j\right)}\left(\omega=0\right)$ \\
		\hline\hline
		$\hat{\ell}+\hat{j}\in\mathbb{N}$ OR $\hat{\ell}-\hat{j}\in\mathbb{N}$ & \Gape[8pt]{Vanishing} \\
		\hline
		$2\hat{\ell}\in\mathbb{N}$ AND $\hat{\ell}\pm\hat{j}\notin\mathbb{N}$ & \Gape[9pt]{Running} \\
		\hline
		$2\hat{\ell}\notin\mathbb{N}$ AND $\hat{\ell}\pm\hat{j}\notin\mathbb{N}$ & \Gape[10pt]{Non-vanishing and Non-running} \\
		\hline
	\end{tabular}
	\caption[Behavior of static Love numbers for the third class of $p$-form perturbations of the higher-dimensional Schwarzschild-Tangherlini black hole.]{Behavior of static Love numbers for the third class of $p$-form perturbations of the higher-dimensional Schwarzschild-Tangherlini black hole, for which $\hat{j}=\frac{j}{d-3}$ is neither an integer nor a half-integer. This class contains the static magnetic susceptibilities ($j=1$) in $d\ge6$. For generic $\hat{\ell}=\frac{\ell}{d-3}$, the static Love numbers for $p$-form perturbations in this class are non-zero and non-running, while they exhibit a classical RG flow for $2\hat{\ell}\in\mathbb{N}$. They are now zero along the two branches of non-integer $\hat{\ell}$ cases $\hat{\ell}+\hat{j}\in\mathbb{N}$ or $\hat{\ell}-\hat{j}\in\mathbb{N}$. As we will see later, the static magnetic-type tidal Love numbers also behave as prescribed here, with the resonant conditions for vanishing Love numbers mimicking those for the static magnetic susceptibilities, for which $j=1$.}
	\label{tbl:Staticp3LNs_SchwarzschildDd}
\end{table}

Let us comment a bit more on what happens to the radial wavefunction for the various behaviors of the static Love numbers. First of all, for generic $\hat{\ell}$ and $\hat{j}$, the source/response split of the radial wavefunction can be performed by means of analytically continuing the hypergeometric function at large distances,
\be
	R^{\left(j\right)}_{\omega\ell,\mathbf{m}} = \bar{R}_{\ell,\mathbf{m}}^{\left(j\right)}\left(\omega\right)\rho^{\hat{\ell}}\left[Z^{\left(j\right)\text{source}}_{\omega\ell,\mathbf{m}}\left(\rho\right) + k_{\ell}^{\left(j\right)}\left(\omega\right)\left(\frac{\rho_{s}}{\rho}\right)^{2\hat{\ell}+1}Z^{\left(j\right)\text{response}}_{\omega\ell,\mathbf{m}}\left(\rho\right)\right] \,,
\ee
with $\bar{R}_{\ell,\mathbf{m}}\left(\omega\right)^{\left(j\right)}$ the strengths of the multipole moments of the perturbing source and
\be\ba
	Z^{\left(j\right)\text{source}}_{\omega\ell,\mathbf{m}}\left(\rho\right) &= \left(1-\frac{\rho_{s}}{\rho}\right)^{\hat{\ell}}\left(\frac{\rho-\rho_{s}}{\rho}\right)^{\sigma\hat{j}-i\beta\omega/2} \\
	&\quad\times {}_2F_1\left(-\hat{\ell}-\sigma\hat{j},-\hat{\ell}-\sigma\hat{j}+i\beta\omega;-2\hat{\ell};\frac{\rho_{s}}{\rho_{s}-\rho}\right) \,, \\
	Z^{\left(j\right)\text{response}}_{\omega\ell,\mathbf{m}}\left(\rho\right) &= \left(1-\frac{\rho_{s}}{\rho}\right)^{-\hat{\ell}-1}\left(\frac{\rho-\rho_{s}}{\rho}\right)^{\sigma\hat{j}-i\beta\omega/2} \\
	&\quad\times {}_2F_1\left(\hat{\ell}+1-\sigma\hat{j},\hat{\ell}+1-\sigma\hat{j}+i\beta\omega;2\hat{\ell}+2;\frac{\rho_{s}}{\rho_{s}-\rho}\right) \,.
\ea\ee
Therefore, the ``source'' and ``response'' parts consist in general of two infinite series in a large distance expansion. For generic $\hat{\ell}$, these series are non-overlapping, hence the non-vanishing and non-running Love numbers. For $2\hat{\ell}\in\mathbb{N}$, the ``source'' series begins overlapping with the ``response'' series. One way to see that this introduces a logarithmic running is to set $2\hat{\ell} = n-\varepsilon$ and expand the wavefunction around small $\varepsilon$. One then observes that both the ``source'' series and the Love numbers develop single poles but in precisely such a way that two poles cancel each, leaving a total wavefunction with no poles but involving logarithms coming from terms of the form $\rho^{2\hat{\ell}+1} = \rho^{n+1}\left(1-\varepsilon\log \rho\right) + \mathcal{O}\left(\varepsilon^2\right)$. From the worldline EFT side, the Love numbers get renormalized from diagrams of the form
\be
	\delta\phi_{\mathbf{a}}^{\left(s\right)} \supset \vcenter{\hbox{\begin{tikzpicture}
				\begin{feynman}
					\vertex[dot] (a0);
					\vertex[below=1cm of a0] (p1);
					\vertex[above=1cm of a0] (p2);
					\vertex[right=0.4cm of a0, blob] (gblob){};							
					\vertex[right=1.5cm of p1] (b1);
					\vertex[right=1.5cm of p2] (b2);
					\vertex[right=1.19cm of p2] (b22){$\times$};
					\vertex[above=0.7cm of a0] (g1);
					\vertex[above=0.4cm of a0] (g2);
					\vertex[right=0.05cm of a0] (gdtos){$\vdots$};
					\vertex[below=0.7cm of a0] (gN);
					\vertex[left=0.2cm of gN] (gN1);
					\vertex[left=0.2cm of g1] (g11);
					\diagram*{
						(p1) -- [double,double distance=0.5ex] (p2),
						(g1) -- [photon] (gblob),
						(g2) -- [photon] (gblob),
						(gN) -- [photon] (gblob),
						(b1) -- (gblob) -- (b2),
					};
					\draw [decoration={brace}, decorate] (gN1.south west) -- (g11.north west)
					node [pos=0.55, left] {\(2\hat{\ell}+1\)};
				\end{feynman}
	\end{tikzpicture}}} \,,
\ee
coming from the $\big(2\hat{\ell}+1\big)$'th relativistic correction to the Newtonian source~\cite{Porto:2016pyg,Charalambous:2021mea,Charalambous:2022rre,Ivanov:2022hlo}. To be more explicit, whenever $2\hat{\ell}\in\mathbb{N}$, the radial wavefunction reads
\be\ba
	{}&R^{\left(j\right)}_{\omega\ell,\mathbf{m}} = \frac{\bar{R}_{\ell,\mathbf{m}}^{\left(j\right)}\left(\omega\right)\rho_{s}^{\hat{\ell}}}{\Gamma\left(1-i\beta\omega\right)}\left(\frac{\rho-\rho_{s}}{\rho}\right)^{\sigma\hat{j}-i\beta\omega/2} \\
	&\times\bigg\{ \left(\frac{\rho-\rho_{s}}{\rho_{s}}\right)^{\hat{\ell}}\sum_{k=0}^{2\hat{\ell}}\frac{(-\hat{\ell}-\sigma\hat{j})_{k}}{(\hat{\ell}+1+\sigma\hat{j}-i\beta\omega-k)_{k}}\frac{(2\hat{\ell}-k)!}{(2\hat{\ell})!k!}\left(-x\right)^{-k} \\
	&+\beta_{\ell}^{\left(j\right)}\left(\omega\right)\left(\frac{\rho-\rho_{s}}{\rho_{s}}\right)^{-\hat{\ell}-1} \sum_{k=0}^{\infty}\frac{(\hat{\ell}+1-\sigma\hat{j})_{k}}{(-\hat{\ell}+\sigma\hat{j}-i\beta\omega-k)_{k}}\frac{(2\hat{\ell}+1)!}{(2\hat{\ell}+1+k)!k!}x^{-k} \\
	&\times\bigg[\log x + \psi\left(k+1\right) + \psi(2\hat{\ell}+2+k) - \psi(\hat{\ell}+1-\sigma\hat{j}+k) - \psi(-\hat{\ell}+\sigma\hat{j}-i\beta\omega-k) \bigg] \bigg\} \,,
\ea\ee
where we have also identified the relevant $\beta$-function associated with the running Love numbers,
\be
	\beta_{\ell}^{\left(j\right)}\left(\omega\right) = \frac{dk_{\ell}^{\left(j\right)}\left(\omega\right)}{d\log L} = \frac{\left(-1\right)^{2\hat{\ell}+1}}{(2\hat{\ell})!(2\hat{\ell}+1)!}\frac{\Gamma(\hat{\ell}+1-\sigma\hat{j})\Gamma(\hat{\ell}+1+\sigma\hat{j}-i\beta\omega)}{\Gamma(-\hat{\ell}-\sigma\hat{j})\Gamma(-\hat{\ell}+\sigma\hat{j}-i\beta\omega)} \,.
\ee

Last, for the resonant conditions $\hat{\ell}+\sigma\hat{j}\in\mathbb{N}$, we see that the $\beta$-function above vanishes, collapsing the radial wavefunction to a (quasi-)polynomial. More generally, even for $2\hat{j}\notin\mathbb{N}$, we see that the hypergeometric function in Eq.~\eqref{eq:RNearZoneJ} reduces to a polynomial,
\be\ba
	R^{\left(j\right)}_{\omega\ell,\mathbf{m}}\bigg|_{\hat{\ell}+\sigma\hat{j} \in \mathbb{N}} &= \bar{R}^{\left(j\right)\text{in}}_{\ell,\mathbf{m}}\left(\omega\right)\left(\frac{x}{1+x}\right)^{-i\beta\omega/2}\left(1+x\right)^{-\sigma\hat{j}}\sum_{n=0}^{\hat{\ell}+\sigma\hat{j}}\begin{pmatrix} \hat{\ell}+\sigma\hat{j} \\ n \end{pmatrix} \frac{\left(\hat{\ell}+1-\sigma\hat{j}\right)_{n}}{\left(1-i\beta\omega\right)_{n}}x^{n} \,.
\ea\ee
It is this (quasi-)polynomial behavior that is characteristic of the vanishing of the Love numbers.

\subsubsection{Tidal Love numbers}
For the gravitational (spin-$2$) response of the Schwarzschild-Tangherlini black hole, most of the analysis turns out to be exactly the same as the $p$-form perturbation analysis above for particular values of $p$. More specifically, after performing the field redefinitions
\be
	\Phi^{\left(\text{T}\right)}_{\ell,\mathbf{m}} = \frac{\Psi^{\left(\text{T}\right)}_{\ell,\mathbf{m}}}{r^{\frac{d-2}{2}}} \quad\text{and}\quad \Phi^{\left(\text{RW}\right)}_{\ell,\mathbf{m}} = \frac{\Psi^{\left(\text{RW}\right)}_{\ell,\mathbf{m}}}{r^{\frac{d-2}{2}}} \,,
\ee
the equation of motion for the tensor modes becomes identical to the equation of motion for the scalar field perturbations, Eq.~\eqref{eq:FullpEOMSchwarzschild} with $j=0$. The equation of motion for the spin-$2$ magnetic-type (Regge-Wheeler) modes also takes the form of the $p$-form perturbations equations of motion Eq.~\eqref{eq:FullpEOMSchwarzschild}, now corresponding to the value $j=d-2$. Interestingly, the magnetic-type and tensor-type gravitational perturbations of the higher-dimensional Schwarzschild-Tangherlini black hole obey the same equations of motion as the co-exact $p$-form and co-exact $\left(p-1\right)$-form modes, respectively, for $p=d-2$. The corresponding static magnetic-type and static tensor-type Love number are therefore captured by the general expression in Eq.~\eqref{eq:StaticpLNs_SchwarzschildDd} with $j=d-2$ and $j=0$ respectively.

Consequently, the tensor-type tidal Love numbers and the scalar Love numbers of the higher-dimensional Schwarzschild-Tangherlini black hole are exactly the same and behave the same way as the Love numbers of the first class of $p$-form perturbations, see Table~\ref{tbl:Staticp1LNs_SchwarzschildDd}. Similarly, the magnetic-type Love numbers of the higher-dimensional Schwarzschild-Tangherlini black hole behave the same way as the second class of $p$-form perturbations for $d=5$ and the same way as the third class of $p$-form perturbations for $d\ge6$, see Table~\ref{tbl:Staticp2LNs_SchwarzschildDd} and Table~\ref{tbl:Staticp3LNs_SchwarzschildDd} respectively.

Let us also note that, since the master variables entering the gravitational perturbations equations of motion are built at most from derivatives of the actual fields in terms of which the response problem is defined, the response coefficients in front of the decaying branches of these master variables are proportional to the actual response coefficients we are looking for, namely,
\be
	k^{\left(\text{T}\right)}_{\ell}\left(\omega\right) = k^{\mathcal{T},\left(2\right)}_{\ell}\left(\omega\right) \,,\quad k^{\left(\text{RW}\right)}_{\ell}\left(\omega\right) = -\frac{\ell+d-2}{\ell-1}k^{\mathcal{B},\left(2\right)}_{\ell}\left(\omega\right) \,.
\ee

Unfortunately, we have not been able to find a useful near-zone truncation of the Zerilli equation of motion. At least for static perturbations, the Zerilli equation has been shown in Refs.~\cite{Kodama:2003jz,Ishibashi:2003ap} to reduce to a hypergeometric differential equation after performing a particular Darboux transformation, see also Refs.~\cite{Hui:2020xxx,Glampedakis:2017rar}. Using this fact, the authors in Ref.~\cite{Hui:2020xxx} have been able to extract the corresponding static electric-type tidal Love numbers to be
\be
	k^{\left(\text{Z}\right)}_{\ell} = \frac{\hat{\ell}}{\hat{\ell}+1}\frac{\Gamma^{2}(\hat{\ell})\Gamma^{2}(\hat{\ell}+2)}{2\pi\,\Gamma(2\hat{\ell}+1)\Gamma(2\hat{\ell}+2)}\tan\pi\hat{\ell} = \frac{\left(\ell+d-3\right)\left(\ell+d-2\right)}{\ell\left(\ell-1\right)}k_{\ell}^{\mathcal{E},\left(2\right)} \,
\ee
where in the second equality we have demonstrated how the static response coefficients of the Zerilli modes are related to the actual response coefficients associated with fields in terms of which the response problem is defined~\cite{Hui:2020xxx}.

\subsection{Scalar and tensor Love numbers of Reissner-Nordstr\"{o}m black holes}
\label{sec:LNs_RNDd}

Next, we consider the higher-dimensional electrically charged Reissner-Nordstr\"{o}m black hole, Eq.~\eqref{eq:RNDdGeometry}. To avoid dealing with coupled differential equations, we will focus to spin-$0$ scalar mode and spin-$2$ tensor mode perturbations. The equations of motion for the scalar field perturbations have the same form (see Eq.~\eqref{eq:V0S}). As for the gravitational tensor modes, even though there are no tensor modes for the gauge field perturbations to couple to the gravitational tensor modes, one needs to supplement with the contribution of the background electromagnetic field which comes from the Maxwell action,
\be\ba
	S^{\left(1\right)}_{\text{full}} &= \int d^{d}x\,\sqrt{-g^{\text{full}}}\left[-\frac{1}{4}F^{\text{full}}_{\mu\nu}F^{\text{full}\,\mu\nu}\right] \\
	&\supset \int d^{d}x\,\sqrt{-g}\,\left[2\pi GF_{\mu\nu}F^{\mu\nu}\left(h_{\mu\nu}h^{\mu\nu}-\frac{1}{2}h^2\right)\right] \\
	&\supset \sum_{\ell,\mathbf{m}}\int d^2x\,\sqrt{-g^{\left(2\right)}}\,r^{d-2}\left[ 2\pi G F_{\mu\nu}F^{\mu\nu}\left|h^{\left(\text{T}\right)}_{\ell,\mathbf{m}}\right|^2 \right] \,,
\ea\ee
as well as an additional contribution from the Einstein-Hilbert action due to a non-zero background energy-momentum tensor,
\be\ba
	S^{\left(\text{gr}\right)}_{\text{full}} &= \int d^{d}x\,\sqrt{-g^{\text{full}}}\left[\frac{1}{16\pi G}R^{\text{full}}\right] \\
	&\supset \int d^{d}x\,\sqrt{-g}\,\left[-\frac{R}{2}\left(h_{\mu\nu}h^{\mu\nu}-\frac{1}{2}h^2\right)\right] \\
	&\supset \sum_{\ell,\mathbf{m}}\int d^2x\,\sqrt{-g^{\left(2\right)}}\,r^{d-2}\left[ -2\pi G\frac{d-4}{d-2} F_{\mu\nu}F^{\mu\nu}\left|h^{\left(\text{T}\right)}_{\ell,\mathbf{m}}\right|^2 \right] \,.
\ea\ee
Taking these into account, the equations of motion for the spin-$0$ scalar and spin-$2$ tensor modes in the background of a Reissner-Nordstr\"{o}m black hole turn out to be exactly the same~\cite{Ishibashi:2011ws,Kodama:2003kk,Pereniguez:2021xcj}. We can then follow the same footsteps as for the higher-dimensional Schwarzschild-Tangherlini black hole. We employ the near-zone splitting analogous to the $j=0$ Eq.~\eqref{eq:Vp_SchwarzschildD},
\be\label{eq:V0V1_RNDd}
	\begin{gathered}
		\mathbb{O}^{\left(0\right)}_{\text{full}} = \partial_{\rho}\,\Delta\,\partial_{\rho} + V_0 + \epsilon\,V_1 \,, \\
		V_0 = -\frac{\left(\rho_{+}-\rho_{-}\right)^2}{4\Delta}\beta^2\partial_{t}^2 \,,\quad V_1 = -\frac{r^2\rho^2-r_{+}^2\rho_{+}^2}{\left(d-3\right)^2\Delta}\partial_{t}^2 \,,
	\end{gathered}
\ee
where we have again introduced $\rho = r^{d-3}$, and $\beta = \frac{2r_{+}}{d-3}\frac{\rho_{+}}{\rho_{+}-\rho_{-}}$ is the inverse surface gravity of the $d$-dimensional Reissner-Nordstr\"{o}m black hole. The leading order near-zone radial solution that is ingoing at the future event horizon has the same form,
\be
	R^{\left(0\right)}_{\omega\ell,\mathbf{m}} = \bar{R}^{\left(0\right)\text{in}}_{\ell,\mathbf{m}}\left(\omega\right)\left(\frac{x}{1+x}\right)^{-i\beta\omega/2}{}_2F_1\left(\hat{\ell}+1,-\hat{\ell};1-i\beta\omega;-x\right) \,,
\ee
where now $x=\frac{\rho-\rho_{+}}{\rho_{+}-\rho_{-}}$, and the corresponding dissipative response coefficients and Love numbers are extracted to be
\be\label{eq:ResponseCoefficientsRND}
	\begin{gathered}
		k^{\left(0\right)}_{\ell}\left(\omega\right) = k^{\left(0\right)\text{Love}}_{\ell}\left(\omega\right) + ik^{\left(0\right)\text{diss}}_{\ell}\left(\omega\right) \,, \\\\
		\ba
			k^{\left(0\right)\text{diss}}_{\ell}\left(\omega\right) &= A_{\ell}\left(\omega\right)\sinh\pi\beta\omega + \mathcal{O}\left(\beta^3\omega^3\right)\,, \\
			k^{\left(0\right)\text{Love}}_{\ell}\left(\omega\right) &= A_{\ell}\left(\omega\right)\tan\pi\hat{\ell}\cosh\pi\beta\omega + \mathcal{O}\left(\beta^2\omega^2\right) \,,
		\ea \\\\
		A_{\ell}\left(\omega\right) = \frac{\Gamma^2(\hat{\ell}+1)\left|\Gamma(\hat{\ell}+1-i\beta\omega)\right|^2}{2\pi\Gamma(2\hat{\ell}+1)\Gamma(2\hat{\ell}+2)}\left(\frac{\rho_{+}-\rho_{-}}{\rho_{s}}\right)^{2\hat{\ell}+1} \,.
	\end{gathered}
\ee
Again, these vanish for integer $\hat{\ell}$, now even beyond the static limit but always at leading order in the near-zone expansion, as emphasized here by the $\mathcal{O}\left(\beta^2\omega^2\right)$ corrections that enter at higher near-zone orders, while for other values of the orbital number they exhibit the expected behavior based on power counting arguments; they logarithmically run for half-integer $\hat{\ell}$ and they are non-zero and non-running for $2\hat{\ell}\notin\mathbb{N}$~\cite{Charalambous:2022rre}.
\section{Love symmetries for $p$-form and gravitational perturbations in higher dimensions}
\label{sec:LoveSymmetryDd}

As demonstrated in the last section, the static black hole Love numbers exhibit towers of resonant conditions for which they vanish. From the worldline EFT side, this raises naturalness concerns~\cite{Porto:2016zng} and calls upon the existence of enhanced symmetries~\cite{tHooft:1979rat}, outputting these vanishings as selection rules. From the microscopic theory side, these enhanced symmetries are not expected to be exact isometries of the full background geometry but, rather, approximate symmetries manifesting themselves in the appropriate domain.

In view of the worldline EFT definition of the Love numbers and its accuracy regime, it is natural to expect these enhanced symmetries to be linked to the near-zone expansion we have employed in our microscopic computations. Furthermore, it is a well established result that only the static Love numbers are the culprit of such fine-tuning issues, while the dynamical Love numbers are in general non-zero and accordingly logarithmically running, in line with Wilsonian naturalness arguments~\cite{Charalambous:2021mea,Saketh:2023bul,Perry:2023wmm}. Consequently, it should suffice to seek for these enhanced symmetries at leading order in the near-zone approximation.

Indeed, it has been signified that there exist near-zone truncations of the equations of motion that give rise to globally defined $\SL$ symmetries, dubbed ``Love symmetries'', whose global structure allows to employ (highest-weight) representation theory arguments and precisely output the seemingly fine-tuned properties of the static Love numbers~\cite{Charalambous:2021kcz,Charalambous:2022rre,Charalambous:2023jgq}. This has been demonstrated for scalar, electromagnetic and gravitational perturbations of the $d=4$ Kerr-Newman black hole and for scalar perturbations of the $d$-dimensional Schwarzschild-Tangherlini black hole in Refs.~\cite{Charalambous:2021kcz,Charalambous:2022rre} and the $d=5$ doubly-rotating Myers-Perry black hole in Ref.~\cite{Charalambous:2023jgq}. Other than these, the Love symmetry proposal for higher-spin perturbations of higher-dimensional black holes has not been investigated, which is the scope of the current section.

Despite the intricate structure of the black hole Love numbers in higher spacetime dimensions, Love symmetry turns out to still exist independently of the value of the rescaled orbital number $\hat{\ell}$. There are now two sets of Love symmetry generators, one for each sign $\sigma=+1$ or $\sigma=-1$ that characterizes the near-zone split in Eq.~\eqref{eq:Vp_SchwarzschildD}. The two Love symmetries are generated by
\be\label{eq:SL2R_SchwarzschildDd}
	\begin{gathered}
		L_0^{\left(\sigma,j\right)} = -\beta\,\partial_{t} - \sigma\hat{j} \,, \\
		L_{\pm1}^{\left(\sigma,j\right)} = e^{\pm t/\beta}\left[\mp\sqrt{\Delta}\,\partial_{\rho} + \partial_{\rho}\left(\sqrt{\Delta}\right)\beta\,\partial_{t} + \sigma\hat{j}\sqrt{\frac{\rho-\rho_{+}}{\rho-\rho_{-}}} \right] \,,
	\end{gathered}
\ee
and satisfy the $\SL$ algebra,
\be
	\left[L_{m}^{\left(\sigma,j\right)},L_{n}^{\left(\sigma,j\right)}\right] = \left(m-n\right)L_{m+n}^{\left(\sigma,j\right)} \,,\quad m,n=0,\pm1 \,,
\ee
while the corresponding Casimir is given by
\be\ba
	\mathcal{C}_2^{\left(\sigma,j\right)} &= \left(L_0^{\left(\sigma,j\right)}\right)^2 - \frac{1}{2}\left(L_{+1}^{\left(\sigma,j\right)}L_{-1}^{\left(\sigma,j\right)} + L_{-1}^{\left(\sigma,j\right)}L_{+1}^{\left(\sigma,j\right)}\right) \\
	&= \partial_{\rho}\,\Delta\,\partial_{\rho} - \frac{\left(\rho_{+}-\rho_{-}\right)^2}{4\Delta}\beta^2\partial_{t}^2 + \hat{j}\frac{\rho_{+}-\rho_{-}}{\rho-\rho_{-}}\left(\sigma\beta\,\partial_{t} + \hat{j}\right) \,,
\ea\ee
which exactly matches the leading order near-zone radial operators for the various cases encountered up until now, i.e. with Eq.~\eqref{eq:Vp_SchwarzschildD} for the spin-$0$, $p$-form and spin-$2$ perturbations of the Schwarzschild-Tangherlini black hole and with Eq.~\eqref{eq:V0V1_RNDd} for the spin-$0$ scalar mode and spin-$2$ tensor mode perturbations of the Reissner-Nordstr\"{o}m black hole. To investigate the regularity of the generators in Eq.~\eqref{eq:SL2R_SchwarzschildDd} at the future or the past event horizons, we need to study their near-horizon behavior after transitioning to advanced ($+$) or retarded ($-$) null coordinates $\left(t_{\pm},r,\theta^{A}\right)$ respectively. Although there is no useful closed form for the null coordinates in generic spacetime dimensionality $d\ge4$, we can still study the near-horizon behavior thanks to Eq.~\eqref{eq:AdvancedRetarded_SchwazrschildDd}-\eqref{eq:Tortoise_SchwazrschildDd}. Doing this, one then immediately sees that the generators in Eq.~\eqref{eq:SL2R_SchwarzschildDd} are indeed regular at both the future and the past event horizons.

Separable solutions $\Phi^{\left(j\right)}_{\omega\ell,\mathbf{m}}$ of the near-zone equations of motion are then observed to furnish representations of the Love symmetry,
\be
	\mathcal{C}_2^{\left(\sigma,j\right)}\Phi^{\left(j\right)}_{\omega\ell,\mathbf{m}} = \hat{\ell}(\hat{\ell}+1)\,\Phi^{\left(j\right)}_{\omega\ell,\mathbf{m}} \,,\quad L_0^{\left(\sigma,j\right)}\Phi^{\left(j\right)}_{\omega\ell,\mathbf{m}} = (i\beta\omega-\sigma\hat{j})\,\Phi^{\left(j\right)}_{\omega\ell,\mathbf{m}} \,.
\ee
The static solution, in particular, has an $L_0$-eigenvalue
\be
	L_0^{\left(\sigma,j\right)}\Phi^{\left(j\right)}_{\omega=0,\ell,\mathbf{m}} = -\sigma\hat{j}\,\Phi^{\left(j\right)}_{\omega=0,\ell,\mathbf{m}} \,.
\ee
One important difference compared to the four-dimensional case is that the Casimir eigenvalue is now in general non-integer unless $\hat{\ell}\in\mathbb{N}$. Furthermore, the weight of the static solution is also not integer unless $\hat{j}=0$ or $\hat{j}=1$, i.e. $j=0$ or $j=d-3$ respectively.

\subsection{Scalar/tensor perturbations of Reissner-Nordstr\"{o}m black holes}
Let us begin with the cases where $j=0$. These capture the spin-$0$ scalar and spin-$2$ tensor modes of the Reissner-Nordstr\"{o}m black hole perturbations. The vector fields generating the Love $\SL$ symmetry simplify to
\be
	L_0 = -\beta\,\partial_{t} \,,\quad L_{\pm1} = e^{\pm t/\beta}\left[\mp\sqrt{\Delta}\,\partial_{\rho} + \partial_{\rho}\left(\sqrt{\Delta}\right)\beta\,\partial_{t} \right] \,,
\ee
which have the exact same form as the ones presented in Refs.~\cite{Bertini:2011ga,Charalambous:2021kcz,Charalambous:2022rre} for the higher-dimensional Schwarzschild-Tangherlini black hole, here extended to the case of the higher-dimensional Reissner-Nordstr\"{o}m black holes. For $d=4$, this reduces to the $\SL$ symmetry found in Ref.~\cite{Kim:2012mh}. Analogously to the $d=4$ examples, let us construct the highest-weight representation with weight $h=-\hat{\ell}$, starting from the primary state $\upsilon_{-\hat{\ell},0}$, satisfying~\cite{Charalambous:2021kcz,Charalambous:2022rre}
\be
	L_{+1}\upsilon_{-\hat{\ell},0} = 0 \,,\quad L_0\upsilon_{-\hat{\ell},0} = -\hat{\ell}\,\upsilon_{-\hat{\ell},0} \quad\Rightarrow\quad \upsilon_{-\hat{\ell},0} = \left(-e^{+t/\beta}\sqrt{\Delta}\right)^{\hat{\ell}} \,,
\ee
where the spherical symmetry of the background geometry has allowed us to focus on axisymmetric ($\mathbf{m}=\mathbf{0}$) perturbations without loss of generality. This state is always regular at the future event horizon, while it is regular at the past event horizon as long as $e^{+t/\beta}\sqrt{\Delta}\sim e^{t_{-}/\beta} \left(r-r_{+}\right)$ is not raised to any negative power. The descendants,
\be
	\upsilon_{-\hat{\ell},n} = \left(L_{-1}\right)^{n} \upsilon_{-\hat{\ell},0} \,,
\ee
are also always regular at the future event horizon and have $L_0$-eigenvalues
\be
	L_0\upsilon_{-\hat{\ell},n} = (n-\hat{\ell})\,\upsilon_{-\hat{\ell},n} \,.
\ee
We see, therefore, the qualitative new feature in $d>4$, compared to $d=4$, that the static solution does not in general belong to a highest-weight representation of the Love $\SL$ symmetry. In particular, the static scalar/tensor mode solution $\Phi_{\omega=0,\ell,\mathbf{m}}$ that is regular at the horizon is an element of the above highest-weight representation \text{if and only if}
\be
	\hat{\ell} \in \mathbb{N} \,,
\ee
which indeed captures the resonant conditions for which the static scalar, and tensor-type tidal, Love numbers of the Reissner-Nordstr\"{o}m black hole vanish. In these cases, the static solution regular at the horizon is identified with the zero $L_0$-eigenvalue descendant
\be
	\text{If $\hat{\ell}\in\mathbb{N}$: }\quad \Phi^{\left(0\right)}_{\omega=0,\ell,\mathbf{m}=\mathbf{0}} \propto \upsilon_{-\hat{\ell},\hat{\ell}} = \left(L_{-1}\right)^{\hat{\ell}}\upsilon_{-\hat{\ell},0} \,.
\ee
Since this is the $\hat{\ell}$'th descendant in a highest-weight representation, it is annihilated by $\left(L_{+1}\right)^{\hat{\ell}+1}$,
\be
	\left(L_{+1}\right)^{\hat{\ell}+1}\Phi^{\left(0\right)}_{\omega=0,\ell,\mathbf{m}=\mathbf{0}} = 0 \quad \text{if $\hat{\ell}\in\mathbb{N}$} \,.
\ee
Using the fact that, for an arbitrary time-independent function $F\left(\rho\right)$,
\be
	\left(L_{+1}\right)^{n}F\left(\rho\right) = \left(-e^{t/\beta}\sqrt{\Delta}\right)^{n}\frac{d^{n}}{d\rho^{n}}F\left(\rho\right) \,,
\ee
we see then the highest-weight property immediately implies a polynomial form in $\rho$,
\be
	\text{If $\hat{\ell}\in\mathbb{N}$ }\Rightarrow \Phi^{\left(0\right)}_{\omega=0,\ell,\mathbf{m}=\mathbf{0}} \propto \upsilon_{-\hat{\ell},\hat{\ell}} = \sum_{n=0}^{\hat{\ell}}c_{n}\rho^{n} = c_{\hat{\ell}}r^{\ell} + \dots + c_0 \,.
\ee
Compared to explicit microscopic computations for the regular solution of the static Klein-Gordon equation, the polynomial above corresponds to the Legendre polynomial of degree $\hat{\ell}$~\cite{Kol:2011vg}. More importantly, the absence of terms $\propto r^{-\left(\ell+d-3\right)}$ is precisely indicative of the vanishing of the corresponding static Love number.

\begin{figure}[t]
	\centering
	\begin{tikzpicture}
		\node at (0,0) (uml4) {$\upsilon_{-\hat{\ell},2\hat{\ell}}$};
		\node at (0,1) (uml3) {$\upsilon_{-\hat{\ell},\hat{\ell}}$};
		\node at (0,2) (uml2) {$\upsilon_{-\hat{\ell},2}$};
		\node at (0,3) (uml1) {$\upsilon_{-\hat{\ell},1}$};
		\node at (0,4) (uml0) {$\upsilon_{-\hat{\ell},0}$};
		
		\draw [snake=zigzag] (1,-0.1) -- (5,-0.1);
		\draw (1,0) -- (5,0);
		\draw (1,1) -- (5,1);
		\node at (3,1.5) (up) {$\vdots$};
		\node at (3,0.5) (um) {$\vdots$};
		\draw (1,2) -- (5,2);
		\draw (1,3) -- (5,3);
		\draw (1,4) -- (5,4);
		\draw [snake=zigzag] (1,4.1) -- (5,4.1);
		
		\draw[blue] [->] (2.5,2) -- node[left] {$L_{+1}$} (2.5,3);
		\draw[blue] [->] (2,3) -- node[left] {$L_{+1}$} (2,4);
		\draw[red] [<-] (4,3) -- node[right] {$L_{-1}$} (4,4);
		\draw[red] [<-] (3.5,2) -- node[right] {$L_{-1}$} (3.5,3);
	\end{tikzpicture}
	\caption[The finite-dimensional highest-weight representation of $\SL$ whose elements solve the leading order near-zone equations of motion for a massless scalar field in the $d$-dimensional Reissner-Nordstr\"{o}m black hole background with integer rescaled multipolar index $\hat{\ell}=\frac{\ell}{d-3}$ and contains the regular static solution.]{The finite-dimensional highest-weight representation of $\SL$ whose elements solve the leading order near-zone equations of motion for a massless scalar field in the $d$-dimensional Reissner-Nordstr\"{o}m black hole background with integer rescaled multipolar index $\hat{\ell}=\frac{\ell}{d-3}$ and contains the regular static solution $\Phi_{\omega=0,\ell,\mathbf{m}=\mathbf{0}}^{\left(0\right)\text{regular}}\propto \upsilon_{-\hat{\ell},\hat{\ell}}$.}
	\label{fig:HWSL2R0_SchwarzschildDd}
\end{figure}
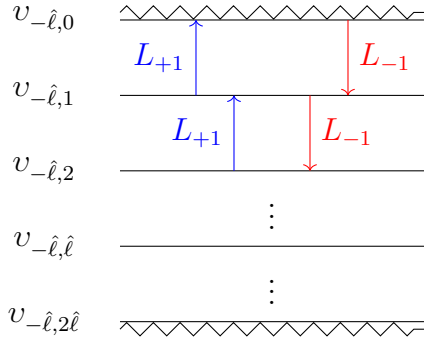

The same conclusion can be drawn from the lowest-weight representation of weight $\bar{h}=+\hat{\ell}$. Starting from the lowest-weight state
\be
	L_{-1}\bar{\upsilon}_{+\hat{\ell},0} = 0 \,,\quad L_0\bar{\upsilon}_{+\hat{\ell},0} = +\hat{\ell}\,\bar{\upsilon}_{+\hat{\ell},0} \quad \Rightarrow \quad \bar{\upsilon}_{+\hat{\ell},0} = \left(+e^{-t/\beta}\sqrt{\Delta}\right)^{\hat{\ell}} \,,
\ee
which is also a solution of the leading order near-zone massless Klein-Gordon equation with rescaled multipolar index $\hat{\ell}$ that is regular on both the future and the past event horizons, the ascendants
\be
	\bar{\upsilon}_{+\hat{\ell},n} = \left(L_{+1}\right)^{n}\bar{\upsilon}_{+\hat{\ell},0}
\ee
have an $L_0$-charge
\be
	L_0\bar{\upsilon}_{+\hat{\ell},n} = (\hat{\ell}-n)\,\bar{\upsilon}_{+\hat{\ell},n} \,,
\ee
and they are all regular at the past event horizon, while they are regular at the future event horizon only for $n<2\hat{\ell}+1$. A regular static solution then belongs to this representation if and only $\hat{\ell}\in\mathbb{N}$, in which case it is identified with the ascendant with zero $L_0$-eigenvalue,
\be
	\text{If $\hat{\ell}\in\mathbb{N}$: } \quad \Phi^{\left(0\right)}_{\omega=0,\ell,\mathbf{m}=\mathbf{0}} \propto \bar{\upsilon}_{+\hat{\ell},-\hat{\ell}} = \left(-L_{+1}\right)^{\hat{\ell}}\bar{\upsilon}_{+\hat{\ell},0} \,.
\ee
Since the regular static solution is unique, we see then that the highest-weight and lowest-weight representations are in fact identical and, hence, this is a finite $(2\hat{\ell}+1)$-dimensional representation of $\SL$ (see Figure~\ref{fig:HWSL2R0_SchwarzschildDd}),
\be
	\text{If $\hat{\ell}\in\mathbb{N}$ }\Rightarrow \bar{\upsilon}_{+\hat{\ell},0} = \upsilon_{-\hat{\ell},2\hat{\ell}} \,.
\ee

This property can be traced back to the time-reversal symmetry of the background. Solutions of the leading order near-zone equations of motion regular at the future event horizon belong to a highest-weight representation, while the corresponding solutions regular at the past event horizon belong to a lowest-weight representation. Indeed, the $t\rightarrow-t$ symmetry ensures that static scalar perturbations regular at the future event horizon will also be regular at the past event horizon and therefore, the two representations overlap to furnish the finite-dimensional representation of the Love $\SL$ symmetry we just saw.

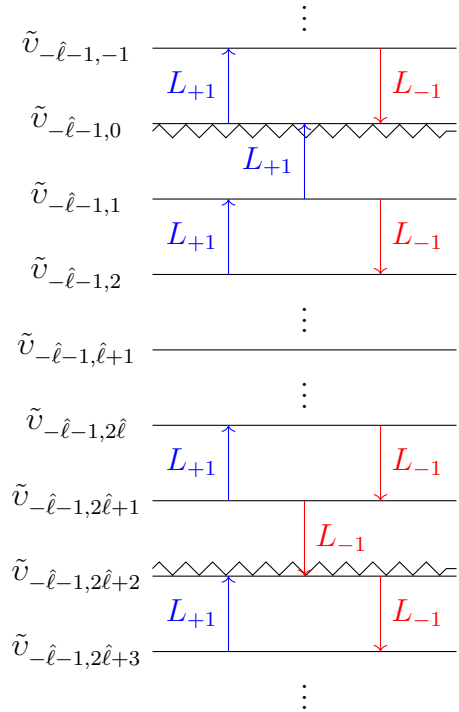
\begin{figure}[t]
	\centering
	\begin{tikzpicture}
		\node at (0,-1) (uml8) {$\tilde{\upsilon}_{-\hat{\ell}-1,2\hat{\ell}+3}$};
		\node at (0,0) (uml7) {$\tilde{\upsilon}_{-\hat{\ell}-1,2\hat{\ell}+2}$};
		\node at (0,1) (uml6) {$\tilde{\upsilon}_{-\hat{\ell}-1,2\hat{\ell}+1}$};
		\node at (0,2) (uml5) {$\tilde{\upsilon}_{-\hat{\ell}-1,2\hat{\ell}}$};
		\node at (0,3) (uml4) {$\tilde{\upsilon}_{-\hat{\ell}-1,\hat{\ell}+1}$};
		\node at (0,4) (uml3) {$\tilde{\upsilon}_{-\hat{\ell}-1,2}$};
		\node at (0,5) (uml2) {$\tilde{\upsilon}_{-\hat{\ell}-1,1}$};
		\node at (0,6) (uml1) {$\tilde{\upsilon}_{-\hat{\ell}-1,0}$};
		\node at (0,7) (uml0) {$\tilde{\upsilon}_{-\hat{\ell}-1,-1}$};
		
		\node at (3,-1.5) (umm) {$\vdots$};
		\draw (1,-1) -- (5,-1);
		\draw [snake=zigzag] (1,0.1) -- (5,0.1);
		\draw (1,0) -- (5,0);
		\draw (1,1) -- (5,1);
		\draw (1,2) -- (5,2);
		\node at (3,2.5) (um) {$\vdots$};
		\draw (1,3) -- (5,3);
		\node at (3,3.5) (up) {$\vdots$};
		\draw (1,4) -- (5,4);
		\draw (1,5) -- (5,5);
		\draw (1,6) -- (5,6);
		\draw [snake=zigzag] (1,5.9) -- (5,5.9);
		\draw (1,7) -- (5,7);
		\node at (3,7.5) (upp) {$\vdots$};
		
		\draw[blue] [->] (2,-1) -- node[left] {$L_{+1}$} (2,0);
		\draw[red] [<-] (4,-1) -- node[right] {$L_{-1}$} (4,0);
		\draw[red] [<-] (3,0) -- node[right] {$L_{-1}$} (3,1);
		\draw[blue] [->] (2,1) -- node[left] {$L_{+1}$} (2,2);
		\draw[red] [<-] (4,1) -- node[right] {$L_{-1}$} (4,2);
		\draw[blue] [->] (2,4) -- node[left] {$L_{+1}$} (2,5);
		\draw[red] [<-] (4,4) -- node[right] {$L_{-1}$} (4,5);
		\draw[blue] [->] (3,5) -- node[left] {$L_{+1}$} (3,6);
		\draw[blue] [->] (2,6) -- node[left] {$L_{+1}$} (2,7);
		\draw[red] [<-] (4,6) -- node[right] {$L_{-1}$} (4,7);
	\end{tikzpicture}
	\caption[The infinite-dimensional representation of $\SL$ whose elements solve the leading order near-zone equations of motion for a massless scalar field in the $d$-dimensional Reissner-Nordstr\"{o}m black hole background with integer rescaled multipolar index $\hat{\ell}=\frac{\ell}{d-3}$ and contains the singular static solution.]{The infinite-dimensional representation of $\SL$ whose elements solve the leading order near-zone equations of motion for a massless scalar field in the $d$-dimensional Reissner-Nordstr\"{o}m black hole background with integer rescaled multipolar index $\hat{\ell}=\frac{\ell}{d-3}$ and contains the singular static solution $\Phi_{\omega=0,\ell,\mathbf{m}=\mathbf{0}}^{\left(0\right)\text{singular}}\propto \tilde{\upsilon}_{-\hat{\ell}-1,\hat{\ell}+1}$.}
	\label{fig:SingSL2RSchwarzschild}
\end{figure}

It turns out that the Love $\SL$ symmetry offers representation theory arguments around the running of the Love numbers as well. The absence of running for $\hat{\ell}\in\mathbb{N}$ is algebraically realized from the fact that the singular static solution belongs to the representation shown in Figure~\ref{fig:SingSL2RSchwarzschild}\footnote{More details on how this representation is constructed can be found in Section 4 of Ref.~\cite{Charalambous:2022rre}.}. This is indeed distinguishable from the regular static solution at any point since the two obey the locally distinguishable annihilation conditions
\be
	\text{If $\hat{\ell}\in\mathbb{N}$} \Rightarrow
	\begin{cases}
		\text{Regular static solution: } \left(L_{+1}\right)^{\hat{\ell}+1}\Phi_{\omega=0,\ell,\mathbf{m}}^{\left(0\right)\text{regular}} = 0 \\
		\text{Singular static solution: } L_{-1}\left(L_{+1}\right)^{\hat{\ell}+1}\Phi_{\omega=0,\ell,\mathbf{m}}^{\left(0\right)\text{singular}} = 0
	\end{cases} \,.
\ee
Compared to previous analyses of $\SL$ modules, Figure~\ref{fig:HWSL2R0_SchwarzschildDd} and Figure~\ref{fig:SingSL2RSchwarzschild} are the type-``$[\circ]$'' and type-``$\circ]\circ[\circ$'' representations $U(-\hat{\ell},-\hat{\ell}\,)$ and $U(\hat{\ell}+1,\hat{\ell}+1)$ respectively in the notation of Ref.~\cite{Howe1992} and the representations $D(\,2\hat{\ell}\,)$ and $D^{+-}(\,2\hat{\ell}\,)$ respectively in the language of Refs.~\cite{Miller1968,Miller1970}.

On the other hand, for $\hat{\ell}\notin\mathbb{N}$, regular and singular static solutions belong to the same standard $\SL$ representations $W(4\hat{\ell}(\hat{\ell}+1),0)$ (\cite{Howe1992}) or $D(\hat{\ell},0)$ (\cite{Miller1968,Miller1970}). The absence of any local algebraic criteria from $\SL$ modules of the Love symmetry would then suggest that running Love numbers are expected to arise in all of these situations. While this is consistent with the cases for which $\hat{\ell}\in\mathbb{N}+\frac{1}{2}$, the vanishing RG flow for the cases for which $2\hat{\ell}\notin\mathbb{N}$ can only be retrieved after combining with power-counting arguments~\cite{Charalambous:2022rre}.

\subsection{First class of $p$-form perturbations of Schwarzschild-Tangherlini black holes}
We now consider the case of the first class of $p$-form perturbations of the Schwarzschild-Tangherlini black hole, for which $j=d-3$ and $\rho_{-}=0$, $\rho_{+}=\rho_{s}$. The Love symmetries generators in Eq.~\eqref{eq:SL2R_SchwarzschildDd} then read
\be\label{eq:SL2Rp1_SchwarzschildDd}
	\begin{gathered}
		L_0^{\left(\sigma,j=d-3\right)} = -\beta\,\partial_{t} - \sigma \,, \\
		L_{\pm1}^{\left(\sigma,j=d-3\right)} = e^{\pm t/\beta}\left[\mp\sqrt{\Delta}\,\partial_{\rho} + \partial_{\rho}\left(\sqrt{\Delta}\right)\beta\,\partial_{t} + \sigma\sqrt{\frac{\rho-\rho_{s}}{\rho}} \right] \,.
	\end{gathered}
\ee
The fact that the $L_0$-eigenvalues only get shifted by integer amounts allows to carry the previous analysis in exactly the same way. The primary state of the highest-weight representation with weight $h=-\hat{\ell}$ is given by
\be
	\begin{gathered}
		L_{+1}^{\left(\sigma,j=d-3\right)}\upsilon_{-\hat{\ell},0}^{\left(\sigma,j=d-3\right)} = 0 \,,\quad L_0^{\left(\sigma,j=d-3\right)}\upsilon_{-\hat{\ell},0}^{\left(\sigma,j=d-3\right)} = -\hat{\ell}\,\upsilon_{-\hat{\ell},0}^{\left(\sigma,j=d-3\right)} \,, \\
		\Rightarrow \quad \upsilon_{-\hat{\ell},0}^{\left(\sigma,j=d-3\right)} = \rho^{\sigma}\left(-e^{+t/\beta}\sqrt{\Delta}\right)^{\hat{\ell}-\sigma} \,.
	\end{gathered}
\ee
and, along with its descendants,
\be
	\upsilon_{-\hat{\ell},n}^{\left(\sigma,j=d-3\right)} = \left(L_{-1}^{\left(\sigma,j=d-3\right)}\right)^{n} \upsilon_{-\hat{\ell},0}^{\left(\sigma,j=d-3\right)} \,,
\ee
they furnish a representation of the Love $\SL$ symmetry spanned by states that are regular at the future event horizon. For the regular static solution to belong to this representation, we must therefore have
\be
	\hat{\ell}-\sigma \in\mathbb{N} \Leftrightarrow \hat{\ell}\in\mathbb{N} \,.
\ee
These are again the exact resonant conditions for which the static electric-type Love numbers of the Schwarzschild-Tangherlini black hole vanish. The regular static solution is the $(\hat{\ell}-\sigma)$'th descendant and the highest-weight property
\be
	\left(L_{+1}^{\left(\sigma,j=d-3\right)}\right)^{\hat{\ell}-\sigma+1}\Phi_{\omega=0,\ell,\mathbf{m}}^{\left(\sigma,j=d-3\right)} = 0 \quad\text{if $\hat{\ell}\in\mathbb{N}$} \,,
\ee
immediately implies the following polynomial form
\be
	\text{If $\hat{\ell}\in\mathbb{N}$ }\Rightarrow \Phi_{\omega=0,\ell,\mathbf{m}=\mathbf{0}}^{\left(\sigma,j=d-3\right)} \propto \upsilon_{-\hat{\ell},\hat{\ell}-\sigma}^{\left(\sigma,j=d-3\right)} = \rho^{\sigma}\sum_{n=0}^{\hat{\ell}-\sigma}c_{n}\rho^{n} = c_{\hat{\ell}-\sigma}r^{\ell} + \dots + c_0 r^{\sigma\left(d-3\right)} \,,
\ee
with no relevant response modes present and, hence, vanishing static responses.

\begin{figure}[t]
	\centering
	\begin{tikzpicture}
		\node at (0,0) (uml4) {$\upsilon_{-\hat{\ell},2\hat{\ell}}^{\left(\sigma,j=d-3\right)}$};
		\node at (0,1) (uml3) {$\upsilon_{-\hat{\ell},\hat{\ell}-\sigma}^{\left(\sigma,j=d-3\right)}$};
		\node at (0,2) (uml2) {$\upsilon_{-\hat{\ell},2}^{\left(\sigma,j=d-3\right)}$};
		\node at (0,3) (uml1) {$\upsilon_{-\hat{\ell},1}^{\left(\sigma,j=d-3\right)}$};
		\node at (0,4) (uml0) {$\upsilon_{-\hat{\ell},0}^{\left(\sigma,j=d-3\right)}$};
		
		\draw [snake=zigzag] (1,-0.1) -- (5,-0.1);
		\draw (1,0) -- (5,0);
		\draw (1,1) -- (5,1);
		\node at (3,1.5) (up) {$\vdots$};
		\node at (3,0.5) (um) {$\vdots$};
		\draw (1,2) -- (5,2);
		\draw (1,3) -- (5,3);
		\draw (1,4) -- (5,4);
		\draw [snake=zigzag] (1,4.1) -- (5,4.1);
		
		\draw[blue] [->] (2.6,2) -- node[left] {$L_{+1}^{\left(\sigma,j=d-3\right)}$} (2.6,3);
		\draw[blue] [->] (2.3,3) -- node[left] {$L_{+1}^{\left(\sigma,j=d-3\right)}$} (2.3,4);
		\draw[red] [<-] (3.7,3) -- node[right] {$L_{-1}^{\left(\sigma,j=d-3\right)}$} (3.7,4);
		\draw[red] [<-] (3.4,2) -- node[right] {$L_{-1}^{\left(\sigma,j=d-3\right)}$} (3.4,3);
	\end{tikzpicture}
	\caption[The finite-dimensional highest-weight representation of $\SL$ whose elements solve the leading order near-zone equations of motion for a first class $p$-form perturbation of the $d$-dimensional Schwarzschild-Tangherlini black hole with integer rescaled multipolar index $\hat{\ell}=\frac{\ell}{d-3}$ and contains the regular static solution.]{The finite-dimensional highest-weight representation of $\SL$ whose elements solve the leading order near-zone equations of motion for a first class $p$-form perturbation of the $d$-dimensional Schwarzschild-Tangherlini black hole with integer rescaled multipolar index $\hat{\ell}=\frac{\ell}{d-3}$ and contains the regular static solution.}
	\label{fig:HWSL2Rp1_SchwarzschildDd}
\end{figure}
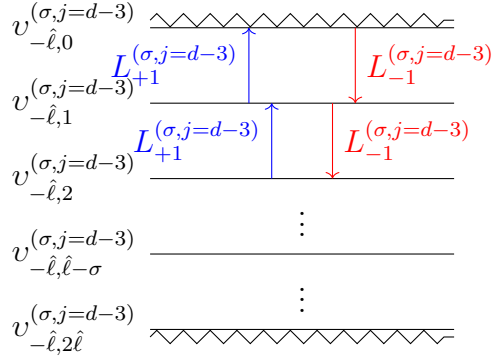

As before, studying the lowest-weight representation with weight $\bar{h}=+\hat{\ell}$ that is spanned by ascendants,
\be
	\bar{\upsilon}_{+\hat{\ell},n}^{\left(\sigma,j=d-3\right)}= \left(-L_{+1}^{\left(\sigma,j=d-3\right)}\right)^{n}\bar{\upsilon}_{+\hat{\ell},0}^{\left(\sigma,j=d-3\right)} \,, \\
\ee
of the lowest-weight vector
\be
	\begin{gathered}
		L_{-1}^{\left(\sigma,j=d-3\right)}\bar{\upsilon}_{+\hat{\ell},0}^{\left(\sigma,j=d-3\right)} = 0 \,,\quad L_0^{\left(\sigma,j=d-3\right)}\bar{\upsilon}_{+\hat{\ell},0}^{\left(\sigma,j=d-3\right)} = +\hat{\ell}\,\bar{\upsilon}_{+\hat{\ell},0}^{\left(\sigma,j=d-3\right)} \,, \\
		\Rightarrow \quad \bar{\upsilon}_{+\hat{\ell},0}^{\left(\sigma,j=d-3\right)} = \rho^{-\sigma}\left(+e^{-t/\beta}\sqrt{\Delta}\right)^{\hat{\ell}+\sigma} \,,
	\end{gathered}
\ee
reveals the static regular solution with vanishing static Love numbers is also the $(\hat{\ell}+\sigma)$'th ascendant and, therefore, this representation is in fact the finite $(2\hat{\ell}+1)$-dimensional type-``$[\circ]$'' representation of the Love $\SL$ symmetry (see Figure~\ref{fig:HWSL2Rp1_SchwarzschildDd}), while the singular static solution belongs to the locally distinguishable type-``$\circ]\circ[\circ$'' representation of Figure~\ref{fig:SingSL2RSchwarzschild}.

\subsection{Second class of $p$-form perturbations of Schwarzschild-Tangherlini black holes}
Next, for the case of the second class of $p$-form perturbations of the Schwarzschild-Tangherlini black hole, for which $j=\frac{d-3}{2}$ and which only emerges in odd spacetime dimensionalities, the Love symmetries generators in Eq.~\eqref{eq:SL2R_SchwarzschildDd} become
\be\label{eq:SL2Rp2_SchwarzschildDd}
	\begin{gathered}
		L_0^{\left(\sigma,j=d-3\right)} = -\beta\,\partial_{t} - \frac{\sigma}{2} \,, \\
		L_{\pm1}^{\left(\sigma,j=d-3\right)} = e^{\pm t/\beta}\left[\mp\sqrt{\Delta}\,\partial_{\rho} + \partial_{\rho}\left(\sqrt{\Delta}\right)\beta\,\partial_{t} + \frac{\sigma}{2}\sqrt{\frac{\rho-\rho_{s}}{\rho}} \right] \,.
	\end{gathered}
\ee
The $L_0$-eigenvalues now only get shifted by half-integer amounts. The previous analysis can then be applied in an exactly analogous manner to reveal that the static solution regular at the future event horizon now belongs to a highest-weight representation \textit{if and only if} $\hat{\ell}$ is half-integer which captures all the resonant condition of vanishing static Love numbers for this class of perturbations, a result inferred by the polynomial form of the solution implied by the highest-weight property. Similarly to the first class of $p$-form perturbations, this particular highest-weight representation is in fact the finite $(2\hat{\ell}+1)$-dimensional type-``$[\circ]$'' representation of the Love $\SL$ symmetry obtained from the one in Figure~\ref{fig:HWSL2Rp1_SchwarzschildDd} after replacing $\sigma\rightarrow\frac{\sigma}{2}$, while the singular static solution belongs to the locally distinguishable type-``$\circ]\circ[\circ$'' representation of the corresponding form shown in Figure~\ref{fig:SingSL2RSchwarzschild}.

\subsection{Third class of $p$-form perturbations of Schwarzschild-Tangherlini black holes}
Last, for the third class of $p$-form perturbations of the Schwarzschild-Tangherlini black hole, for which $2\hat{j}\notin\mathbb{N}$, things are a bit more interesting. Explicitly, the Love symmetries generators in Eq.~\eqref{eq:SL2R_SchwarzschildDd} are given by
\be
	\begin{gathered}
		L_0^{\left(\sigma,j\right)} = -\beta\,\partial_{t} - \sigma\hat{j} \,, \\
		L_{\pm1}^{\left(\sigma,j\right)} = e^{\pm t/\beta}\left[\mp\sqrt{\Delta}\,\partial_{\rho} + \partial_{\rho}\left(\sqrt{\Delta}\right)\beta\,\partial_{t} + \sigma\hat{j}\sqrt{\frac{\rho-\rho_{s}}{\rho}} \right] \,,
	\end{gathered}
\ee
and we see that the $L_0$-eigenvalues now get shifted by non-integer amounts. The highest-weight representation with weight $h=-\hat{\ell}$ is spanned by descendants,
\be
	\upsilon_{-\hat{\ell},n}^{\left(\sigma,j\right)} = \left(L_{-1}^{\left(\sigma,j\right)}\right)^{n} \upsilon_{-\hat{\ell},0}^{\left(\sigma,j\right)} \,,
\ee
of the primary state $\upsilon_{-\hat{\ell},0}^{\left(\sigma,j\right)}$ relevant for the third class $p$-form perturbations, satisfying
\be
	\begin{gathered}
		L_{+1}^{\left(\sigma,j\right)}\upsilon_{-\hat{\ell},0}^{\left(\sigma,j\right)} = 0 \,,\quad L_0^{\left(\sigma,j\right)}\upsilon_{-\hat{\ell},0}^{\left(\sigma,j\right)} = -\hat{\ell}\,\upsilon_{-\hat{\ell},0}^{\left(\sigma,j\right)} \,, \\
		\Rightarrow \quad \upsilon_{-\hat{\ell},0}^{\left(\sigma,j\right)} = \rho^{\sigma\hat{j}}\left(-e^{+t/\beta}\sqrt{\Delta}\right)^{\hat{\ell}-\sigma\hat{j}} \,.
	\end{gathered}
\ee
These states are always regular at the future event horizon and their $L_0$-eigenvalues are given by
\be
	L_0^{\left(\sigma,j\right)}\upsilon_{-\hat{\ell},n}^{\left(\sigma,j\right)} = (n-\hat{\ell})\,\upsilon_{-\hat{\ell},n}^{\left(\sigma,j\right)} \,.
\ee
As before, we encounter the new feature in $d>4$ that the static solution does not in general belong to a highest-weight representation of the Love $\SL$ symmetries. For this to happen, there are some resonant conditions that need to be satisfied. In particular, the static solution $\Phi^{\left(j\right)}_{\omega=0,\ell,\mathbf{m}}$ that is regular at the horizon is an element of the above highest-weight representation \text{if and only if}
\be
	\hat{\ell}-\sigma\hat{j} \in \mathbb{N} \,.
\ee
In $d>4$, this only covers one branch of the resonant conditions for which the static Love numbers of this class of $p$-form perturbations vanish (see Table~\ref{tbl:Staticp3LNs_SchwarzschildDd}). Nevertheless, the second branch of these resonant conditions is captured by the second Love symmetry, corresponding to the opposite sign $\sigma$. However, we will see momentarily that the second branch also arises from the lowest-weight representation.

Similar to the $d=4$ cases, the highest-weight property implies a polynomial form. In particular, from the fact that, for arbitrary purely radial functions $F\left(\rho\right)$,
\be
	\left(L_{+1}^{\left(\sigma,j\right)}\right)^{n}\left[\left(\rho-\rho_{-}\right)^{\sigma\hat{j}}F\left(\rho\right)\right] = \left(-e^{+t/\beta}\sqrt{\Delta}\right)^{n}\left(\rho-\rho_{-}\right)^{\sigma\hat{j}}\frac{d^{n}}{d\rho^{n}}F\left(\rho\right) \,,
\ee
the annihilation condition $(L_{+1}^{\left(\sigma,j\right)})^{\hat{\ell}-\sigma\hat{j}+1}\Phi^{\left(j\right)}_{\omega=0,\ell,\mathbf{m}}\bigg|_{\hat{\ell}-\sigma\hat{j}\in\mathbb{N}} = 0$ implies
\be
	\text{If $\hat{\ell}-\sigma\hat{j}\in\mathbb{N}$ }\Rightarrow \Phi_{\omega=0,\ell,\mathbf{m}=\mathbf{0}}^{\left(\sigma,j\right)} \propto \upsilon_{-\hat{\ell},\hat{\ell}-\sigma\hat{j}}^{\left(\sigma,j\right)} = \rho^{\sigma\hat{j}}\sum_{n=0}^{\hat{\ell}-\sigma\hat{j}}c_{n}\rho^{n} = c_{\hat{\ell}-\sigma\hat{j}}r^{\ell} + \dots + c_0 r^{\sigma j} \,,
\ee
which indeed has the appropriate polynomial form from which to infer the vanishing of the static Love numbers by the absence of a response mode.

As for the lowest-weight representation of the Love $\SL$ symmetry for sign $\sigma$ with weight $\bar{h} = +\hat{\ell}$, the lowest-weight vector $\bar{\upsilon}_{+\hat{\ell},0}^{\left(\sigma,j\right)}$ is found to be
\be
	\begin{gathered}
		L_{-1}^{\left(\sigma,j\right)}\bar{\upsilon}_{+\hat{\ell},0}^{\left(\sigma,j\right)} = 0 \,,\quad L_0^{\left(\sigma,j\right)}\bar{\upsilon}_{+\hat{\ell},0}^{\left(\sigma,j\right)} = +\hat{\ell}\,\bar{\upsilon}_{+\hat{\ell},0}^{\left(\sigma,j\right)} \,, \\
		\Rightarrow \quad \bar{\upsilon}_{+\hat{\ell},0}^{\left(\sigma,j\right)} = \rho^{-\sigma\hat{j}}\left(+e^{-t/\beta}\sqrt{\Delta}\right)^{\hat{\ell}+\sigma\hat{j}} \,,
	\end{gathered}
\ee
and is always regular at the past event horizon, while it is regular at the future event horizon as long as $e^{-t/\beta}\sqrt{\Delta}\sim e^{-t_{+}/\beta} \left(r-r_{+}\right)$ is not raised to any negative power. Its ascendants,
\be
	\bar{\upsilon}_{+\hat{\ell},n}^{\left(\sigma,j\right)} = \left(-L_{+1}^{\left(\sigma,j\right)}\right)^{n} \bar{\upsilon}_{+\hat{\ell},0}^{\left(\sigma,j\right)} \,,
\ee
share the same boundary conditions and their charge under $L_0$ is
\be
	L_0^{\left(\sigma,j\right)}\bar{\upsilon}_{+\hat{\ell},n}^{\left(\sigma,j\right)} = (\hat{\ell}-n)\,\bar{\upsilon}_{+\hat{\ell},n}^{\left(\sigma,j\right)} \,.
\ee
For the static solution regular at the horizon to belong to this representation, we must therefore have
\be
	\hat{\ell}+\sigma\hat{j}\in\mathbb{N} \,,
\ee
and the lowest-weight property implies an analogous polynomial form of the solution with no decaying mode.

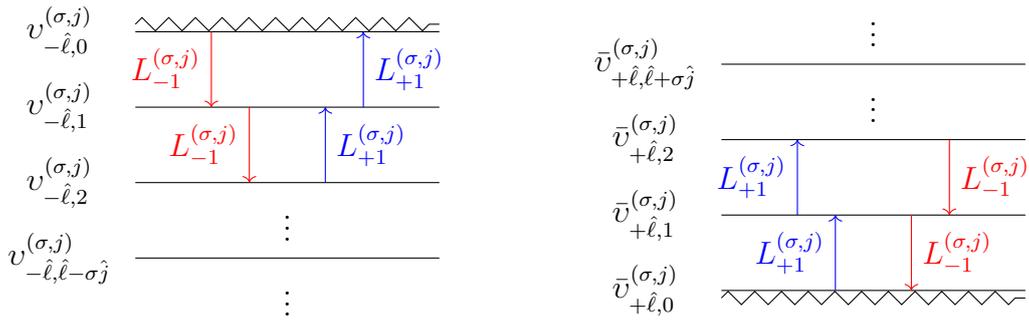
\begin{figure}[t]
	\centering
	\begin{subfigure}[b]{0.49\textwidth}
		\centering
		\begin{tikzpicture}
			\node at (0,1) (uml3) {$\upsilon_{-\hat{\ell},\hat{\ell}-\sigma\hat{j}}^{\left(\sigma,j\right)}$};
			\node at (0,2) (uml2) {$\upsilon_{-\hat{\ell},2}^{\left(\sigma,j\right)}$};
			\node at (0,3) (uml1) {$\upsilon_{-\hat{\ell},1}^{\left(\sigma,j\right)}$};
			\node at (0,4) (uml0) {$\upsilon_{-\hat{\ell},0}^{\left(\sigma,j\right)}$};
			
			\draw (1,1) -- (5,1);
			\node at (3,1.5) (up) {$\vdots$};
			\node at (3,0.5) (um) {$\vdots$};
			\draw (1,2) -- (5,2);
			\draw (1,3) -- (5,3);
			\draw (1,4) -- (5,4);
			\draw [snake=zigzag] (1,4.1) -- (5,4.1);
			
			\draw[red] [<-] (2.5,2) -- node[left] {$L_{-1}^{\left(\sigma,j\right)}$} (2.5,3);
			\draw[red] [<-] (2,3) -- node[left] {$L_{-1}^{\left(\sigma,j\right)}$} (2,4);
			\draw[blue] [->] (4,3) -- node[right] {$L_{+1}^{\left(\sigma,j\right)}$} (4,4);
			\draw[blue] [->] (3.5,2) -- node[right] {$L_{+1}^{\left(\sigma,j\right)}$} (3.5,3);
		\end{tikzpicture}
		\caption{The highest-weight $\SL$ representation that contains the regular static solution along the $\hat{\ell}-\sigma\hat{j}\in\mathbb{N}$ branch.}
	\end{subfigure}
	\hfill
	\begin{subfigure}[b]{0.49\textwidth}
		\centering
		\begin{tikzpicture}
			\node at (0,3) (upll) {$\bar{\upsilon}_{+\hat{\ell},\hat{\ell}+\sigma\hat{j}}^{\left(\sigma,j\right)}$};
			\node at (0,2) (upl2) {$\bar{\upsilon}_{+\hat{\ell},2}^{\left(\sigma,j\right)}$};
			\node at (0,1) (upl1) {$\bar{\upsilon}_{+\hat{\ell},1}^{\left(\sigma,j\right)}$};
			\node at (0,0) (upl0) {$\bar{\upsilon}_{+\hat{\ell},0}^{\left(\sigma,j\right)}$};
			
			\draw [snake=zigzag] (1,-0.1) -- (5,-0.1);
			\draw (1,0) -- (5,0);
			\draw (1,1) -- (5,1);
			\draw (1,2) -- (5,2);
			\node at (3,2.5) (up) {$\vdots$};
			\draw (1,3) -- (5,3);
			\node at (3,3.5) (um) {$\vdots$};
			
			\draw[blue] [->] (2.5,0) -- node[left] {$L_{+1}^{\left(\sigma,j\right)}$} (2.5,1);
			\draw[blue] [->] (2,1) -- node[left] {$L_{+1}^{\left(\sigma,j\right)}$} (2,2);
			\draw[red] [<-] (4,1) -- node[right] {$L_{-1}^{\left(\sigma,j\right)}$} (4,2);
			\draw[red] [<-] (3.5,0) -- node[right] {$L_{-1}^{\left(\sigma,j\right)}$} (3.5,1);
		\end{tikzpicture}
		\caption{The lowest-weight $\SL$ representation that contains the regular static solution along the $\hat{\ell}+\sigma\hat{j}\in\mathbb{N}$ branch.}
	\end{subfigure}
	\caption[The infinite-dimensional highest-weight and lowest-weight representations of $\SL$ whose elements solve the leading order near-zone equations of motion for the third class $p$-form perturbations of the $d$-dimensional Schwarzschild-Tangherlini black hole with rescaled orbital numbers satisfying $\hat{\ell}\pm\hat{j}\in\mathbb{N}$ and contain the regular static solution.]{The infinite-dimensional highest-weight and lowest-weight representations of $\SL$ whose elements solve the leading order near-zone equations of motion for the third class $p$-form perturbations of the $d$-dimensional Schwarzschild-Tangherlini black hole with rescaled orbital numbers satisfying $\hat{\ell}\pm\hat{j}\in\mathbb{N}$ and contain the regular static solution.}
	\label{fig:HWSL2p3_SchwarzschildDd}
\end{figure}

We now see an interesting new feature compared to the case of the first and second classes of $p$-form perturbations of the higher-dimensional Schwarzschild-Tangherlini black hole. To begin with, for the first and second classes of $p$-form perturbations, for which $2\hat{j}$ is an integer, the highest-weight representation encountered before turned out to in fact be the finite $(2\hat{\ell}+1)$-dimensional type-``$[\circ]$'' representation whenever the static solution was one of its elements. As for the third class $p$-form perturbations, for which $2\hat{j}$ is not an integer in $d>4$, the lowest-weight representation captures the second branch of resonant conditions associated with vanishing static Love numbers, namely, the branch with $\hat{\ell}+\sigma\hat{j}\in\mathbb{N}$, see Figure~\ref{fig:HWSL2p3_SchwarzschildDd}. The highest-weight and lowest-weight representations associated with the current third class of $p$-form perturbations are non-overlapping and become infinite-dimensional Verma modules. In contrast to the Verma modules encountered in the four-dimensional Kerr-Newman Love multiplets, the regular at the horizon static solution is capable of belonging to either of the two, highest-weight or lowest-weight, modules, depending on which branch of the resonant conditions, $\hat{\ell}-\hat{j}\in\mathbb{N}$ or $\hat{\ell}+\hat{j}\in\mathbb{N}$ respectively, is encountered. The corresponding singular static solutions, however, still belong to the locally distinguishable type-``$\circ]\circ[\circ$'' representation.

\subsection{Near-zone Witt algebras}
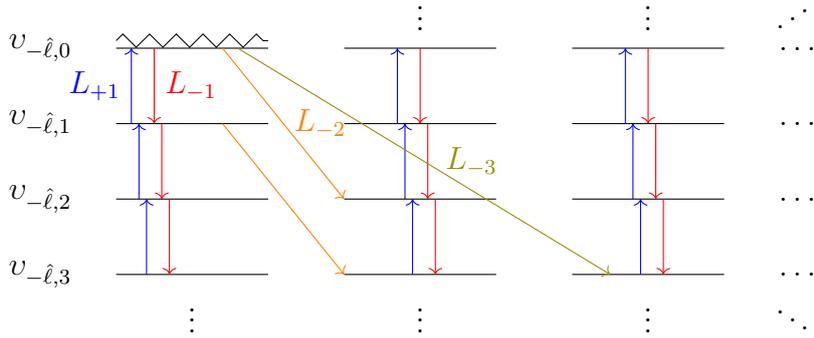
\begin{figure}[t]
	\centering
	\begin{tikzpicture}
		\node at (0,1) (uml3) {$\upsilon_{-\hat{\ell},3}$};
		\node at (0,2) (uml2) {$\upsilon_{-\hat{\ell},2}$};
		\node at (0,3) (uml1) {$\upsilon_{-\hat{\ell},1}$};
		\node at (0,4) (uml0) {$\upsilon_{-\hat{\ell},0}$};
		
		\node at (2,0.5) (um) {$\vdots$};
		\draw (1,1) -- (3,1);
		\draw (1,2) -- (3,2);
		\draw (1,3) -- (3,3);
		\draw (1,4) -- (3,4);
		\draw [snake=zigzag] (1,4.1) -- (3,4.1);
		
		\node at (5,4.5) (um) {$\vdots$};
		\draw (4,4) -- (6,4);
		\draw (4,3) -- (6,3);
		\draw (4,2) -- (6,2);
		\draw (4,1) -- (6,1);
		\node at (5,0.5) (um) {$\vdots$};
		
		\node at (8,4.5) (um) {$\vdots$};
		\draw (7,4) -- (9,4);
		\draw (7,3) -- (9,3);
		\draw (7,2) -- (9,2);
		\draw (7,1) -- (9,1);
		\node at (8,0.5) (um) {$\vdots$};

		\node at (10,1) (um) {$\dots$};
		\node at (10,2) (um) {$\dots$};
		\node at (10,3) (um) {$\dots$};
		\node at (10,4) (um) {$\dots$};
		\node at (9.9,0.5) (um) {$\ddots$};
		\node at (9.9,4.5) (um) {\reflectbox{$\ddots$}};
		
		\draw[blue] [->] (1.4,1) -- node[left] {} (1.4,2);
		\draw[blue] [->] (1.3,2) -- node[left] {} (1.3,3);
		\draw[blue] [->] (1.2,3) -- node[left] {$L_{+1}$} (1.2,4);
		\draw[red] [<-] (1.5,3) -- node[right] {$L_{-1}$} (1.5,4);
		\draw[red] [<-] (1.6,2) -- node[right] {} (1.6,3);
		\draw[red] [<-] (1.7,1) -- node[right] {} (1.7,2);
		
		\draw[red] [<-] (8.0,3) -- node[right] {} (8.0,4);
		\draw[red] [<-] (8.1,2) -- node[right] {} (8.1,3);
		\draw[red] [<-] (8.2,1) -- node[right] {} (8.2,2);
		\draw[blue] [->] (7.9,1) -- node[left] {} (7.9,2);
		\draw[blue] [->] (7.8,2) -- node[left] {} (7.8,3);
		\draw[blue] [->] (7.7,3) -- node[left] {} (7.7,4);
		
		\draw[red] [<-] (5.0,3) -- node[right] {} (5.0,4);
		\draw[red] [<-] (5.1,2) -- node[right] {} (5.1,3);
		\draw[red] [<-] (5.2,1) -- node[right] {} (5.2,2);
		\draw[blue] [->] (4.9,1) -- node[left] {} (4.9,2);
		\draw[blue] [->] (4.8,2) -- node[left] {} (4.8,3);
		\draw[blue] [->] (4.7,3) -- node[left] {} (4.7,4);
		
		\draw[orange] [<-] (4.0,2) -- node[right] {$L_{-2}$} (2.4,4);
		\draw[orange] [<-] (4.0,1) -- node[right] {} (2.4,3);

		\draw[olive] [<-] (7.5,1) -- node[right] {$\,\,L_{-3}$} (2.6,4);
	\end{tikzpicture}
	\caption[The Witt algebra extension of the highest-weight representation of $\SL$ for a massless scalar field in the $d$-dimensional Reissner-Nordstr\"{o}m black hole background with regular boundary conditions at the future event horizon.]{The Witt algebra extension of the highest-weight representation of $\SL$ for a massless scalar field in the $d$-dimensional Reissner-Nordstr\"{o}m black hole background with regular boundary conditions at the future event horizon.}
	\label{fig:HWWitt0_RNDd}
\end{figure}

It turns out that the aforementioned near-zone $\SL$ Love symmetries can be infinitely extended to full Virasoro algebras, as was first noted for the cases of spin-$0$ scalar perturbations in Ref.~\cite{Ortin:2012mt}. Indeed, consider the following generators
\be\ba\label{eq:WittLove}
	{}&L_{m}^{\left(\sigma,j\right)} = -\frac{e^{mt/\beta}}{\left(\sqrt{\Delta}\right)^{m}}\bigg[\frac{\left(\rho_{+}-\rho\right)^{m}-\left(\rho_{-}-\rho\right)^{m}}{\left(\rho_{+}-\rho_{-}\right)}\,\Delta\,\partial_{\rho} + \frac{\left(\rho_{+}-\rho\right)^{m}+\left(\rho_{-}-\rho\right)^{m}}{2}\,\beta\,\partial_{t} \\
	&\quad\quad\quad\quad\quad\quad\quad\quad + \sigma\hat{j}\left(\frac{1+m}{2}\left(\rho_{+}-\rho\right)^{m}+\frac{1-m}{2}\left(\rho_{-}-\rho\right)^{m}\right)\bigg] \\
	&= -\frac{e^{mt/\beta}}{\left(\sqrt{\Delta}\right)^{\left|m\right|}}\bigg[\text{sign}\left\{m\right\}\frac{\left(\rho_{+}-\rho\right)^{\left|m\right|}-\left(\rho_{-}-\rho\right)^{\left|m\right|}}{\left(\rho_{+}-\rho_{-}\right)}\,\Delta\,\partial_{\rho} + \frac{\left(\rho_{+}-\rho\right)^{\left|m\right|}+\left(\rho_{-}-\rho\right)^{\left|m\right|}}{2}\,\beta\,\partial_{t} \\
	&\quad\quad\quad\quad\quad\quad\quad\quad + \sigma\hat{j}\left(\frac{1+\left|m\right|}{2}\left(\rho_{+}-\rho\right)^{\left|m\right|}+\frac{1-\left|m\right|}{2}\left(\rho_{-}-\rho\right)^{\left|m\right|}\right)\bigg] \,.
\ea\ee
For $m=-1,0,+1$, these reduce to the $\SL$ Love symmetry generators in Eq.~\eqref{eq:SL2R_SchwarzschildDd}. For generic $m\in\mathbb{Z}$ these are the unique extension of the $\SL$ Love symmetry generators, up to automorphisms\footnote{These automorphisms contain, besides the standard rescalings $L_{m}\rightarrow \alpha^{m}L_{m}$, with $\alpha\in\mathbb{R}$, the model-specific scalar shifts
\begin{equation*}
	L_{m} \rightarrow L_{m} + \frac{\gamma}{\left(\sqrt{\Delta}\right)^{m}} \left[\frac{1+m}{2}\left(\rho_{+}-\rho\right)^{m}-\frac{1-m}{2}\left(\rho_{-}-\rho\right)^{m} - \frac{\rho-\rho_{+}}{\rho_{+}-\rho_{-}}\left[\left(\rho_{+}-\rho\right)^{m}-\left(\rho_{-}-\rho\right)^{m}\right]\right] \,,
\end{equation*}
with $\gamma\in\mathbb{R}$, that affect only the $\left|m\right|\ge2$ generators.}. They satisfy a centerless Virasoro (Witt) algebra,
\be
	\left[L_{m}^{\left(\sigma,j\right)},L_{n}^{\left(\sigma,j\right)}\right] = \left(m-n\right)L_{m+n}^{\left(\sigma,j\right)} \,,\quad m,n\in\mathbb{Z}
\ee
and are regular at the future event horizon for $m \le +1$, while they are regular at the past event horizon for $m \ge -1$. Therefore, only the $\SL$ Love part with $-1 \le m \le +1$ is globally defined, preserving the boundary conditions near the horizon. This makes the interpretation of the states arising from actions of the generators $L_{m}$ with $m\ne -1,0,+1$, somewhat unclear. For instance, the globally defined descendant $L_{-1}^{N}\upsilon_{-\hat{\ell},0}$ of the highest-weight $\SL$ multiplet in the scalar perturbations case ($j=0$) gets supplemented by an infinite number of level-$N$ states of the form
\be
	\upsilon_{-\hat{\ell},\left\{n_{\mathbb{Z}}\right\}}^{\left[N\right]} = \sum_{\substack{m=-\infty \\m\ne+1}}^{\infty}L_{m}^{n_{m}}\upsilon_{-\hat{\ell};0} \quad \text{such that $\sum_{\substack{m=-\infty \\m\ne+1}}^{\infty}m\,n_{m} = -N$}\,,\quad n_{m}\in\mathbb{N} \,.
\ee
These are to be contrasted with the textbook Verma modules of the Virasoro algebra which are defined such that $L_{m}\upsilon_{h;0} = 0$, $\forall m>0$. In the current centerless case, this only contains the trivial singlet with $h=0$. Furthermore, the above level-$N$ states are in general not regular at the future or the past event horizon. Nevertheless, focusing to the part of this representation that contains only states that are regular at the future event horizon prescribes the inclusion of the following finitely-many descendants at the $N$'th level
\be
	\upsilon_{-\hat{\ell},n_1n_2\dots n_{k}}^{\left[N\right]} = \prod_{m=1}^{k}L_{-m}^{n_{m}}\upsilon_{-\hat{\ell},0} \,,\quad \text{such that $\sum_{m=1}^{k}m\,n_{m} = N$} \,.
\ee
For example, at level $N=1$ one still has the single state $L_{-1}\upsilon_{-\hat{\ell},0}$, while, at levels $N=2$, $N=3$ and $N=4$, one now has two, three and five possible independent descendants respectively,
\be\ba
	\text{Level $N=1$: }& L_{-1}\upsilon_{-\hat{\ell},0} \,, \\
	\text{Level $N=2$: }& L_{-1}^2\upsilon_{-\hat{\ell},0} \,, \quad L_{-2}\upsilon_{-\hat{\ell},0} \,, \\
	\text{Level $N=3$: }& L_{-1}^3\upsilon_{-\hat{\ell},0} \,, \quad L_{-1}L_{-2}\upsilon_{-\hat{\ell},0} \,, \quad L_{-3}\upsilon_{-\hat{\ell},0} \,, \\
	\text{Level $N=4$: }& L_{-1}^4\upsilon_{-\hat{\ell},0} \,,\quad L_{-1}^2L_{-2}\upsilon_{-\hat{\ell},0} \,, \quad L_{-1}L_{-3}\upsilon_{-\hat{\ell},0} \,, \quad L_{-2}^2\upsilon_{-\hat{\ell},0} \,, \quad L_{-4}\upsilon_{-\hat{\ell},0} \,,
\ea\ee
and so on. However, in contrast to the states $L_{-1}^{N}\upsilon_{-\hat{\ell},0}$, the descendants that arise from actions of $L_{-m}$ with $m\ge2$ are not subjected to any annihilation condition following from the highest-weight property. This can be visualized by $L_{-1}$ and $L_{+1}$ being a vertical descender and a vertical ascender in the highest-weight ladder respectively, while $L_{-m}$ with $m\ge2$ act as diagonal descenders that can never reach the highest-weight state $\upsilon_{-\hat{\ell},0}$ by vertically climbing up the ladder (see Figure~\ref{fig:HWWitt0_RNDd}). It would be interesting to investigate whether these new descendants have any physical significance which we leave for future work. At first sight, they do not look relevant for solving the near-zone equations of motion since, for instance, they do not commute with the $\SL$ Casimir and, hence, they are not solutions.


\section{Beyond general-relativistic black holes}
\label{sec:LoveSymmetryModGR}

As a last investigation, we will perform a study similar to the Riemann-cubed paradigm in four spacetime dimensions in Ref.~\cite{Charalambous:2022rre} and compute the static scalar Love numbers for some higher-derivative theories of gravity. We will focus to the $\alpha^{\prime}$-corrected gravitational actions of string theory and extract the leading order static scalar susceptibilities for the simplest case of the corresponding modified Schwarzschild-Tangherlini black holes. These consist of the Callan-Myers-Perry black hole of bosonic/heterotic string theory~\cite{Callan:1988hs,Myers:1998gt} and the type-II superstring theory $\alpha^{\prime3}$-corrections to the Schwarzschild-Tangherlini black hole~\cite{Myers:1987qx}. We will then attempt to find sufficient geometric conditions for the existence of near-zone $\SL$ symmetries, which will turn out to come hand-in-hand with vanishing Love numbers for the corresponding black hole geometries.

The full radial equation of motion for the static scalar field spherical harmonics modes $\Phi_{\ell,\mathbf{m}}\left(r\right)$ reads\footnote{We remind here that we work in the coordinate system specified by the line element in Eq.~\eqref{eq:SphericallySymemtricBHDd}, with $r$ an areal radius coordinate.}
\be
	\left[f_{r}\partial_{x}^2 + \frac{x^{2\left(d-4\right)}}{2f_{t}}\partial_{x}\left(\frac{f_{t}f_{r}}{x^{2\left(d-4\right)}}\right)\partial_{x}\right]\Phi_{\ell,\mathbf{m}} = \frac{\ell\left(\ell+d-3\right)}{x^2}\Phi_{\ell,\mathbf{m}} \,,
\ee
where we have introduced the variable
\be
	x = \frac{r_{\text{h}}}{r} \,,
\ee
with $r_{\text{h}}$ the radial location of the event horizon at all orders in $\alpha^{\prime}$. We will treat this equation perturbatively around $\alpha^{\prime}=0$, with the order parameter being denoted by $\lambda$ and which is proportional to the appropriate power of $\alpha^{\prime}$ for each situation we will examine. The scalar field is expanded as
\be
	\Phi_{\ell,\mathbf{m}}\left(x\right) = r_{\text{h}}^{\ell}\bar{\mathcal{E}}_{\ell,\mathbf{m}}\left[\Phi_{\ell}^{\left(0\right)}\left(x\right) + \lambda\,\Phi^{\left(1\right)}_{\ell} + \mathcal{O}\left(\lambda^2\right) \right] \,,
\ee
with the zeroth order solution regular at the horizon $x=1$ given by the general-relativistic static scalar field profile,
\be
	\Phi_{\ell}^{\left(0\right)}\left(x\right) = \frac{\Gamma^2(\hat{\ell}+1)}{\Gamma(2\hat{\ell}+1)}\,{}_2F_1\left(\hat{\ell}+1,-\hat{\ell};1;1-\frac{1}{x^{d-3}}\right) \,,
\ee
and with the higher-order terms chosen to grow at infinity slower than the leading order solution,
\be
	\lim_{x\rightarrow0}x^{\ell}\Phi_{\ell}^{\left(n\right)} = 0 \quad \text{for $n>0$} \,.
\ee

\subsection{Bosonic/Heterotic string theory Callan-Myers-Perry black hole}
The Callan-Myers-Perry black hole describes the leading stringy corrections to the Schwarzschild-Tangherlini black hole in heterotic/bosonic string theory~\cite{Callan:1988hs,Myers:1998gt}, see also Refs.~\cite{Moura:2006pz,Moura:2011rr}. The gravitational action is $\alpha^{\prime}$-corrected by a Riemann-squared term\footnote{From the world-sheet perspective, this is a $1$-loop correction, while, withing the framework of EFT corrections to General Relativity, these enter at $2$-loop order~\cite{Donoghue:2017pgk,tHooft:1974toh,Goroff:1985th}. More generally, a Riemann-to-the-$k$'th-power correction enters as a $\left(k-1\right)$-loop order correction to the Einstein-Hilbert action.},
\be\label{eq:BosonicHeteroticStringAction}
	\begin{gathered}
		S_{\text{gr}} = \frac{1}{16\pi G}\int d^{d}x\,\sqrt{-g}\left[R - \frac{4}{d-2}\left(\partial\phi\right)^2 + \lambda\,e^{-4\phi/\left(d-2\right)}Y(\tilde{R})\right] \,, \\
		Y\left(R\right) = \frac{1}{2}R_{\mu\nu\rho\sigma}R^{\mu\nu\rho\sigma} \,,\quad \tilde{R}_{\mu\nu}^{\quad\rho\sigma} = R_{\mu\nu}^{\quad\rho\sigma} - \delta_{[\mu}^{[\rho}\nabla_{\nu]}\nabla^{\sigma]}\phi \,,
	\end{gathered}
\ee
where we also included the dilaton term and we are working in the Einstein-frame. The string coupling parameter above is equal to $\lambda=\frac{\alpha^{\prime}}{2}$ for the bosonic and $\lambda=\frac{\alpha^{\prime}}{4}$ for the heterotic string theory. The $d$-dimensional Callan-Myers-Perry black hole geometry has a constant dilaton and is given by~\cite{Callan:1988hs,Myers:1998gt}
\be
	\begin{gathered}
		ds^2 = -f\left(r\right)dt^2 + \frac{dr^2}{f\left(r\right)} + r^2d\Omega_{d-2}^2 \,, \\
		f\left(r\left(x\right)\right) = \left(1-x^{d-3}\right)\left[ 1 - 	\frac{\left(d-3\right)\left(d-4\right)}{2}\frac{\lambda}{r_{\text{h}}^2}\,x^{d-3}\frac{1-x^{d-1}}{1-x^{d-3}} \right] + \mathcal{O}\left(\lambda^2\right) \,.
	\end{gathered}
\ee
The event horizon $r_{\text{h}}$ is related to the ADM mass $M$, as encoded in the Schwarzschild radius $r_{s}$, of the black hole according to
\be
	r_{\text{h}} = r_{s}\left(1+\frac{d-4}{2}\frac{\lambda}{r_{s}^2}\right) + \mathcal{O}\left(\lambda^2\right) \,.
\ee
The black hole solution built perturbatively in $\alpha^{\prime}$ is valid only in regions where $r^2\gg\alpha^{\prime}$. For our purposes, it is sufficient to require that the gravitational radius of the black hole is much bigger than the string length, $r_{s}^2\gg\alpha^{\prime}$.

Let us now look at some specific examples for the $\alpha^{\prime}$-corrected static scalar field perturbations of the Callan-Myers-Perry black hole. The corrections to the coefficients that appear in front the decaying terms that go like $x^{\ell+d-3} \sim r^{-\left(\ell+d-3\right)}$ will be denoted by $\varkappa_{\ell}^{\left(1\right)}$. For $d=5$ and $\ell=2$ or $\ell=4$, we find
\be
	\begin{gathered}
		\ba
			\Phi^{\left(1\right)}_{\ell=2} &= -\frac{7}{2} + x^2 + 8\ln x + 2\left(1-\frac{2}{x^2}\right)\left(\text{Li}_2\left(1-x^2\right)-\frac{\pi^2}{6}\right) \,, \\
			\Phi^{\left(1\right)}_{\ell=4} &= -\frac{35}{x^2} + \frac{80}{3} + x^2 - 36\left(1-\frac{2}{x^2}\right)\ln x \\
			&\quad- 6\left(1-\frac{6}{x^2}+\frac{6}{x^4}\right)\left(\text{Li}_2\left(1-x^2\right)-\frac{\pi^2}{6}\right) \,,
		\ea \\\\
		\varkappa^{\left(1\right)}_{\ell=2} = -\lambda\left(\frac{1}{18} + \frac{2}{3}\ln x\right) \,,\quad \varkappa^{\left(1\right)}_{\ell=4} = -\lambda\left(\frac{43}{3600} + \frac{2}{5}\ln x\right) \,.
\end{gathered}
\ee
We see that these cases give rise to logarithmically running Love numbers, the value of the constant in front of the logarithms being identified with the corresponding $\beta$-function. An example of non-running static scalar Love numbers is the $d=6$, $\ell=3$ case
\be\ba
	\Phi^{\left(1\right)}_{\ell=3} &= \frac{9}{x^2} - \frac{45}{x} + \frac{63}{2} \\
	&\quad- 15\left(1-\frac{2}{x^3}\right)\left(\ln\left(1-x^3\right) + \frac{3}{2}x^2{}_2F_1\left(1,\frac{2}{3};\frac{5}{3};x^3\right)\right) \,, \\
	\varkappa^{\left(1\right)}_{\ell=3} &= -\lambda\frac{5}{2} \,.
\ea\ee
The running/non-running is in fact in accordance with power counting arguments. Indeed, following the arguments used in Section 3 of Ref.~\cite{Charalambous:2022rre}, one expects to find a non-vanishing RG flow if
\be
	1 \le 2\hat{\ell}+1-\frac{2}{d-3} \in \mathbb{N} \,,
\ee
otherwise, the natural expectation is some non-zero and non-running scalar Love number. This is indeed in accordance with our above results, i.e. the above condition is satisfied for $d=5$ and $\ell=2$ or $\ell=4$, while, for $d=6$ and $\ell=3$, it is not. It appears, therefore, that the $\alpha^{\prime}$-corrected Riemann-squared action for the bosonic/heterotic string theory does not exhibit any further seemingly fine-tuned behavior with respect to the black hole response problem\footnote{It should be noted here that black holes in the presence of stringy corrections still exhibit fine-tuning in that the Love numbers are expressed in terms of the string length scale $l_{s}\sim \sqrt{\alpha^{\prime}}$, rather than the natural (much larger) length scale of the Schwarzschild radius $r_{s}\gg l_{s}$. This ``zeroth'' order fine-tuning is what is addressed by the selection rules arising from the representation theory arguments around the near-zone Love symmetries we saw in Section~\ref{sec:LoveSymmetryDd}. We thank Mikhail Ivanov for pointing out this fact.}.

\subsection{Type-II superstring theory black holes}
Next, we consider the $\alpha^{\prime3}$-corrected black hole in type-II superstring theory. The type-II superstring theory effective action arising from tree-level amplitudes for four-graviton scattering, up to and including terms at eighth order in the graviton and dilaton momenta, is quartic in the Riemann tensor and, in the Einstein frame, is given by~\cite{Myers:1987qx} (see also Ref.~\cite{Chen:2021qrz})
\be\label{eq:TypeIISuperstringAction}
	\begin{gathered}
		S = \frac{1}{16\pi G}\int d^{d}x\,\sqrt{-g}\left[R - \frac{4}{d-2}\left(\partial\phi\right)^2 + \lambda\,e^{-12\phi/\left(d-2\right)}Y(\tilde{R})\right] \,, \\
		Y\left(R\right) = 2R_{\mu\nu\rho\sigma}R_{\kappa\quad\lambda}^{\,\,\,\,\,\nu\rho}R^{\mu\alpha\beta\kappa}R^{\lambda\quad\sigma}_{\,\,\,\,\alpha\beta} + R_{\mu\nu\rho\sigma}R_{\kappa\lambda}^{\quad\rho\sigma}R^{\mu\alpha\beta\kappa}R^{\lambda\quad\nu}_{\,\,\,\,\alpha\beta} \,,
	\end{gathered}
\ee
where $\lambda = \frac{1}{16}\zeta\left(3\right)\alpha^{\prime3}$ is the string coupling parameter and $\tilde{R}_{\mu\nu}^{\quad\rho\sigma}$ is given in Eq.~\eqref{eq:BosonicHeteroticStringAction}.

The asymptotically flat and electrically neutral black hole solution of this theory now has a non-constant dilaton and its geometry in the Einstein frame reads~\cite{Myers:1987qx,Chen:2021qrz}
\be
	\begin{gathered}
		ds^2 = -f_{t}\left(r\right)dt^2 + \frac{dr^2}{f_{r}\left(r\right)} + r^2d\Omega_{d-2}^2 \,, \\
		f_{t}\left(r\left(x\right)\right) = \left(1-x^{d-3}\right)\left[1+2\frac{\lambda}{r_{\text{h}}^6}\mu\left(x\right)\right] \,, \quad f_{r}\left(r\left(x\right)\right) = \left(1-x^{d-3}\right)\left[1-2\frac{\lambda}{r_{\text{h}}^6}\varepsilon\left(x\right)\right] \,, \\
		\mu\left(x\right) = -\varepsilon\left(x\right) - C_{d}\,x^{3\left(d-1\right)} \,,\quad \varepsilon\left(x\right) = D_{d}\,x^{3\left(d-1\right)} + E_{d}\,x^{d-3}\frac{1-x^{2d}}{1-x^{d-3}} \,,
	\end{gathered}
\ee
where the constants $C_{d}$, $D_{d}$ and $E_{d}$ are given by
\be\ba
	C_{d} &= \frac{2}{3}\left(d-1\right)\left(d-3\right)\left(2\,d^3-10\,d^2+6\,d+15\right) \,, \\
	D_{d} &= -\frac{1}{24}\left(d-3\right)\left(52\,d^4-375\,d^3+758\,d^2-117\,d-570\right) \,, \\
	E_{d} &= \frac{1}{24}\left(d-3\right)\left(20\,d^4-225\,d^3+946\,d^2-1779\,d+1290\right) \,.
\ea\ee

The power counting arguments of Ref.~\cite{Charalambous:2022rre} now imply that one expects logarithmically running scalar Love numbers whenever
\be
	3\le2\hat{\ell}+1-\frac{6}{d-3}\in\mathbb{N}
\ee
at $\alpha^{\prime3}$ order. This means that logarithms appear first at orbital number $\ell=d$. To avoid cumbersome expressions at very high multipolar orders, we focus here to $d=4$. Then, one expects to find non-running and non-vanishing static Love numbers for $\ell=2,3$ but, for $\ell\ge4$, one should be faced with RG-flowing static responses. Indeed, for $\ell=2,3$, we find no logs,
\be
	\begin{gathered}
		\ba
			\Phi^{\left(1\right)}_{\ell=2} &= -\frac{5}{2x} + \frac{5}{6} - \frac{1619}{4200}x^3 - \frac{1619}{2800}x^4 - \frac{1619}{2450}x^5 - \frac{1619}{2352}x^6 - \frac{3153}{784}x^7 + \frac{71}{32}x^8 \,, \\
			\Phi^{\left(1\right)}_{\ell=3} &= -\frac{15}{4x^2} + \frac{3}{x} - \frac{3}{8} - \frac{6693}{19600}x^4 - \frac{6693}{9800}x^5 - \frac{4533}{784}x^6 + \frac{12861}{1960}x^7 - \frac{213}{160}x^8 \,, \\
		\ea \\\\
		\varkappa^{\left(1\right)}_{\ell=2} = -\lambda\frac{1619}{4200} \,,\quad \varkappa^{\left(1\right)}_{\ell=3} = -\lambda\frac{6693}{19600} \,,
	\end{gathered}
\ee
while, for $\ell=4$,
\be
	\begin{gathered}
		\ba
			{}&\Phi^{\left(1\right)}_{\ell=4} = \frac{39195}{x^3} - \frac{480155}{7 x^2} + \frac{2214665}{63 x} -\frac{306241}{63} + 28 x - \frac{14}{9} x^2 + \frac{4}{9}x^3 - \frac{1}{2}x^4 \\
			&- \frac{102407}{12348}x^5 + \frac{327629}{24696} x^6 - \frac{7901}{1372}x^7 + \frac{71}{112}x^8 + \frac{1400}{3x^2}\left(1-\frac{2}{x}\right)\left(5-\frac{42}{x}+\frac{42}{x^2}\right)\ln x \\
			&-560\left(1-\frac{20}{x}+\frac{90}{x^2}-\frac{140}{x^3}+\frac{70}{x^4}\right)\left(\text{Li}_2\left(x\right)+\ln\left(1-x\right)\ln x\right) \,,
		\ea \\\\
		\varkappa^{\left(1\right)}_{\ell=4} = -\frac{3947599}{555660}+\frac{8}{9}\ln x \,,
	\end{gathered}
\ee
as expected. Consequently, similar to the Callan-Myers-Perry black hole, the $\alpha^{\prime3}$-corrected Schwarzschild-Tangherlini black hole of type-II superstring theory does not seem to exhibit any fine-tuned scalar Love numbers.

\subsection{A sufficient geometric constraint for the existence of near-zone symmetries}
\label{sec:LoveModGRConstraints}
As we just saw, neither the Callan-Myers-Perry black hole of bosonic/heterotic string theory nor the $\alpha^{\prime3}$-corrected Schwarzschild-Tangherlini black hole of type-II superstring theory demonstrate any superficially unnatural black hole scalar Love numbers. This is expected to be accompanied with the absence of a Love symmetry structure.

Having these results as explicit counterexamples, we will attempt now to extract sufficient geometric conditions for the existence of Love symmetry beyond general-relativistic black hole configurations by studying a massless scalar field in the background of a generalized spherically symmetric black hole geometry, Eq.~\eqref{eq:SphericallySymemtricBHDd} with $f_{t}\left(r\right)\ne f_{r}\left(r\right)$. After the field redefinition $\Phi_{\ell,\mathbf{m}}=\Psi_{\ell,\mathbf{m}}^{\left(0\right)}/r^{\frac{d-2}{2}}$, the full massless Klein-Gordon equation derived in Section~\ref{sec:EOM_Seq0} becomes
\be
	\begin{gathered}
		\mathbb{O}_{\text{full}}^{\left(0\right)}\Phi_{\ell,\mathbf{m}} = \hat{\ell}(\hat{\ell}+1)\Phi_{\ell,\mathbf{m}} \,, \\
		\mathbb{O}_{\text{full}}^{\left(0\right)} = \partial_{\rho}\,\Delta_{r}\,\partial_{\rho} + \frac{\Delta_{r}^2}{2\Delta_{t}}\left(\frac{\Delta_{t}}{\Delta_{r}}\right)^{\prime}\partial_{\rho} - \frac{r^2\rho^2}{\left(d-3\right)^2\Delta_{t}}\,\partial_{t}^2 \,,
	\end{gathered}
\ee
where $\rho=r^{d-3}$ and $\hat{\ell}=\ell/\left(d-3\right)$ as before, $\Delta_{t}\equiv\rho^2f_{t}$, $\Delta_{r}\equiv\rho^2f_{r}$ and primes denote derivatives with respect to $\rho$.

Similar to the corresponding analysis in four dimensions presented Ref.~\cite{Charalambous:2022rre}, the following near-zone approximation turns out to be the only possible candidate for enjoying a globally defined $\SL$ symmetry,
\be\label{eq:NZModGRDd}
	\mathbb{O}_{\text{NZ}}^{\left(0\right)} = \partial_{\rho}\,\Delta_{r}\,\partial_{\rho} + \frac{\Delta_{r}^2}{2\Delta_{t}}\left(\frac{\Delta_{t}}{\Delta_{r}}\right)^{\prime}\partial_{\rho} - \frac{r_{\text{h}}^{2\left(d-2\right)}}{\left(d-3\right)^2\Delta_{t}}\,\partial_{t}^2 \,.
\ee
The associated vector fields generating the near-zone $\SL$ algebra,
\be\label{eq:SL2RModGRDd}
	\begin{gathered}
		L_0 = -\beta\,\partial_{t} \,,\quad
		L_{\pm1} = e^{\pm t/\beta}\left[\mp\sqrt{\Delta_{r}}\,\partial_{\rho} + \sqrt{\frac{\Delta_{r}}{\Delta_{t}}}\partial_{\rho}\left(\sqrt{\Delta_{t}}\right)\beta\,\partial_{t}\right] \,,
	\end{gathered}
\ee
with $\beta$ the inverse surface gravity in Eq.~\eqref{eq:betaDd}, are regular at both the future and the past event horizon and give rise to a Casimir operator that matches this near-zone truncation of the Klein-Gordon operator if and only if\footnote{This is the solution to a particular differential equation that is outputted by the requirement that the Love symmetry vector fields satisfy the $\SL$ algebra and that their Casimir produces a consistent near-zone truncation of the massless Klein-Gordon operator.}
\be\label{eq:SL2RModGRConstraintDd}
	\Delta_{r}\left(\rho\right) = \Delta_{t}\left(\rho\right)\frac{4\Delta_{t}\left(\rho\right)+\left(\frac{\beta_{s}}{\beta}\rho_{\text{h}}\right)^2}{\Delta_{t}^{\prime2}\left(\rho\right)} \,,
\ee
where we have defined the inverse surface gravity for the Schwarzschild-Tangherlini black hole $\beta_{s}=\frac{2r_{\text{h}}}{d-3}$, or, at the level of the functions $f_{t}\left(r\right)$ and $f_{r}\left(r\right)$ themselves,
\be
	f_{r}\left(r\right) = f_{t}\left(r\right) \frac{\left(d-3\right)^2r^{2\left(d-4\right)}}{\left(r^{2\left(d-3\right)}f_{t}\left(r\right)\right)^{\prime2}} \left[4r^{2\left(d-3\right)}f_{t}\left(r\right) + \left(\frac{\beta_{s}}{\beta}\right)^2r_{\text{h}}^{2\left(d-3\right)}\right] \,.
\ee
For the case of $f_{t}=f_{r}$, the above condition tells us that the general-relativistic Reissner-Nordstr\"{o}m geometry is the only acceptable black hole solution, which already rules out the Callan-Myers-Perry black hole solution. Furthermore, plugging in the explicit $\alpha^{\prime3}$-corrections to the Schwarzschild-Tangherlini black hole in type-II superstring theory reveals that the above condition is again not satisfied, in accordance with the explicit computations of the static scalar Love numbers. Of course, this does not rule out all black hole solutions of string theory. An explicit counterexample is the STU black hole of supergravity~\cite{Cvetic:1996kv} which satisfies the above geometric constraint and Love symmetry has indeed been shown to exist for the more general rotating STU black hole configuration~\cite{Cvetic:2021vxa}. Furthermore, even though the geometric condition derived here sets a sufficient constraint on the existence of Love symmetry, this needs not be a necessary constraint as well. In particular, we have only examined the case of a massless scalar field minimally coupled to pure gravity which may very well not be a good representative of the modified theory of gravity under study.

One can also check that the above near-zone $\SL$ implies the vanishing of static Love numbers when $\hat{\ell}\in\mathbb{N}$. Using the same symmetry argument of the regular static solution being an element of a highest-weight representation of this $\SL$, we obtain $\left(L_{+1}\right)^{\hat{\ell}+1}\Phi_{\omega=0,\ell,\mathbf{m}} = 0$ if and only if $\hat{\ell}$ is an integer. From the fact that
\be
	\left(L_{+1}\right)^{n}F\left(\rho\right) = \left(-e^{t/\beta}\sqrt{\Delta_{t}}\right)^{n}\left[\sqrt{\frac{\Delta_{r}}{\Delta_{t}}}\frac{d}{d\rho}\right]^{n}F\left(\rho\right)
\ee
we see that the corresponding static solution is a polynomial but this time in the variable $\tilde{\rho}$, defined as
\be\label{eq:tilderho}
	d\tilde{\rho} \equiv \sqrt{\frac{\Delta_{t}}{\Delta_{r}}} \,d\rho \Rightarrow \tilde{\rho} = \sqrt{\Delta_{t}+\left(\frac{\beta_{s}}{2\beta}\rho_{\text{h}}\right)^2} +\tilde{\rho}_{\text{h}} - \frac{\beta_{s}}{2\beta}\rho_{\text{h}} \,,
\ee
where $\tilde{\rho}_{\text{h}}$ is an integration constant indicating the location of the event horizon in this new radial coordinate,
\be
	\text{If $\hat{\ell}\in\mathbb{N}$: }\Rightarrow \Phi_{\omega=0,\ell,\mathbf{m}}\left(r\right) = \sum_{n=0}^{\hat{\ell}}c_{n}^{\left(\mathbf{m}\right)}\tilde{\rho}^{n}\left(r\right)
\ee
Asymptotically, $\tilde{\rho}\rightarrow\rho$ due to the asymptotic flatness of $f_{t}$. Expanding this polynomial in $\tilde{\rho}$ at large distance in the initial radial variable $\rho$, one would observe the appearance of an $\rho^{-\hat{\ell}-1}=r^{-\ell-d+3}$ term. However, this term is a relativistic correction in the profile of the ``source'' part of the solution, rather than a response effect from induced multipole moments. Indeed, if the geometric condition in Eq.~\eqref{eq:SL2RModGRConstraintDd} for the existence of a near-zone $\SL$ symmetry is satisfied, we arrive at a situation practically identical to the case of spin-$0$ scalar mode perturbations of the Reissner-Nordstr\"{o}m black hole, Eq.~\eqref{eq:V0V1_RNDd}, when working with the variable $\tilde{\rho}$. More explicitly, the full radial Klein-Gordon operator reads,
\be
	\mathbb{O}_{\text{full}}^{\left(0\right)} = \partial_{\tilde{\rho}}\,\Delta_{t}\,\partial_{\tilde{\rho}} - \frac{r^{2\left(d-2\right)}}{\left(d-3\right)^2\Delta_{t}}\,\partial_{t}^2 \,,
\ee
and $\Delta_{t}$ is a quadratic polynomial in $\tilde{\rho}$,
\be\label{eq:DeltaTModGR}
	\Delta_{t} = \left(\tilde{\rho}-\tilde{\rho}_{+}\right)\left(\tilde{\rho}-\tilde{\rho}_{-}\right) \,,
\ee
where we have denoted the locations of the outer (event), and inner (Cauchy) horizons as
\be
	\tilde{\rho}_{\pm} = \tilde{\rho}_{\text{h}} - \frac{1\mp1}{2}\frac{\beta_{s}}{\beta}\rho_{\text{h}} \,.
\ee
Matching onto the worldline EFT can be achieved by solving the equations motion after analytically continuing the orbital number to perform the source/response split of the scalar field, and only in the end sending $\ell$ to take its physical integer values~\cite{LeTiec:2020spy,LeTiec:2020bos,Chia:2020yla,Charalambous:2021mea,Creci:2021rkz,Ivanov:2022hlo}. Doing this, we see that the ``response'' part of the static scalar field is singular at the horizon when $\hat{\ell}\in\mathbb{N}$ and is therefore absent, while the ``source'' part becomes a polynomial of degree $\hat{\ell}$ in $\tilde{\rho}$. Consequently, the corresponding static Love numbers vanish identically and we see again how a polynomial form of the solution is indicative of this vanishing. For generic $\hat{\ell}$, the procedure just described gives the following static scalar Love numbers,
\be
	k_{\ell}^{\left(0\right)} = \frac{\Gamma^4(\hat{\ell}+1)}{2\pi\,\Gamma(2\hat{\ell}+1)\Gamma(2\hat{\ell}+2)}\tan\pi\hat{\ell}\,\left(\frac{\beta_{s}}{\beta}\frac{\rho_{\text{h}}}{\rho_{s}}\right)^{2\hat{\ell}+1} \,,
\ee
where the last factor in the parenthesis is just $\left(\tilde{\rho}_{+}-\tilde{\rho}_{-}\right)/\rho_{s}$. These are exactly the same as the static scalar Love numbers for the higher-dimensional Reissner-Nordstr\"{o}m black hole obtained in Section~\ref{sec:LNs_RNDd}.

One can also apply the same analysis for the $p$-form perturbations equations of motion, Eq.~\eqref{eq:Vpj}. In fact, the equations of motion in the background of a generic electrically neutral black hole geometry can be collectively written as
\be
	\begin{gathered}
		\mathbb{O}_{\text{full}}^{\left(j\right)}\Phi^{\left(j\right)}_{\ell,\mathbf{m}} = \hat{\ell}(\hat{\ell}+1)\Phi^{\left(j\right)}_{\ell,\mathbf{m}} \,, \\
		\mathbb{O}_{\text{full}}^{\left(j\right)} = \partial_{\rho}\,\Delta_{r}\,\partial_{\rho} + \frac{\Delta_{r}^2}{2\Delta_{t}}\left(\frac{\Delta_{t}}{\Delta_{r}}\right)^{\prime}\partial_{\rho} - \frac{r^2\rho^2}{\left(d-3\right)^2\Delta_{t}}\,\partial_{t}^2 + U^{\left(j\right)}\left(\rho\right) \,, \\
	\end{gathered}
\ee
with the reduced potential given by
\be
	U^{\left(j\right)}\left(\rho\right) = \frac{\hat{j}}{d-3}rD_{a}r^{a} - \hat{j}(1-\hat{j})\left(1-r_{a}r^{a}\right) \,.
\ee
Introducing the coordinate $\tilde{\rho}$ in the same way as before, i.e. $d\tilde{\rho} = \sqrt{\frac{\Delta_{t}}{\Delta_{r}}}\,d\rho$, the radial operator is brought to the suggestive form
\be
	\mathbb{O}_{\text{full}}^{\left(j\right)} = \partial_{\tilde{\rho}}\,\Delta_{t}\,\partial_{\tilde{\rho}} - \frac{r^2\rho^2}{\left(d-3\right)^2\Delta_{t}}\,\partial_{t}^2 + U^{\left(j\right)}\left(\rho\right)
\ee
regardless of what the geometry is.

Motivated by the results for the Schwarzschild-Tangherlini black hole in Section~\ref{sec:LoveSymmetryDd}, we expect that a non-zero spin will not affect the vector part of the candidate near-zone symmetry generators. In other words, the previous geometric condition is expected to still be an outcome of this analysis. Assuming this is indeed the case, one can go ahead and see whether there is any additional geometric constraint that arises for $p\ne0$. It turns out that there is one additional geometric constraint which will completely fix the geometry. To see this, it is instructive to first express everything using the independent variable $\tilde{\rho}$, namely, rewrite
\be
	f_{r} = \Delta_{t}\left(\frac{1}{\rho}\frac{d\rho}{d\tilde{\rho}}\right)^2 \,.
\ee
Introducing the variable
\be
	u\left(\tilde{\rho}\right) = \left[\rho\left(\tilde{\rho}\right)\right]^{-\hat{j}} \,,
\ee
the reduced potential function then takes the form
\be
	U^{\left(j\right)} = -\left[\frac{1}{u}\frac{d}{d\tilde{\rho}}\left(\Delta_{t}\frac{du}{d\tilde{\rho}}\right) + \hat{j}(1-\hat{j})\right] \,.
\ee
At this point we have made no assumption on the explicit form of $\Delta_{t}$. Assuming that the vector part of the candidate Love symmetry generators is the same as for the scalar response problem, i.e. that $\Delta_{t}$ is a quadratic polynomial in $\tilde{\rho}$, the conditions that these generators form an $\SL$ algebra whose quadratic Casimir operator produces a consistent near-zone truncation of the equations of motion then primarily imply that
\be
	\begin{gathered}
		L_0 = -\beta\,\partial_{t} - \gamma \,, \\
		L_{\pm1} = e^{\pm t/\beta}\left[\mp\sqrt{\Delta_{t}}\,\partial_{\tilde{\rho}} + \partial_{\tilde{\rho}}\left(\sqrt{\Delta_{t}}\right)\beta\,\partial_{t} + \gamma\sqrt{\frac{\tilde{\rho}-\tilde{\rho}_{+}}{\tilde{\rho}-\tilde{\rho}_{-}}}\right] \,, \\
		U^{\left(j\right)}_{\SL} = \frac{\tilde{\rho}_{+}-\tilde{\rho}_{-}}{\tilde{\rho}-\tilde{\rho}_{-}}\gamma^2 \,,
	\end{gathered}
\ee
for some constant $\gamma$. Matching the two potentials then gives a differential equation for $u\left(\tilde{\rho}\right)$,
\be
	\left[\frac{d}{d\tilde{\rho}}\Delta_{t}\frac{d}{d\tilde{\rho}} +\frac{\tilde{\rho}_{+}-\tilde{\rho}_{-}}{\tilde{\rho}-\tilde{\rho}_{-}}\gamma^2\right] u = -\hat{j}(1-\hat{j})\,u \,.
\ee
This can be analytically solved in terms of Euler's hypergeometric functions. In fact, after introducing the variable $x=\frac{\tilde{\rho}-\tilde{\rho}_{+}}{\tilde{\rho}_{+}-\tilde{\rho}_{-}}$, this differential equation is exactly the same as the static problem for $p$-form perturbations of the higher-dimensional Schwarzschild-Tangherlini black hole after the replacements $\hat{j}\rightarrow\gamma$ and $\hat{\ell}\rightarrow-\hat{j}$ in Eq.~\eqref{eq:NZRadialScharzschildDd}. The general solution is, therefore,
\be\ba
	u\left(\tilde{\rho}\right) &= c_1\left(\frac{\tilde{\rho}-\tilde{\rho}_{-}}{\tilde{\rho}_{+}-\tilde{\rho}_{-}}\right)^{+\gamma}{}_2F_1\left(1-\hat{j}+\gamma,\hat{j}+\gamma;1;-\frac{\tilde{\rho}-\tilde{\rho}_{+}}{\tilde{\rho}_{+}-\tilde{\rho}_{-}}\right) \\
	&+c_2\left(\frac{\tilde{\rho}-\tilde{\rho}_{-}}{\tilde{\rho}_{+}-\tilde{\rho}_{-}}\right)^{-\gamma}{}_2F_1\left(1-\hat{j}-\gamma,\hat{j}-\gamma;1;-\frac{\tilde{\rho}-\tilde{\rho}_{+}}{\tilde{\rho}_{+}-\tilde{\rho}_{-}}\right) \,.
\ea\ee
Expanding around large distances,
\be\ba
	u\left(\tilde{\rho}\right) \xrightarrow{\tilde{\rho}\rightarrow\infty} \left(c_1+c_2\right)&\bigg[\frac{\Gamma(2\hat{j}-1)}{\Gamma(\hat{j}+\gamma)\Gamma(\hat{j}-\gamma)}\left(\frac{\tilde{\rho}-\tilde{\rho}_{+}}{\tilde{\rho}_{+}-\tilde{\rho}_{-}}\right)^{\hat{j}-1} \\
	&+ \frac{\Gamma(1-2\hat{j})}{\Gamma(1-\hat{j}+\gamma)\Gamma(1-\hat{j}-\gamma)}\left(\frac{\tilde{\rho}-\tilde{\rho}_{+}}{\tilde{\rho}_{+}-\tilde{\rho}_{-}}\right)^{-\hat{j}}\bigg] \,,
\ea\ee
we see that the asymptotic flatness condition, which implies $\tilde{\rho}\rightarrow\rho$, along with the fact that $u=\rho^{-\hat{j}}$, fixes the constant $\gamma$ to be
\be
	\gamma = \pm\hat{j} \,.
\ee
One then observes that the Love symmetry generators are exactly the same as for the general-relativistic Schwarzschild-Tangherlini black hole in Eq.~\ref{eq:SL2R_SchwarzschildDd}. More explicitly, the relation between $\rho$ and $\tilde{\rho}$ is required to be
\be
	\rho = \tilde{\rho}-\tilde{\rho}_{-} \,,
\ee
which immediately implies
\be
	f_{r}=f_{t} = 1 - \frac{\rho_{\text{h}}}{\rho} \,.
\ee
In other words, the Schwarzschild-Tangherlini black hole is the \textit{only} possible isolated asymptotically flat and electrically neutral black hole that exhibits the Love symmetry beyond the scalar response problem, at least within the regime of the current assumptions. Based on this, it is tempting to conclude that the most general theory of gravity whose black hole response problem can admit the Love symmetry beyond the scalar response problem has an action of the form
\be
	S^{\left(\text{gr}\right)} = \frac{1}{16\pi G}\int d^{d}x\sqrt{-g}\,R\,f\left(R_{\rho\sigma\mu\nu}\right)
\ee
for arbitrary\footnote{A minimal requirement here is that $f\left(R_{\rho\sigma\mu\nu}=0\right)$ is finite, i.e. that flat Minkowski spacetime is a solution of this theory and, hence, asymptotically flat solutions exist.} functions $f\left(R_{\rho\sigma\mu\nu}\right)$ of the Riemann tensor. This is the most general class of theories of gravity that admits Ricci-flat vacuum solutions, a special subclass of which is $f\left(R\right)$ gravity, but which does not include Lovelock gravity beyond General Relativity.

However, the existence of Love symmetry can easily be seen to be perturbation dependent. Take for instance the $p$-form perturbation problem of the higher-dimensional electrically charged Reissner-Nordstr\"{o}m black holes, with $p\ne1$. The above analysis then shows that only the scalar response problem can enjoy an enhanced Love symmetry, while, for example, the $2$-form response problem for a $6$-dimensional Reissner-Nordstr\"{o}m black hole will not have this property, in stark contrast to the $2$-form response problem of the $6$-dimensional Schwarzschild-Tangherlini black hole. We, therefore, have an explicit illustration within General Relativity itself where Love symmetry does not exist for asymptotically flat black holes. Let it be noted here that examples of non-zero Love numbers have also been reported for black holes in the presence of non-zero cosmological constant, see e.g. Refs~\cite{Emparan:2017qxd,Nair:2024mya}.
\section{Summary and Discussion}
\label{sec:Discussion}

In this work, we have studied the response problem for higher-dimensional spherically symmetric black holes under higher spin perturbations. After identifying the relevant master variables for each type of perturbation, we extended the work of Ref.~\cite{Hui:2020xxx} to include, besides spin-$0$ (massless scalar), spin-$1$ (electromagnetic) and spin-$2$ (gravitational) perturbations, the case of $p$-form perturbations of the Schwarzschild-Tangherlini black holes and computed the associated Love numbers within the near-zone regime. We were able to write down the static Love numbers in terms of two parameters: the multipolar order $\ell$ and the $SO\left(d-1\right)$ sector index $j$, see Eq.~\eqref{eq:StaticpLNs_SchwarzschildDd}. Similar to previous works around the response problem of the Schwarzschild-Tangherlini black hole~\cite{Kol:2011vg,Hui:2020xxx}, we find that the static Love numbers are in general in accordance with Wilsonian naturalness arguments, except for discrete towers of resonant conditions for which they vanish; these have been categorized into three classes with the corresponding behaviors given in Table~\ref{tbl:Staticp1LNs_SchwarzschildDd}, Table~\ref{tbl:Staticp2LNs_SchwarzschildDd} and Table~\ref{tbl:Staticp3LNs_SchwarzschildDd}. Furthermore, we have rigorously derived the static Love numbers associated with spin-$0$ scalar and spin-$2$ tensor-type tidal perturbations of the higher-dimensional Reissner-Nordstr\"{o}m black hole, first proposed in Ref.~\cite{Pereniguez:2021xcj}, and following the same pattern as with the corresponding cases for the Schwarzschild-Tangherlini black hole. As a byproduct of employing the near-zone scheme, we were also able to extract the spin-$0$ scalar and spin-$2$ tensor-type tidal dissipation numbers of the higher-dimensional Reissner-Nordstr\"{o}m black hole at leading order in the frequency $\omega$ of the perturbation in Eq.~\eqref{eq:ResponseCoefficientsRND}, i.e. the viscosity coefficients entering at linear order in $\omega$ for spherically symmetric and non-rotating backgrounds.

In regards to the seemingly fine-tuned resonant conditions of vanishing static Love numbers, we have identified them with selection rules outputted from enhanced ``Love'' symmetries. These are globally defined $\SL$ symmetries manifesting in the near-zone region, the vanishing of static Love numbers arising from the fact that the associated perturbations belong to a highest-weight representation of the corresponding Love symmetry. Interestingly, the Love symmetries have unique extensions to centerless Virasoro algebras, the implications of which are still poorly understood and left for future work.

We have furthermore investigated the response problem for black holes in modified theories of gravity, with explicit calculations for the static Love numbers of the Callan-Myers-Perry black hole of bosonic/heterotic string theory~\cite{Callan:1988hs,Myers:1998gt} and the $\alpha^{\prime3}$-corrected Schwarzschild-Tangherlini black hole of type-II superstring theory~\cite{Myers:1987qx}. Similar calculations were also performed in four spacetime dimensions; for $d=4$, the leading pure gravity modifications enter through Riemann-cubed corrections~\cite{Charalambous:2022rre,DeLuca:2022tkm}, while higher derivative modifications have also been considered in the literature~\cite{Cardoso:2017cfl,Cardoso:2018ptl,Katagiri:2023yzm}. While the existence of Love symmetries is not necessarily a general-relativistic effect, see e.g. Ref.~\cite{Cvetic:2021vxa} for the case of the STU black hole, it appears to be in $1$-to-$1$ correspondence with the emergence of magic zeroes with respect to the black hole response problem. We have further explored this by extracting sufficient geometric constraints for the existence of $\SL$ symmetries in the near-zone region with mixed results. For one, they suggest that the most general asymptotically flat and spherically symmetric black hole exhibiting Love symmetries under all type of perturbations studied so far is an isolated Ricci-flat solution, namely, the Schwarzschild-Tangherlini black hole. On the other hand, we came across with a simple, yet explicit, example where general-relativistic black holes have non-zero static Love numbers and exhibit no near-zone enhanced symmetries: the $p$-form response problem of the Reissner-Nordstr\"{o}m black hole, for any $p\ge2$, in stark contrast to the corresponding response problem of the electrically neutral Schwarzschild-Tangherlini black hole.


At this point, it is useful to note some features of the near-zone Love symmetry proposal compared to exact static symmetry proposals addressing the vanishing of the black hole static Love numbers~\cite{Hui:2021vcv,Hui:2022vbh,Berens:2022ebl,Katagiri:2022vyz,BenAchour:2022uqo,BenAchour:2023dgj}. The exact static symmetry proposals have the appealing feature of acting directly at IR level, while the Love symmetries have the unconventional feature of UV/IR mixing~\cite{Charalambous:2021kcz,Charalambous:2022rre}, being able to map a state outside of the validity of the near-zone. Indeed, the states in the highest-weight multiplets of the Love symmetries are compatible with the near-zone conditions for non-zero frequencies only in the near-extremal limit $r_{+}/\beta\ll1$. Nevertheless, both types of proposals have been assigned geometric interpretations, based on an underlying AdS structure. Near-zone global symmetries can be understood as approximate isometries of the black hole geometry, in the sense that they are exact isometries of ``subtracted geometries''~\cite{Charalambous:2022rre,Charalambous:2023jgq}, i.e. effective black hole geometries that preserve the thermodynamic properties of the black hole but subtract information about its surroundings~\cite{Cvetic:2011hp,Cvetic:2011dn}. At the same time, the ladder symmetry structure of the static response problem~\cite{Hui:2021vcv,Hui:2022vbh,Berens:2022ebl,Katagiri:2022vyz,Sharma:2024hlz}, for instance, has itself been attributed to conformal Killing vectors of the same subtracted geometries~\cite{Hui:2022vbh,Cardoso:2017qmj}. The procedure of studying static perturbations of the black hole via a near-zone expansion is also subtly different from the exact static analysis. As opposed to setting $\omega=0$ from the beginning, static perturbations within the near-zone regime are realized through the $\omega\rightarrow0$ limit. Related to this, Refs.~\cite{Nair:2022xfm,Chakraborty:2023zed} have emphasized that the $\omega=0$ and the phenomenologically more interesting $\omega\rightarrow0$ Love numbers are in general not the same. Even though such discontinuities appear to be irrelevant in the black hole limit~\cite{Chakraborty:2023zed}, the emergence of enhanced symmetries within the near-zone regime could allow to study their breaking for ultra-compact horizonless bodies and extract phenomenologically relevant properties.

The exact nature of the near-zone Love symmetries is still unclear. They generally fall into the category of near-zone $\SL$ symmetries that are more familiarly encountered within the context of the non-extremal Kerr/CFT correspondence~\cite{Castro:2010fd,Krishnan:2010pv,Chen:2010zwa,Lowe:2011aa,Bertini:2011ga,Kim:2012mh,Charalambous:2021kcz,Cvetic:2021vxa,Charalambous:2022rre,Charalambous:2023jgq}. Due to the ambiguity in choosing a consistent near-zone truncation of the equations of motion, one can construct infinitely many such $\SL$ structures, one for each near-zone truncation, whose generators, however, are in general only locally defined and, hence, not able to address the vanishings of the Love numbers via representation theory arguments. This requirement, i.e. the global definiteness of the near-zone $\SL$ symmetry, turns out to always single out only two of these infinitely-many near-zone truncations~\cite{Charalambous:2021kcz,Charalambous:2022rre,Charalambous:2023jgq}. Their global structure then allows to employ highest-weight representation theory arguments and extract the seemingly fine-tuned properties of the static Love numbers as selection rules~\cite{Charalambous:2021kcz,Charalambous:2022rre,Charalambous:2023jgq}. In fact, both $\SL$ Love symmetries turn out be a subset of a larger symmetry structure, e.g. $\SL\ltimes\hat{U}\left(1\right)$ for $d=4$~\cite{Charalambous:2021kcz,Charalambous:2022rre} or $\SL\ltimes\hat{U}\left(1\right)^2$ for $d=5$~\cite{Charalambous:2023jgq} rotating black holes, while all the other near-zone $\SL$ structures that are only locally definable can themselves be realized as particular local diffeomorphisms of the Love symmetries, see e.g. Appendices D of Ref.~\cite{Charalambous:2022rre} and Ref.~\cite{Charalambous:2023jgq}.

To be more concrete, the Love symmetry for a rotating black hole spans the whole $2$-d conformal group, i.e. it is actually an $\SL\times\overline{\SL}$ symmetry, with the second $\overline{\SL}$ factor being only locally defined. In the spirit of the non-extremal Kerr/CFT correspondence in four spacetime dimensions~\cite{Castro:2010fd,Krishnan:2010pv,Chen:2010zwa,Lowe:2011aa}, the regime where the Love symmetry emerges is dual to a $\text{CFT}_2$ thermal state, with the left-movers being at zero temperature and the right-movers being at fixed non-zero temperature $T_{\text{R}}$. The non-zero temperature of the right-movers is what makes the second $\overline{\SL}$ factor non globally defined. More generally, the temperature $T_{\text{L}}$ of the left-movers can be changed by performing local diffeomorphisms of the form $t\rightarrow t+\tau\beta\left(\phi-\Omega t\right)$, in Boyer-Lindquist coordinates $\left(t,r,\theta,\phi\right)$, with $\beta$ and $\Omega$ the inverse surface gravity and angular velocity of the black hole respectively. Under such transformations, the Love $\SL\times\overline{\SL}$ gets mapped to a different $\SL\times\overline{\SL}$ structure that corresponds to a $\text{CFT}_2$ thermal state with $T_{\text{L}}=\frac{\tau}{2\pi}\ne0$, deeming the first $\SL$ only locally definable as well. Then, all near-horizon $\SL\times\overline{\SL}$ enhancements can be nicely captured in an infinite-dimensional extension of the Love symmetry; in four spacetime dimensions, this larger algebraic structure is an $\left(\SL_{\text{Love}}\ltimes\hat{U}\left(1\right)\right)\times\left(\overline{\SL}_{\text{Love}}\ltimes\hat{U}\left(1\right)\right)$ structure, equipped with local tempral diffeomorphisms of the form just described~\cite{Charalambous:2021kcz,Charalambous:2022rre}.

It is suspected that these persisting $\SL$ structures are, in fact, remnants of the enhanced isometry of the near-horizon throat of extremal black holes~\cite{Bardeen:1999px,Kunduri:2007vf}. This connection is most clear for the case of the non-rotating Reissner-Nordstr\"{o}m black hole. As was demonstrated explicitly in Ref.~\cite{Charalambous:2022rre}, appropriately taking the extremal limit of the associated globally defined Love symmetry generators recovers the exact Killing vectors generating the $\SL$ isometry subgroup of the near-horizon throat. However, the existence of enhanced $\SL$ isometry subgroups for extremal black holes appears to be a universal, theory-independent, phenomenon~\cite{Kunduri:2007vf}, as opposed to the existence of the Love symmetry as we saw more explicitly in Section~\ref{sec:LoveSymmetryModGR}. Nevertheless, there exist another observation that seems to interpolate between the extremal and non-extremal conformal structures. The crucial relevant difference between black holes in generic modified theories of gravity and black hole geometries such as those of General Relativity is that there are no near-horizon modes of extremal black holes that can propagate in the ``far-horizon'' region. For general-relativistic black holes, on the other hand, it was remarked in Refs.~\cite{Charalambous:2022rre,Charalambous:2023jgq} that static axisymmetric modes do survive beyond the near-horizon regime. This accidental robustness of the near-horizon symmetry of extremal black holes can be traced back to the fact that the full discriminant function determining the locations of the horizons remains a quadratic polynomial. In fact, the sufficient geometric condition in Eq.~\eqref{eq:SL2RModGRConstraintDd} for the existence of Love symmetry for scalar perturbations of spherically symmetric black holes in a generic theory of gravity precisely implies this form of the discriminant function, see Eq.~\eqref{eq:DeltaTModGR}. Related to this, it would be particularly interesting to seek a connection between the globally defined near-zone symmetries of non-extremal black holes and the accidental symmetry found in Ref.~\cite{Porfyriadis:2021psx}, see also Refs~\cite{Hadar:2020kry,Horowitz:2022leb,Horowitz:2023xyl}, which maps perturbations of exactly extremal black holes to perturbations of near-extremal black holes.

Another possible application of near-zone symmetries such as Love symmetries is within the context of asymptotic symmetries, the classic asymptotically flat paradigm being the infinite-dimensional BMS group at null infinity~\cite{Bondi:1962px,Sachs:1962zza}. More importantly, the near-horizon asymptotic symmetries were also found to be extended, spanning an infinite-dimensional BMS-like algebra~\cite{Donnay:2015abr,Donnay:2016ejv,Penna:2015gza,Grumiller:2019fmp,Donnay:2018ckb,Donnay:2020yxw}. It would then be interesting to explore whether near-zone symmetries can enter as interpolators between near-horizon and near-null-infinity symmetries, since the near-zone region is itself extending beyond the near-horizon regime and has a non-empty overlap with the far-zone region.


Last, it is interesting to comment on what happens to dynamic and non-linear responses. In general, dynamical Love numbers are non-zero and logarithmically running, in accordance with naturalness expectations~\cite{Charalambous:2021mea,Perry:2023wmm}, see also Refs.~\cite{Chakraborty:2023zed,DeLuca:2022xlz}. However, recent works on non-linear responses reveal that non-linear static Love numbers also appear to exhibit fine-tuned properties in some occasions, see e.g. Refs.~\cite{Gurlebeck:2015xpa,DeLuca:2023mio,Riva:2023rcm}. It remains to be seen whether Love symmetries play any role in addressing these types of magic zeroes associated to the response problem, a task left for future work.

\paragraph{Acknowledgments}
I thank Sergei Dubovsky and Laura Donnay for helpful discussions. I am particularly grateful to Mikhail Ivanov for detailed feedback on the draft. PC is supported by the European Research Council (ERC) Project 101076737 -- CeleBH. Views and opinions expressed are however those of the author only and do not necessarily reflect those of the European Union or the European Research Council. Neither the European Union nor the granting authority can be held responsible for them.

\appendix
\section{Useful formulae involving the $\Gamma$-function and Euler's hypergeometric function}
\label{app:2F1Gamma}

In this Appendix, we enumerate a number of useful formulae relevant in solving the near-zone equations of motion and extracting the conservative and dissipative pieces of the response coefficients. All of these formulae can be found in the NIST Digital Library of Mathematical Functions~\cite{NIST}.

\subsection{$\Gamma$-function}
\label{sec:GammaFunction}

We begin with the (complete) $\Gamma$-function, defined by Euler's integral,
\be
	\Gamma\left(z\right) = \int_0^{\infty}dt\,t^{z-1}e^{-t} \,,\quad \text{Re}\left\{z\right\}>0 \,,
\ee
and serving as an extension of the familiar factorial function, satisfying the recurrence relation $\Gamma\left(z+1\right)=z\Gamma\left(z\right)$. For positive integer arguments, it is just the usual factorial offset by one unit,
\be
	\Gamma\left(n\right) = \left(n-1\right)! \,,\quad n=1,2,\dots \,.
\ee
The $\Gamma$-function can also be analytically continued to $\text{Re}\left\{z\right\}\le0$. For example, this can be done by the mirror/reflection formula
\be\label{eq:GammaMirrorFormula}
	\Gamma\left(z\right)\Gamma\left(1-z\right) = \frac{\pi}{\sin\pi z} \,,
\ee
which is particularly useful when studying the behavior of the response coefficients as it allows to explicitly reveal the vanishing or running of the Love numbers when sending the orbital number to range in its physical integer values.

The $\Gamma$-function is a meromorphic function with no roots and with simple poles at non-positive integers, with residue
\be
	\underset{z=-n}{\text{Res}}\Gamma\left(z\right) = \frac{\left(-1\right)^{n}}{n!} \,,\quad n=0,1,2,\dots \,.
\ee
Its logarithmic derivative defines the digamma or $\psi$-function,
\be
	\psi\left(z\right) = \frac{\Gamma^{\prime}\left(z\right)}{\Gamma\left(z\right)}
\ee
which is a meromorpic function with simple poles of residue $-1$ at semi-negative integers, satisfying the recursion relation $\psi\left(z+1\right) = \psi\left(z\right) + z^{-1}$. 

Another useful $\Gamma$-function identity is the Legendre duplication formula,
\be
	\Gamma\left(z\right)\Gamma\left(z+\frac{1}{2}\right) = 2^{1-2z}\sqrt{\pi}\Gamma\left(2z\right) \,,
\ee
which is a special case of the Gauss multiplication formula,
\be
	\prod_{k=1}^{n}\Gamma\left(z+\frac{k-1}{n}\right) = n^{\frac{1}{2}-nz}\left(2\pi\right)^{\frac{n-1}{2}}\Gamma\left(nz\right) \,.
\ee
The Legendre duplication formula helps comparing the expressions of the response coefficients written in this work with other works in the literature. Last, it is sometimes convenient to employ the identity
\be
	\left|\Gamma\left(n+1+ix\right)\right|^2 = \frac{\pi x}{\sin\pi x}\prod_{k=1}^{n}\left(k^2+x^2\right) \,,\quad n\in\mathbb{N}\,,\quad x\in\mathbb{R}
\ee
to write the Love numbers in a more practical manner.

\subsection{Euler's hypergeometric function}
\label{sec:2F1Function}

Euler's hypergeometric function is characterized by $2+1$ parameters, $a$, $b$ and $c$, and is defined on the disk $\left|z\right|<1$ by the series
\be
	{}_2F_1\left(a,b;c;z\right) = \sum_{k=0}^{\infty}\frac{\left(a\right)_{k}\left(b\right)_{k}}{\left(c\right)_{k}}\frac{z^{k}}{k!} \,,
\ee
where $\left(a\right)_{k} = \frac{\Gamma\left(a+k\right)}{\Gamma\left(a\right)}$ is the Pochhammer symbol, sometimes also referred to as the rising factorial. It is one of the independent solutions expandable as a Frobenius series around $z=0$ of the hypergeometric differential equation
\be
	\left[z\left(1-z\right)\frac{d^2}{dz^2}+\left[c-\left(a+b+1\right)z\right]\frac{d}{dz} - ab\right]\,y\left(z\right) = 0 \,,
\ee
given that $c$ is not a non-positive integer. Useful transformation properties within the principal branch $\left|\text{Arg}\left(1-z\right)\right|<\pi$ involve Euler's transformation,
\be
	{}_2F_1\left(a,b;c;z\right) = \left(1-z\right)^{c-a-b}{}_2F_1\left(c-a,c-b;c;z\right) \,,
\ee
and the two Pfaff transformations,
\be\ba
	{}_2F_1\left(a,b;c;z\right) &= \left(1-z\right)^{-a}{}_2F_1\left(a,c-b;c;\frac{z}{z-1}\right) \\
	&= \left(1-z\right)^{-b}{}_2F_1\left(c-a,b;c;\frac{z}{z-1}\right) \,.
\ea\ee
The hypergeometric function can be analytically continued to $\left|z\right|>1$ via
\be\ba\label{eq:2F1LargeX}
	{}_2F_1\left(a,b;c;z\right) &= \frac{\Gamma\left(c\right)\Gamma\left(b-a\right)}{\Gamma\left(b\right)\Gamma\left(c-a\right)}\left(-z\right)^{-a}{}_2F_1\left(a,a-c+1;a-b+1;\frac{1}{z}\right) \\
	&+\frac{\Gamma\left(c\right)\Gamma\left(a-b\right)}{\Gamma\left(a\right)\Gamma\left(c-b\right)}\left(-z\right)^{-b}{}_2F_1\left(b,b-c+1;b-a+1;\frac{1}{z}\right) \,,
\ea\ee
which is valid for $\left|\text{Arg}\left(-z\right)\right|<\pi$, e.g. for negative real arguments such as the hypergeometric functions encountered in this work. This analytic continuation formula is particularly useful when extracting the source/response splitting of the profiles of the black hole perturbations and, subsequently, the response coefficients.

The hypergeometric function is analytic for all $a,b\in\mathbb{C}$ but does not exist for non-positive integer values of the parameter $c$ due to the development of simple poles. Nevertheless, the following limit exists
\be
	\lim\limits_{c\rightarrow-n}\frac{{}_2F_1\left(a,b;c;z\right)}{\Gamma\left(c\right)} = \frac{\left(a\right)_{n+1}\left(b\right)_{n+1}}{\left(n+1\right)!}z^{n+1}{}_2F_1\left(a+n+1,b+n+1;n+2;z\right) \,.
\ee
This is relevant when discussing the seemingly diverging behavior of the Love numbers which is compensated by a divergence of the above form in the ``source'' part of the solution with the end result being a regular solution profile involving logarithms that reflect the classical RG flow of the Love numbers.

Last, when $a$ or $b$ is a non-positive integer, the hypergeometric function reduces to a polynomial,
\be
	{}_2F_1\left(-n,b;c;z\right) = \sum_{k=0}^{n}\left(-1\right)^{k}\binom{n}{k}\frac{\left(b\right)_{k}}{\left(c\right)_{k}}\frac{z^{k}}{k!} \,,\quad n=0,1,2,\dots \,,
\ee
as long as $c$ is not a negative integer larger than $n$.

\addcontentsline{toc}{section}{References}
\bibliographystyle{JHEP}
\bibliography{LoveDdHigherS}

\providecommand{\href}[2]{#2}\begingroup\raggedright\begin{thebibliography}{100}

\bibitem{LIGOScientific:2016aoc}
{\scshape LIGO Scientific, Virgo} collaboration, B.~P. Abbott et~al.,
  \emph{{Observation of Gravitational Waves from a Binary Black Hole Merger}},
  \href{https://doi.org/10.1103/PhysRevLett.116.061102}{\emph{Phys. Rev. Lett.}
  {\bfseries 116} (2016) 061102}
  [\href{https://arxiv.org/abs/1602.03837}{{\ttfamily 1602.03837}}].

\bibitem{KAGRA:2021vkt}
{\scshape KAGRA, VIRGO, LIGO Scientific} collaboration, R.~Abbott et~al.,
  \emph{{GWTC-3: Compact Binary Coalescences Observed by LIGO and Virgo during
  the Second Part of the Third Observing Run}},
  \href{https://doi.org/10.1103/PhysRevX.13.041039}{\emph{Phys. Rev. X}
  {\bfseries 13} (2023) 041039}
  [\href{https://arxiv.org/abs/2111.03606}{{\ttfamily 2111.03606}}].

\bibitem{Saleem:2021iwi}
M.~Saleem et~al., \emph{{The science case for LIGO-India}},
  \href{https://doi.org/10.1088/1361-6382/ac3b99}{\emph{Class. Quant. Grav.}
  {\bfseries 39} (2022) 025004}
  [\href{https://arxiv.org/abs/2105.01716}{{\ttfamily 2105.01716}}].

\bibitem{LIGOScientific:2016wof}
{\scshape LIGO Scientific} collaboration, B.~P. Abbott et~al., \emph{{Exploring
  the Sensitivity of Next Generation Gravitational Wave Detectors}},
  \href{https://doi.org/10.1088/1361-6382/aa51f4}{\emph{Class. Quant. Grav.}
  {\bfseries 34} (2017) 044001}
  [\href{https://arxiv.org/abs/1607.08697}{{\ttfamily 1607.08697}}].

\bibitem{Punturo:2010zza}
M.~Punturo et~al., \emph{{The third generation of gravitational wave
  observatories and their science reach}},
  \href{https://doi.org/10.1088/0264-9381/27/8/084007}{\emph{Class. Quant.
  Grav.} {\bfseries 27} (2010) 084007}.

\bibitem{2017arXiv170200786A}
P.~{Amaro-Seoane}, H.~{Audley}, S.~{Babak}, J.~{Baker}, E.~{Barausse},
  P.~{Bender} et~al., \emph{{Laser Interferometer Space Antenna}},
  \href{https://doi.org/10.48550/arXiv.1702.00786}{\emph{arXiv e-prints} (2017)
  arXiv:1702.00786} [\href{https://arxiv.org/abs/1702.00786}{{\ttfamily
  1702.00786}}].

\bibitem{Reitze:2019iox}
D.~Reitze et~al., \emph{{Cosmic Explorer: The U.S. Contribution to
  Gravitational-Wave Astronomy beyond LIGO}}, {\emph{Bull. Am. Astron. Soc.}
  {\bfseries 51} (2019) 035}
  [\href{https://arxiv.org/abs/1907.04833}{{\ttfamily 1907.04833}}].

\bibitem{Love:1912}
A.~E.~H. Love, \emph{{Some Problems of Geodynamics}},
  \href{https://doi.org/10.1038/089471a0}{\emph{Nature} {\bfseries 89} (1912)
  471}.

\bibitem{PoissonWill2014}
E.~Poisson and C.~M. Will, \emph{Gravity: Newtonian, Post-Newtonian,
  Relativistic}. Cambridge University Press, 2014,
  \href{https://doi.org/10.1017/CBO9781139507486}{10.1017/CBO9781139507486}.

\bibitem{LIGOScientific:2017vwq}
{\scshape LIGO Scientific, Virgo} collaboration, B.~P. Abbott et~al.,
  \emph{{GW170817: Observation of Gravitational Waves from a Binary Neutron
  Star Inspiral}},
  \href{https://doi.org/10.1103/PhysRevLett.119.161101}{\emph{Phys. Rev. Lett.}
  {\bfseries 119} (2017) 161101}
  [\href{https://arxiv.org/abs/1710.05832}{{\ttfamily 1710.05832}}].

\bibitem{Raithel:2018ncd}
C.~Raithel, F.~\"Ozel and D.~Psaltis, \emph{{Tidal deformability from GW170817
  as a direct probe of the neutron star radius}},
  \href{https://doi.org/10.3847/2041-8213/aabcbf}{\emph{Astrophys. J. Lett.}
  {\bfseries 857} (2018) L23}
  [\href{https://arxiv.org/abs/1803.07687}{{\ttfamily 1803.07687}}].

\bibitem{Chatziioannou:2020pqz}
K.~Chatziioannou, \emph{{Neutron star tidal deformability and equation of state
  constraints}}, \href{https://doi.org/10.1007/s10714-020-02754-3}{\emph{Gen.
  Rel. Grav.} {\bfseries 52} (2020) 109}
  [\href{https://arxiv.org/abs/2006.03168}{{\ttfamily 2006.03168}}].

\bibitem{Pacilio:2021jmq}
C.~Pacilio, A.~Maselli, M.~Fasano and P.~Pani, \emph{{Ranking Love Numbers for
  the Neutron Star Equation of State: The Need for Third-Generation
  Detectors}},
  \href{https://doi.org/10.1103/PhysRevLett.128.101101}{\emph{Phys. Rev. Lett.}
  {\bfseries 128} (2022) 101101}
  [\href{https://arxiv.org/abs/2104.10035}{{\ttfamily 2104.10035}}].

\bibitem{Flanagan:2007ix}
E.~E. Flanagan and T.~Hinderer, \emph{{Constraining neutron star tidal Love
  numbers with gravitational wave detectors}},
  \href{https://doi.org/10.1103/PhysRevD.77.021502}{\emph{Phys. Rev. D}
  {\bfseries 77} (2008) 021502}
  [\href{https://arxiv.org/abs/0709.1915}{{\ttfamily 0709.1915}}].

\bibitem{Piovano:2022ojl}
G.~A. Piovano, A.~Maselli and P.~Pani, \emph{{Constraining the tidal
  deformability of supermassive objects with extreme mass ratio inspirals and
  semianalytical frequency-domain waveforms}},
  \href{https://doi.org/10.1103/PhysRevD.107.024021}{\emph{Phys. Rev. D}
  {\bfseries 107} (2023) 024021}
  [\href{https://arxiv.org/abs/2207.07452}{{\ttfamily 2207.07452}}].

\bibitem{Cardoso:2017cfl}
V.~Cardoso, E.~Franzin, A.~Maselli, P.~Pani and G.~Raposo, \emph{{Testing
  strong-field gravity with tidal Love numbers}},
  \href{https://doi.org/10.1103/PhysRevD.95.084014}{\emph{Phys. Rev. D}
  {\bfseries 95} (2017) 084014}
  [\href{https://arxiv.org/abs/1701.01116}{{\ttfamily 1701.01116}}].

\bibitem{Franzin:2017mtq}
E.~Franzin, V.~Cardoso, P.~Pani and G.~Raposo, \emph{{Testing strong gravity
  with gravitational waves and Love numbers}},
  \href{https://doi.org/10.1088/1742-6596/841/1/012035}{\emph{J. Phys. Conf.
  Ser.} {\bfseries 841} (2017) 012035}.

\bibitem{Cardoso:2018ptl}
V.~Cardoso, M.~Kimura, A.~Maselli and L.~Senatore, \emph{{Black Holes in an
  Effective Field Theory Extension of General Relativity}},
  \href{https://doi.org/10.1103/PhysRevLett.121.251105}{\emph{Phys. Rev. Lett.}
  {\bfseries 121} (2018) 251105}
  [\href{https://arxiv.org/abs/1808.08962}{{\ttfamily 1808.08962}}].

\bibitem{Katagiri:2023yzm}
T.~Katagiri, H.~Nakano and K.~Omukai, \emph{{Stability of relativistic tidal
  response against small potential modification}},
  \href{https://doi.org/10.1103/PhysRevD.108.084049}{\emph{Phys. Rev. D}
  {\bfseries 108} (2023) 084049}
  [\href{https://arxiv.org/abs/2304.04551}{{\ttfamily 2304.04551}}].

\bibitem{Usman:2018imj}
S.~A. Usman, J.~C. Mills and S.~Fairhurst, \emph{{Constraining the Inclinations
  of Binary Mergers from Gravitational-wave Observations}},
  \href{https://doi.org/10.3847/1538-4357/ab0b3e}{\emph{Astrophys. J.}
  {\bfseries 877} (2019) 82}
  [\href{https://arxiv.org/abs/1809.10727}{{\ttfamily 1809.10727}}].

\bibitem{Xie:2022brn}
Y.~Xie, D.~Chatterjee, G.~Holder, D.~E. Holz, S.~Perkins, K.~Yagi et~al.,
  \emph{{Breaking bad degeneracies with Love relations: Improving
  gravitational-wave measurements through universal relations}},
  \href{https://doi.org/10.1103/PhysRevD.107.043010}{\emph{Phys. Rev. D}
  {\bfseries 107} (2023) 043010}
  [\href{https://arxiv.org/abs/2210.09386}{{\ttfamily 2210.09386}}].

\bibitem{Yagi:2013bca}
K.~Yagi and N.~Yunes, \emph{{I-Love-Q}},
  \href{https://doi.org/10.1126/science.1236462}{\emph{Science} {\bfseries 341}
  (2013) 365} [\href{https://arxiv.org/abs/1302.4499}{{\ttfamily 1302.4499}}].

\bibitem{Yagi:2013awa}
K.~Yagi and N.~Yunes, \emph{{I-Love-Q Relations in Neutron Stars and their
  Applications to Astrophysics, Gravitational Waves and Fundamental Physics}},
  \href{https://doi.org/10.1103/PhysRevD.88.023009}{\emph{Phys. Rev. D}
  {\bfseries 88} (2013) 023009}
  [\href{https://arxiv.org/abs/1303.1528}{{\ttfamily 1303.1528}}].

\bibitem{Yagi:2015pkc}
K.~Yagi and N.~Yunes, \emph{{Binary Love Relations}},
  \href{https://doi.org/10.1088/0264-9381/33/13/13LT01}{\emph{Class. Quant.
  Grav.} {\bfseries 33} (2016) 13LT01}
  [\href{https://arxiv.org/abs/1512.02639}{{\ttfamily 1512.02639}}].

\bibitem{Yagi:2016qmr}
K.~Yagi and N.~Yunes, \emph{{Approximate Universal Relations among Tidal
  Parameters for Neutron Star Binaries}},
  \href{https://doi.org/10.1088/1361-6382/34/1/015006}{\emph{Class. Quant.
  Grav.} {\bfseries 34} (2017) 015006}
  [\href{https://arxiv.org/abs/1608.06187}{{\ttfamily 1608.06187}}].

\bibitem{Goldberger:2004jt}
W.~D. Goldberger and I.~Z. Rothstein, \emph{{An Effective field theory of
  gravity for extended objects}},
  \href{https://doi.org/10.1103/PhysRevD.73.104029}{\emph{Phys. Rev. D}
  {\bfseries 73} (2006) 104029}
  [\href{https://arxiv.org/abs/hep-th/0409156}{{\ttfamily hep-th/0409156}}].

\bibitem{Porto:2005ac}
R.~A. Porto, \emph{{Post-Newtonian corrections to the motion of spinning bodies
  in NRGR}}, \href{https://doi.org/10.1103/PhysRevD.73.104031}{\emph{Phys. Rev.
  D} {\bfseries 73} (2006) 104031}
  [\href{https://arxiv.org/abs/gr-qc/0511061}{{\ttfamily gr-qc/0511061}}].

\bibitem{Porto:2016pyg}
R.~A. Porto, \emph{{The effective field theorist\textquoteright{}s approach to
  gravitational dynamics}},
  \href{https://doi.org/10.1016/j.physrep.2016.04.003}{\emph{Phys. Rept.}
  {\bfseries 633} (2016) 1} [\href{https://arxiv.org/abs/1601.04914}{{\ttfamily
  1601.04914}}].

\bibitem{Levi:2015msa}
M.~Levi and J.~Steinhoff, \emph{{Spinning gravitating objects in the effective
  field theory in the post-Newtonian scheme}},
  \href{https://doi.org/10.1007/JHEP09(2015)219}{\emph{JHEP} {\bfseries 09}
  (2015) 219} [\href{https://arxiv.org/abs/1501.04956}{{\ttfamily
  1501.04956}}].

\bibitem{Levi:2018nxp}
M.~Levi, \emph{{Effective Field Theories of Post-Newtonian Gravity: A
  comprehensive review}},
  \href{https://doi.org/10.1088/1361-6633/ab12bc}{\emph{Rept. Prog. Phys.}
  {\bfseries 83} (2020) 075901}
  [\href{https://arxiv.org/abs/1807.01699}{{\ttfamily 1807.01699}}].

\bibitem{Goldberger:2022ebt}
W.~D. Goldberger, \emph{{Effective field theories of gravity and compact binary
  dynamics: A Snowmass 2021 whitepaper}},  in \emph{{2022 Snowmass Summer
  Study}}, 6, 2022, \href{https://arxiv.org/abs/2206.14249}{{\ttfamily
  2206.14249}}.

\bibitem{Fang:2005qq}
H.~Fang and G.~Lovelace, \emph{{Tidal coupling of a Schwarzschild black hole
  and circularly orbiting moon}},
  \href{https://doi.org/10.1103/PhysRevD.72.124016}{\emph{Phys. Rev. D}
  {\bfseries 72} (2005) 124016}
  [\href{https://arxiv.org/abs/gr-qc/0505156}{{\ttfamily gr-qc/0505156}}].

\bibitem{Damour:2009vw}
T.~Damour and A.~Nagar, \emph{{Relativistic tidal properties of neutron
  stars}}, \href{https://doi.org/10.1103/PhysRevD.80.084035}{\emph{Phys. Rev.
  D} {\bfseries 80} (2009) 084035}
  [\href{https://arxiv.org/abs/0906.0096}{{\ttfamily 0906.0096}}].

\bibitem{Binnington:2009bb}
T.~Binnington and E.~Poisson, \emph{{Relativistic theory of tidal Love
  numbers}}, \href{https://doi.org/10.1103/PhysRevD.80.084018}{\emph{Phys. Rev.
  D} {\bfseries 80} (2009) 084018}
  [\href{https://arxiv.org/abs/0906.1366}{{\ttfamily 0906.1366}}].

\bibitem{Gurlebeck:2015xpa}
N.~G\"urlebeck, \emph{{No-hair theorem for Black Holes in Astrophysical
  Environments}},
  \href{https://doi.org/10.1103/PhysRevLett.114.151102}{\emph{Phys. Rev. Lett.}
  {\bfseries 114} (2015) 151102}
  [\href{https://arxiv.org/abs/1503.03240}{{\ttfamily 1503.03240}}].

\bibitem{Bicak:1977}
J.~Bičák and L.~Dvořák, \emph{{Stationary electromagnetic fields around
  black holes. I. General solutions and the fields of some special sources near
  a Schwarzschild black hole}},
  \href{https://doi.org/10.1007/BF01587004}{\emph{Czechoslovak Journal of
  Physics} {\bfseries 2} (1977) 127}.

\bibitem{Bicak:1976}
J.~Bičák and L.~Dvořák, \emph{{Stationary electromagnetic fields around
  black holes. II. General solutions and the fields of some special sources
  near a Kerr black hole}},
  \href{https://doi.org/10.1007/BF00766421}{\emph{General Relativity and
  Gravitation} {\bfseries 7} (1976) 959}.

\bibitem{Poisson:2014gka}
E.~Poisson, \emph{{Tidal deformation of a slowly rotating black hole}},
  \href{https://doi.org/10.1103/PhysRevD.91.044004}{\emph{Phys. Rev. D}
  {\bfseries 91} (2015) 044004}
  [\href{https://arxiv.org/abs/1411.4711}{{\ttfamily 1411.4711}}].

\bibitem{Landry:2015zfa}
P.~Landry and E.~Poisson, \emph{{Tidal deformation of a slowly rotating
  material body. External metric}},
  \href{https://doi.org/10.1103/PhysRevD.91.104018}{\emph{Phys. Rev. D}
  {\bfseries 91} (2015) 104018}
  [\href{https://arxiv.org/abs/1503.07366}{{\ttfamily 1503.07366}}].

\bibitem{Pani:2015hfa}
P.~Pani, L.~Gualtieri, A.~Maselli and V.~Ferrari, \emph{{Tidal deformations of
  a spinning compact object}},
  \href{https://doi.org/10.1103/PhysRevD.92.024010}{\emph{Phys. Rev. D}
  {\bfseries 92} (2015) 024010}
  [\href{https://arxiv.org/abs/1503.07365}{{\ttfamily 1503.07365}}].

\bibitem{Pani:2015nua}
P.~Pani, L.~Gualtieri and V.~Ferrari, \emph{{Tidal Love numbers of a slowly
  spinning neutron star}},
  \href{https://doi.org/10.1103/PhysRevD.92.124003}{\emph{Phys. Rev. D}
  {\bfseries 92} (2015) 124003}
  [\href{https://arxiv.org/abs/1509.02171}{{\ttfamily 1509.02171}}].

\bibitem{LeTiec:2020spy}
A.~Le~Tiec and M.~Casals, \emph{{Spinning Black Holes Fall in Love}},
  \href{https://doi.org/10.1103/PhysRevLett.126.131102}{\emph{Phys. Rev. Lett.}
  {\bfseries 126} (2021) 131102}
  [\href{https://arxiv.org/abs/2007.00214}{{\ttfamily 2007.00214}}].

\bibitem{LeTiec:2020bos}
A.~Le~Tiec, M.~Casals and E.~Franzin, \emph{{Tidal Love Numbers of Kerr Black
  Holes}}, \href{https://doi.org/10.1103/PhysRevD.103.084021}{\emph{Phys. Rev.
  D} {\bfseries 103} (2021) 084021}
  [\href{https://arxiv.org/abs/2010.15795}{{\ttfamily 2010.15795}}].

\bibitem{Chia:2020yla}
H.~S. Chia, \emph{{Tidal deformation and dissipation of rotating black holes}},
  \href{https://doi.org/10.1103/PhysRevD.104.024013}{\emph{Phys. Rev. D}
  {\bfseries 104} (2021) 024013}
  [\href{https://arxiv.org/abs/2010.07300}{{\ttfamily 2010.07300}}].

\bibitem{Charalambous:2021mea}
P.~Charalambous, S.~Dubovsky and M.~M. Ivanov, \emph{{On the Vanishing of Love
  Numbers for Kerr Black Holes}},
  \href{https://doi.org/10.1007/JHEP05(2021)038}{\emph{JHEP} {\bfseries 05}
  (2021) 038} [\href{https://arxiv.org/abs/2102.08917}{{\ttfamily
  2102.08917}}].

\bibitem{Ivanov:2022qqt}
M.~M. Ivanov and Z.~Zhou, \emph{{Vanishing of Black Hole Tidal Love Numbers
  from Scattering Amplitudes}},
  \href{https://doi.org/10.1103/PhysRevLett.130.091403}{\emph{Phys. Rev. Lett.}
  {\bfseries 130} (2023) 091403}
  [\href{https://arxiv.org/abs/2209.14324}{{\ttfamily 2209.14324}}].

\bibitem{Ivanov:2022hlo}
M.~M. Ivanov and Z.~Zhou, \emph{{Revisiting the matching of black hole tidal
  responses: A systematic study of relativistic and logarithmic corrections}},
  \href{https://doi.org/10.1103/PhysRevD.107.084030}{\emph{Phys. Rev. D}
  {\bfseries 107} (2023) 084030}
  [\href{https://arxiv.org/abs/2208.08459}{{\ttfamily 2208.08459}}].

\bibitem{Tolman:1939jz}
R.~C. Tolman, \emph{{Static solutions of Einstein's field equations for spheres
  of fluid}}, \href{https://doi.org/10.1103/PhysRev.55.364}{\emph{Phys. Rev.}
  {\bfseries 55} (1939) 364}.

\bibitem{Oppenheimer:1939ne}
J.~R. Oppenheimer and G.~M. Volkoff, \emph{{On massive neutron cores}},
  \href{https://doi.org/10.1103/PhysRev.55.374}{\emph{Phys. Rev.} {\bfseries
  55} (1939) 374}.

\bibitem{Kalogera:1996ci}
V.~Kalogera and G.~Baym, \emph{{The maximum mass of a neutron star}},
  \href{https://doi.org/10.1086/310296}{\emph{Astrophys. J. Lett.} {\bfseries
  470} (1996) L61} [\href{https://arxiv.org/abs/astro-ph/9608059}{{\ttfamily
  astro-ph/9608059}}].

\bibitem{DeLuca:2021ite}
V.~De~Luca and P.~Pani, \emph{{Tidal deformability of dressed black holes and
  tests of ultralight bosons in extended mass ranges}},
  \href{https://doi.org/10.1088/1475-7516/2021/08/032}{\emph{JCAP} {\bfseries
  08} (2021) 032} [\href{https://arxiv.org/abs/2106.14428}{{\ttfamily
  2106.14428}}].

\bibitem{Pani:2019cyc}
P.~Pani and A.~Maselli, \emph{{Love in Extrema Ratio}},
  \href{https://doi.org/10.1142/S0218271819440012}{\emph{Int. J. Mod. Phys. D}
  {\bfseries 28} (2019) 1944001}
  [\href{https://arxiv.org/abs/1905.03947}{{\ttfamily 1905.03947}}].

\bibitem{Pani:2015tga}
P.~Pani, \emph{{I-Love-Q relations for gravastars and the approach to the
  black-hole limit}},
  \href{https://doi.org/10.1103/PhysRevD.95.049902}{\emph{Phys. Rev. D}
  {\bfseries 92} (2015) 124030}
  [\href{https://arxiv.org/abs/1506.06050}{{\ttfamily 1506.06050}}].

\bibitem{Uchikata:2016qku}
N.~Uchikata, S.~Yoshida and P.~Pani, \emph{{Tidal deformability and I-Love-Q
  relations for gravastars with polytropic thin shells}},
  \href{https://doi.org/10.1103/PhysRevD.94.064015}{\emph{Phys. Rev. D}
  {\bfseries 94} (2016) 064015}
  [\href{https://arxiv.org/abs/1607.03593}{{\ttfamily 1607.03593}}].

\bibitem{tHooft:1979rat}
G.~'t~Hooft, \emph{{Naturalness, chiral symmetry, and spontaneous chiral
  symmetry breaking}},
  \href{https://doi.org/10.1007/978-1-4684-7571-5_9}{\emph{NATO Sci. Ser. B}
  {\bfseries 59} (1980) 135}.

\bibitem{Porto:2016zng}
R.~A. Porto, \emph{{The Tune of Love and the Nature(ness) of Spacetime}},
  \href{https://doi.org/10.1002/prop.201600064}{\emph{Fortsch. Phys.}
  {\bfseries 64} (2016) 723}
  [\href{https://arxiv.org/abs/1606.08895}{{\ttfamily 1606.08895}}].

\bibitem{Bardeen:1999px}
J.~M. Bardeen and G.~T. Horowitz, \emph{{The Extreme Kerr throat geometry: A
  Vacuum analog of
  ${\mathrm{AdS}}_{2}{\ifmmode\times\else\texttimes\fi{}\mathrm{S}}^{2}$}},
  \href{https://doi.org/10.1103/PhysRevD.60.104030}{\emph{Phys. Rev. D}
  {\bfseries 60} (1999) 104030}
  [\href{https://arxiv.org/abs/hep-th/9905099}{{\ttfamily hep-th/9905099}}].

\bibitem{Kunduri:2007vf}
H.~K. Kunduri, J.~Lucietti and H.~S. Reall, \emph{{Near-horizon symmetries of
  extremal black holes}},
  \href{https://doi.org/10.1088/0264-9381/24/16/012}{\emph{Class. Quant. Grav.}
  {\bfseries 24} (2007) 4169}
  [\href{https://arxiv.org/abs/0705.4214}{{\ttfamily 0705.4214}}].

\bibitem{Amsel:2009et}
A.~J. Amsel, G.~T. Horowitz, D.~Marolf and M.~M. Roberts, \emph{{Uniqueness of
  Extremal Kerr and Kerr-Newman Black Holes}},
  \href{https://doi.org/10.1103/PhysRevD.81.024033}{\emph{Phys. Rev. D}
  {\bfseries 81} (2010) 024033}
  [\href{https://arxiv.org/abs/0906.2367}{{\ttfamily 0906.2367}}].

\bibitem{Guica:2008mu}
M.~Guica, T.~Hartman, W.~Song and A.~Strominger, \emph{{The Kerr/CFT
  Correspondence}},
  \href{https://doi.org/10.1103/PhysRevD.80.124008}{\emph{Phys. Rev. D}
  {\bfseries 80} (2009) 124008}
  [\href{https://arxiv.org/abs/0809.4266}{{\ttfamily 0809.4266}}].

\bibitem{Lu:2008jk}
H.~Lu, J.~Mei and C.~N. Pope, \emph{{Kerr/CFT Correspondence in Diverse
  Dimensions}},
  \href{https://doi.org/10.1088/1126-6708/2009/04/054}{\emph{JHEP} {\bfseries
  04} (2009) 054} [\href{https://arxiv.org/abs/0811.2225}{{\ttfamily
  0811.2225}}].

\bibitem{Castro:2010fd}
A.~Castro, A.~Maloney and A.~Strominger, \emph{{Hidden Conformal Symmetry of
  the Kerr Black Hole}},
  \href{https://doi.org/10.1103/PhysRevD.82.024008}{\emph{Phys. Rev. D}
  {\bfseries 82} (2010) 024008}
  [\href{https://arxiv.org/abs/1004.0996}{{\ttfamily 1004.0996}}].

\bibitem{Krishnan:2010pv}
C.~Krishnan, \emph{{Hidden Conformal Symmetries of Five-Dimensional Black
  Holes}}, \href{https://doi.org/10.1007/JHEP07(2010)039}{\emph{JHEP}
  {\bfseries 07} (2010) 039} [\href{https://arxiv.org/abs/1004.3537}{{\ttfamily
  1004.3537}}].

\bibitem{Chen:2010zwa}
D.~Chen, P.~Wang and H.~Wu, \emph{{Hidden conformal symmetry of rotating
  charged black holes}},
  \href{https://doi.org/10.1007/s10714-010-1080-7}{\emph{Gen. Rel. Grav.}
  {\bfseries 43} (2011) 181} [\href{https://arxiv.org/abs/1005.1404}{{\ttfamily
  1005.1404}}].

\bibitem{Lowe:2011aa}
D.~A. Lowe and A.~Skanata, \emph{{Generalized Hidden Kerr/CFT}},
  \href{https://doi.org/10.1088/1751-8113/45/47/475401}{\emph{J. Phys. A}
  {\bfseries 45} (2012) 475401}
  [\href{https://arxiv.org/abs/1112.1431}{{\ttfamily 1112.1431}}].

\bibitem{Aminov:2020yma}
G.~Aminov, A.~Grassi and Y.~Hatsuda, \emph{{Black Hole Quasinormal Modes and
  Seiberg\textendash{}Witten Theory}},
  \href{https://doi.org/10.1007/s00023-021-01137-x}{\emph{Annales Henri
  Poincare} {\bfseries 23} (2022) 1951}
  [\href{https://arxiv.org/abs/2006.06111}{{\ttfamily 2006.06111}}].

\bibitem{Bonelli:2021uvf}
G.~Bonelli, C.~Iossa, D.~P. Lichtig and A.~Tanzini, \emph{{Exact solution of
  Kerr black hole perturbations via CFT$_2$ and instanton counting: Greybody
  factor, quasinormal modes, and Love numbers}},
  \href{https://doi.org/10.1103/PhysRevD.105.044047}{\emph{Phys. Rev. D}
  {\bfseries 105} (2022) 044047}
  [\href{https://arxiv.org/abs/2105.04483}{{\ttfamily 2105.04483}}].

\bibitem{Consoli:2022eey}
D.~Consoli, F.~Fucito, J.~F. Morales and R.~Poghossian, \emph{{CFT description
  of BH\textquoteright{}s and ECO\textquoteright{}s: QNMs, superradiance,
  echoes and tidal responses}},
  \href{https://doi.org/10.1007/JHEP12(2022)115}{\emph{JHEP} {\bfseries 12}
  (2022) 115} [\href{https://arxiv.org/abs/2206.09437}{{\ttfamily
  2206.09437}}].

\bibitem{Bautista:2023sdf}
Y.~F. Bautista, G.~Bonelli, C.~Iossa, A.~Tanzini and Z.~Zhou, \emph{{Black Hole
  Perturbation Theory Meets CFT$_2$: Kerr Compton Amplitudes from
  Nekrasov-Shatashvili Functions}},
  \href{https://arxiv.org/abs/2312.05965}{{\ttfamily 2312.05965}}.

\bibitem{Johnson:2019ljv}
M.~D. Johnson et~al., \emph{{Universal interferometric signatures of a black
  hole\textquoteright{}s photon ring}},
  \href{https://doi.org/10.1126/sciadv.aaz1310}{\emph{Sci. Adv.} {\bfseries 6}
  (2020) eaaz1310} [\href{https://arxiv.org/abs/1907.04329}{{\ttfamily
  1907.04329}}].

\bibitem{Himwich:2020msm}
E.~Himwich, M.~D. Johnson, A.~Lupsasca and A.~Strominger, \emph{{Universal
  polarimetric signatures of the black hole photon ring}},
  \href{https://doi.org/10.1103/PhysRevD.101.084020}{\emph{Phys. Rev. D}
  {\bfseries 101} (2020) 084020}
  [\href{https://arxiv.org/abs/2001.08750}{{\ttfamily 2001.08750}}].

\bibitem{Raffaelli:2021gzh}
B.~Raffaelli, \emph{{Hidden conformal symmetry on the black hole photon
  sphere}}, \href{https://doi.org/10.1007/JHEP03(2022)125}{\emph{JHEP}
  {\bfseries 03} (2022) 125}
  [\href{https://arxiv.org/abs/2112.12543}{{\ttfamily 2112.12543}}].

\bibitem{Hadar:2022xag}
S.~Hadar, D.~Kapec, A.~Lupsasca and A.~Strominger, \emph{{Holography of the
  photon ring}}, \href{https://doi.org/10.1088/1361-6382/ac8d43}{\emph{Class.
  Quant. Grav.} {\bfseries 39} (2022) 215001}
  [\href{https://arxiv.org/abs/2205.05064}{{\ttfamily 2205.05064}}].

\bibitem{Chen:2022fpl}
B.~Chen, Y.~Hou and Z.~Hu, \emph{{On emergent conformal symmetry near the
  photon ring}}, \href{https://doi.org/10.1007/JHEP05(2023)115}{\emph{JHEP}
  {\bfseries 05} (2023) 115}
  [\href{https://arxiv.org/abs/2212.02958}{{\ttfamily 2212.02958}}].

\bibitem{Charalambous:2021kcz}
P.~Charalambous, S.~Dubovsky and M.~M. Ivanov, \emph{{Hidden Symmetry of
  Vanishing Love Numbers}},
  \href{https://doi.org/10.1103/PhysRevLett.127.101101}{\emph{Phys. Rev. Lett.}
  {\bfseries 127} (2021) 101101}
  [\href{https://arxiv.org/abs/2103.01234}{{\ttfamily 2103.01234}}].

\bibitem{Charalambous:2022rre}
P.~Charalambous, S.~Dubovsky and M.~M. Ivanov, \emph{{Love symmetry}},
  \href{https://doi.org/10.1007/JHEP10(2022)175}{\emph{JHEP} {\bfseries 10}
  (2022) 175} [\href{https://arxiv.org/abs/2209.02091}{{\ttfamily
  2209.02091}}].

\bibitem{Charalambous:2023jgq}
P.~Charalambous and M.~M. Ivanov, \emph{{Scalar Love numbers and Love
  symmetries of 5-dimensional Myers-Perry black holes}},
  \href{https://doi.org/10.1007/JHEP07(2023)222}{\emph{JHEP} {\bfseries 07}
  (2023) 222} [\href{https://arxiv.org/abs/2303.16036}{{\ttfamily
  2303.16036}}].

\bibitem{Hui:2021vcv}
L.~Hui, A.~Joyce, R.~Penco, L.~Santoni and A.~R. Solomon, \emph{{Ladder
  symmetries of black holes. Implications for Love numbers and no-hair
  theorems}}, \href{https://doi.org/10.1088/1475-7516/2022/01/032}{\emph{JCAP}
  {\bfseries 01} (2022) 032}
  [\href{https://arxiv.org/abs/2105.01069}{{\ttfamily 2105.01069}}].

\bibitem{Hui:2022vbh}
L.~Hui, A.~Joyce, R.~Penco, L.~Santoni and A.~R. Solomon, \emph{{Near-zone
  symmetries of Kerr black holes}},
  \href{https://doi.org/10.1007/JHEP09(2022)049}{\emph{JHEP} {\bfseries 09}
  (2022) 049} [\href{https://arxiv.org/abs/2203.08832}{{\ttfamily
  2203.08832}}].

\bibitem{Berens:2022ebl}
R.~Berens, L.~Hui and Z.~Sun, \emph{{Ladder symmetries of black holes and de
  Sitter space: love numbers and quasinormal modes}},
  \href{https://doi.org/10.1088/1475-7516/2023/06/056}{\emph{JCAP} {\bfseries
  06} (2023) 056} [\href{https://arxiv.org/abs/2212.09367}{{\ttfamily
  2212.09367}}].

\bibitem{Katagiri:2022vyz}
T.~Katagiri, M.~Kimura, H.~Nakano and K.~Omukai, \emph{{Vanishing Love numbers
  of black holes in general relativity: From spacetime conformal symmetry of a
  two-dimensional reduced geometry}},
  \href{https://doi.org/10.1103/PhysRevD.107.124030}{\emph{Phys. Rev. D}
  {\bfseries 107} (2023) 124030}
  [\href{https://arxiv.org/abs/2209.10469}{{\ttfamily 2209.10469}}].

\bibitem{Sharma:2024hlz}
C.~Sharma, R.~Ghosh and S.~Sarkar, \emph{{Symmetries of Love: Ladder Structure
  of Static and Rotating Black Holes}},
  \href{https://arxiv.org/abs/2401.00703}{{\ttfamily 2401.00703}}.

\bibitem{Cardoso:2017qmj}
V.~Cardoso, T.~Houri and M.~Kimura, \emph{{Mass Ladder Operators from Spacetime
  Conformal Symmetry}},
  \href{https://doi.org/10.1103/PhysRevD.96.024044}{\emph{Phys. Rev. D}
  {\bfseries 96} (2017) 024044}
  [\href{https://arxiv.org/abs/1706.07339}{{\ttfamily 1706.07339}}].

\bibitem{BenAchour:2022uqo}
J.~Ben~Achour, E.~R. Livine, S.~Mukohyama and J.-P. Uzan, \emph{{Hidden
  symmetry of the static response of black holes: applications to Love
  numbers}}, \href{https://doi.org/10.1007/JHEP07(2022)112}{\emph{JHEP}
  {\bfseries 07} (2022) 112}
  [\href{https://arxiv.org/abs/2202.12828}{{\ttfamily 2202.12828}}].

\bibitem{BenAchour:2023dgj}
J.~Ben~Achour, E.~R. Livine and D.~Oriti, \emph{{Schr\"odinger symmetry of
  Schwarzschild-(A)dS black hole mechanics}},
  \href{https://doi.org/10.1103/PhysRevD.108.104028}{\emph{Phys. Rev. D}
  {\bfseries 108} (2023) 104028}
  [\href{https://arxiv.org/abs/2302.07644}{{\ttfamily 2302.07644}}].

\bibitem{Kol:2011vg}
B.~Kol and M.~Smolkin, \emph{{Black hole stereotyping: Induced gravito-static
  polarization}}, \href{https://doi.org/10.1007/JHEP02(2012)010}{\emph{JHEP}
  {\bfseries 02} (2012) 010} [\href{https://arxiv.org/abs/1110.3764}{{\ttfamily
  1110.3764}}].

\bibitem{Hui:2020xxx}
L.~Hui, A.~Joyce, R.~Penco, L.~Santoni and A.~R. Solomon, \emph{{Static
  response and Love numbers of Schwarzschild black holes}},
  \href{https://doi.org/10.1088/1475-7516/2021/04/052}{\emph{JCAP} {\bfseries
  04} (2021) 052} [\href{https://arxiv.org/abs/2010.00593}{{\ttfamily
  2010.00593}}].

\bibitem{Ishibashi:2003ap}
A.~Ishibashi and H.~Kodama, \emph{{Stability of higher dimensional
  Schwarzschild black holes}},
  \href{https://doi.org/10.1143/PTP.110.901}{\emph{Prog. Theor. Phys.}
  {\bfseries 110} (2003) 901}
  [\href{https://arxiv.org/abs/hep-th/0305185}{{\ttfamily hep-th/0305185}}].

\bibitem{Martel:2005ir}
K.~Martel and E.~Poisson, \emph{{Gravitational perturbations of the
  Schwarzschild spacetime: A Practical covariant and gauge-invariant
  formalism}}, \href{https://doi.org/10.1103/PhysRevD.71.104003}{\emph{Phys.
  Rev. D} {\bfseries 71} (2005) 104003}
  [\href{https://arxiv.org/abs/gr-qc/0502028}{{\ttfamily gr-qc/0502028}}].

\bibitem{Yoshida:2019tvk}
D.~Yoshida and J.~Soda, \emph{{Quasinormal modes of $p$-forms in spherical
  black holes}}, \href{https://doi.org/10.1103/PhysRevD.99.044054}{\emph{Phys.
  Rev. D} {\bfseries 99} (2019) 044054}
  [\href{https://arxiv.org/abs/1901.07723}{{\ttfamily 1901.07723}}].

\bibitem{Pereniguez:2021xcj}
D.~Pere\~niguez and V.~Cardoso, \emph{{Love numbers and magnetic susceptibility
  of charged black holes}},
  \href{https://doi.org/10.1103/PhysRevD.105.044026}{\emph{Phys. Rev. D}
  {\bfseries 105} (2022) 044026}
  [\href{https://arxiv.org/abs/2112.08400}{{\ttfamily 2112.08400}}].

\bibitem{Ortin:2012mt}
T.~Ortin and C.~S. Shahbazi, \emph{{A Note on the hidden conformal structure of
  non-extremal black holes}},
  \href{https://doi.org/10.1016/j.physletb.2012.08.017}{\emph{Phys. Lett. B}
  {\bfseries 716} (2012) 231}
  [\href{https://arxiv.org/abs/1204.5910}{{\ttfamily 1204.5910}}].

\bibitem{Callan:1988hs}
C.~G. Callan, Jr., R.~C. Myers and M.~J. Perry, \emph{{Black Holes in String
  Theory}}, \href{https://doi.org/10.1016/0550-3213(89)90172-7}{\emph{Nucl.
  Phys. B} {\bfseries 311} (1989) 673}.

\bibitem{Myers:1998gt}
R.~C. Myers, \emph{{Black holes in higher curvature gravity}}, pp.~121--136.
\newblock 11, 1998.
\newblock \href{https://arxiv.org/abs/gr-qc/9811042}{{\ttfamily
  gr-qc/9811042}}.
\newblock 10.1007/978-94-017-0934-7{${}_{}$}8.

\bibitem{Myers:1987qx}
R.~C. Myers, \emph{{Superstring Gravity and Black Holes}},
  \href{https://doi.org/10.1016/0550-3213(87)90402-0}{\emph{Nucl. Phys. B}
  {\bfseries 289} (1987) 701}.

\bibitem{Tangherlini:1963bw}
F.~R. Tangherlini, \emph{{Schwarzschild field in n dimensions and the
  dimensionality of space problem}},
  \href{https://doi.org/10.1007/BF02784569}{\emph{Nuovo Cim.} {\bfseries 27}
  (1963) 636}.

\bibitem{Hollands:2012xy}
S.~Hollands and A.~Ishibashi, \emph{{Black hole uniqueness theorems in higher
  dimensional spacetimes}},
  \href{https://doi.org/10.1088/0264-9381/29/16/163001}{\emph{Class. Quant.
  Grav.} {\bfseries 29} (2012) 163001}
  [\href{https://arxiv.org/abs/1206.1164}{{\ttfamily 1206.1164}}].

\bibitem{Pravda:2004ka}
V.~Pravda, A.~Pravdova, A.~Coley and R.~Milson, \emph{{Bianchi identities in
  higher dimensions}},
  \href{https://doi.org/10.1088/0264-9381/24/6/C01}{\emph{Class. Quant. Grav.}
  {\bfseries 21} (2004) 2873}
  [\href{https://arxiv.org/abs/gr-qc/0401013}{{\ttfamily gr-qc/0401013}}].

\bibitem{Coley:2004jv}
A.~Coley, R.~Milson, V.~Pravda and A.~Pravdova, \emph{{Classification of the
  Weyl tensor in higher dimensions}},
  \href{https://doi.org/10.1088/0264-9381/21/7/L01}{\emph{Class. Quant. Grav.}
  {\bfseries 21} (2004) L35}
  [\href{https://arxiv.org/abs/gr-qc/0401008}{{\ttfamily gr-qc/0401008}}].

\bibitem{Pravdova:2008gp}
A.~Pravdova and V.~Pravda, \emph{{Newman-Penrose formalism in higher
  dimensions: Vacuum spacetimes with a non-twisting multiple WAND}},
  \href{https://doi.org/10.1088/0264-9381/25/23/235008}{\emph{Class. Quant.
  Grav.} {\bfseries 25} (2008) 235008}
  [\href{https://arxiv.org/abs/0806.2423}{{\ttfamily 0806.2423}}].

\bibitem{Durkee:2010xq}
M.~Durkee, V.~Pravda, A.~Pravdova and H.~S. Reall, \emph{{Generalization of the
  Geroch-Held-Penrose formalism to higher dimensions}},
  \href{https://doi.org/10.1088/0264-9381/27/21/215010}{\emph{Class. Quant.
  Grav.} {\bfseries 27} (2010) 215010}
  [\href{https://arxiv.org/abs/1002.4826}{{\ttfamily 1002.4826}}].

\bibitem{Durkee:2010qu}
M.~Durkee and H.~S. Reall, \emph{{Perturbations of higher-dimensional
  spacetimes}},
  \href{https://doi.org/10.1088/0264-9381/28/3/035011}{\emph{Class. Quant.
  Grav.} {\bfseries 28} (2011) 035011}
  [\href{https://arxiv.org/abs/1009.0015}{{\ttfamily 1009.0015}}].

\bibitem{Godazgar:2011sn}
M.~Godazgar, \emph{{The perturbation theory of higher dimensional spacetimes a
  la Teukolsky}},
  \href{https://doi.org/10.1088/0264-9381/29/5/055008}{\emph{Class. Quant.
  Grav.} {\bfseries 29} (2012) 055008}
  [\href{https://arxiv.org/abs/1110.5779}{{\ttfamily 1110.5779}}].

\bibitem{Regge:1957td}
T.~Regge and J.~A. Wheeler, \emph{{Stability of a Schwarzschild singularity}},
  \href{https://doi.org/10.1103/PhysRev.108.1063}{\emph{Phys. Rev.} {\bfseries
  108} (1957) 1063}.

\bibitem{Zerilli:1970se}
F.~J. Zerilli, \emph{{Effective potential for even parity Regge-Wheeler
  gravitational perturbation equations}},
  \href{https://doi.org/10.1103/PhysRevLett.24.737}{\emph{Phys. Rev. Lett.}
  {\bfseries 24} (1970) 737}.

\bibitem{Kodama:2003jz}
H.~Kodama and A.~Ishibashi, \emph{{A Master equation for gravitational
  perturbations of maximally symmetric black holes in higher dimensions}},
  \href{https://doi.org/10.1143/PTP.110.701}{\emph{Prog. Theor. Phys.}
  {\bfseries 110} (2003) 701}
  [\href{https://arxiv.org/abs/hep-th/0305147}{{\ttfamily hep-th/0305147}}].

\bibitem{Kodama:2003kk}
H.~Kodama and A.~Ishibashi, \emph{{Master equations for perturbations of
  generalized static black holes with charge in higher dimensions}},
  \href{https://doi.org/10.1143/PTP.111.29}{\emph{Prog. Theor. Phys.}
  {\bfseries 111} (2004) 29}
  [\href{https://arxiv.org/abs/hep-th/0308128}{{\ttfamily hep-th/0308128}}].

\bibitem{Ishibashi:2011ws}
A.~Ishibashi and H.~Kodama, \emph{{Perturbations and Stability of Static Black
  Holes in Higher Dimensions}},
  \href{https://doi.org/10.1143/PTPS.189.165}{\emph{Prog. Theor. Phys. Suppl.}
  {\bfseries 189} (2011) 165}
  [\href{https://arxiv.org/abs/1103.6148}{{\ttfamily 1103.6148}}].

\bibitem{Chodos:1983zi}
A.~Chodos and E.~Myers, \emph{{Gravitational Contribution to the Casimir Energy
  in Kaluza-Klein Theories}},
  \href{https://doi.org/10.1016/0003-4916(84)90039-3}{\emph{Annals Phys.}
  {\bfseries 156} (1984) 412}.

\bibitem{Higuchi:1986wu}
A.~Higuchi, \emph{{Symmetric Tensor Spherical Harmonics on the $N$ Sphere and
  Their Application to the De Sitter Group SO($N$,1)}},
  \href{https://doi.org/10.1063/1.527513}{\emph{J. Math. Phys.} {\bfseries 28}
  (1987) 1553}.

\bibitem{Camporesi1994}
R.~Camporesi and A.~Higuchi, \emph{The plancherel measure for p-forms in real
  hyperbolic spaces},
  \href{https://doi.org/https://doi.org/10.1016/0393-0440(94)90047-7}{\emph{Journal
  of Geometry and Physics} {\bfseries 15} (1994) 57}.

\bibitem{Lovelock:1971yv}
D.~Lovelock, \emph{{The Einstein tensor and its generalizations}},
  \href{https://doi.org/10.1063/1.1665613}{\emph{J. Math. Phys.} {\bfseries 12}
  (1971) 498}.

\bibitem{Donoghue:2017pgk}
J.~F. Donoghue, M.~M. Ivanov and A.~Shkerin, \emph{{EPFL Lectures on General
  Relativity as a Quantum Field Theory}},
  \href{https://arxiv.org/abs/1702.00319}{{\ttfamily 1702.00319}}.

\bibitem{Saketh:2023bul}
M.~V.~S. Saketh, Z.~Zhou and M.~M. Ivanov, \emph{{Dynamical Tidal Response of
  Kerr Black Holes from Scattering Amplitudes}},
  \href{https://arxiv.org/abs/2307.10391}{{\ttfamily 2307.10391}}.

\bibitem{Creci:2021rkz}
G.~Creci, T.~Hinderer and J.~Steinhoff, \emph{{Tidal response from scattering
  and the role of analytic continuation}},
  \href{https://doi.org/10.1103/PhysRevD.104.124061}{\emph{Phys. Rev. D}
  {\bfseries 104} (2021) 124061}
  [\href{https://arxiv.org/abs/2108.03385}{{\ttfamily 2108.03385}}].

\bibitem{Schwinger:1960qe}
J.~S. Schwinger, \emph{{Brownian motion of a quantum oscillator}},
  \href{https://doi.org/10.1063/1.1703727}{\emph{J. Math. Phys.} {\bfseries 2}
  (1961) 407}.

\bibitem{Keldysh:1964ud}
L.~V. Keldysh, \emph{{Diagram technique for nonequilibrium processes}},
  {\emph{Zh. Eksp. Teor. Fiz.} {\bfseries 47} (1964) 1515}.

\bibitem{Goldberger:2005cd}
W.~D. Goldberger and I.~Z. Rothstein, \emph{{Dissipative effects in the
  worldline approach to black hole dynamics}},
  \href{https://doi.org/10.1103/PhysRevD.73.104030}{\emph{Phys. Rev. D}
  {\bfseries 73} (2006) 104030}
  [\href{https://arxiv.org/abs/hep-th/0511133}{{\ttfamily hep-th/0511133}}].

\bibitem{Goldberger:2019sya}
W.~D. Goldberger and I.~Z. Rothstein, \emph{{An Effective Field Theory of
  Quantum Mechanical Black Hole Horizons}},
  \href{https://doi.org/10.1007/JHEP04(2020)056}{\emph{JHEP} {\bfseries 04}
  (2020) 056} [\href{https://arxiv.org/abs/1912.13435}{{\ttfamily
  1912.13435}}].

\bibitem{Goldberger:2020fot}
W.~D. Goldberger, J.~Li and I.~Z. Rothstein, \emph{{Non-conservative effects on
  spinning black holes from world-line effective field theory}},
  \href{https://doi.org/10.1007/JHEP06(2021)053}{\emph{JHEP} {\bfseries 06}
  (2021) 053} [\href{https://arxiv.org/abs/2012.14869}{{\ttfamily
  2012.14869}}].

\bibitem{Goldberger:2020wbx}
W.~D. Goldberger and I.~Z. Rothstein, \emph{{Horizon radiation reaction
  forces}}, \href{https://doi.org/10.1007/JHEP10(2020)026}{\emph{JHEP}
  {\bfseries 10} (2020) 026}
  [\href{https://arxiv.org/abs/2007.00731}{{\ttfamily 2007.00731}}].

\bibitem{Starobinsky:1973aij}
A.~A. Starobinski{\v{i}}, \emph{{Amplification of waves during reflection from
  a rotating ``black hole''}}, {\emph{Sov. Phys. JETP} {\bfseries 37} (1973)
  28}.

\bibitem{Starobinskil:1974nkd}
A.~A. Starobinski{\v{i}} and S.~M. Churilov, \emph{{Amplification of
  electromagnetic and gravitational waves scattered by a rotating ``black
  hole''}}, {\emph{Sov. Phys. JETP} {\bfseries 65} (1974) 1}.

\bibitem{Maldacena:1997ih}
J.~M. Maldacena and A.~Strominger, \emph{{Universal low-energy dynamics for
  rotating black holes}},
  \href{https://doi.org/10.1103/PhysRevD.56.4975}{\emph{Phys. Rev. D}
  {\bfseries 56} (1997) 4975}
  [\href{https://arxiv.org/abs/hep-th/9702015}{{\ttfamily hep-th/9702015}}].

\bibitem{Glampedakis:2017rar}
K.~Glampedakis, A.~D. Johnson and D.~Kennefick, \emph{{Darboux transformation
  in black hole perturbation theory}},
  \href{https://doi.org/10.1103/PhysRevD.96.024036}{\emph{Phys. Rev. D}
  {\bfseries 96} (2017) 024036}
  [\href{https://arxiv.org/abs/1702.06459}{{\ttfamily 1702.06459}}].

\bibitem{Perry:2023wmm}
M.~Perry and M.~J. Rodriguez, \emph{{Dynamical Love Numbers for Kerr Black
  Holes}},  \href{https://arxiv.org/abs/2310.03660}{{\ttfamily 2310.03660}}.

\bibitem{Bertini:2011ga}
S.~Bertini, S.~L. Cacciatori and D.~Klemm, \emph{{Conformal structure of the
  Schwarzschild black hole}},
  \href{https://doi.org/10.1103/PhysRevD.85.064018}{\emph{Phys. Rev. D}
  {\bfseries 85} (2012) 064018}
  [\href{https://arxiv.org/abs/1106.0999}{{\ttfamily 1106.0999}}].

\bibitem{Kim:2012mh}
Y.-W. Kim, Y.~S. Myung and Y.-J. Park, \emph{{Quasinormal modes and hidden
  conformal symmetry in the Reissner-Nordstr\"om black hole}},
  \href{https://doi.org/10.1140/epjc/s10052-013-2440-8}{\emph{Eur. Phys. J. C}
  {\bfseries 73} (2013) 2440}
  [\href{https://arxiv.org/abs/1205.3701}{{\ttfamily 1205.3701}}].

\bibitem{Howe1992}
R.~Howe and E.~Chye~Tan, \emph{{Non-abelian harmonic analysis : applications of
  $\mathrm{SL}(2,\mathbb{R})$}}, Universitext. Springer-Verlag, New York, 1992.

\bibitem{Miller1968}
W.~Miller, Jr., \emph{{Lie Theory and the Hypergeometric Functions}},
  {\emph{{Journal of Mathematics and Mechanics}} {\bfseries 17} (1968) 1143}.

\bibitem{Miller1970}
W.~Miller, Jr., \emph{{Lie Theory and Some Special Solutions of the
  Hypergeometric Equations}}, \href{https://doi.org/10.1137/0501037}{\emph{SIAM
  Journal on Mathematical Analysis} {\bfseries 1} (1970) 405}
  [\href{https://arxiv.org/abs/https://doi.org/10.1137/0501037}{{\ttfamily
  https://doi.org/10.1137/0501037}}].

\bibitem{Moura:2006pz}
F.~Moura and R.~Schiappa, \emph{{Higher-derivative corrected black holes:
  Perturbative stability and absorption cross-section in heterotic string
  theory}}, \href{https://doi.org/10.1088/0264-9381/24/2/006}{\emph{Class.
  Quant. Grav.} {\bfseries 24} (2007) 361}
  [\href{https://arxiv.org/abs/hep-th/0605001}{{\ttfamily hep-th/0605001}}].

\bibitem{Moura:2011rr}
F.~Moura, \emph{{Scattering of spherically symmetric $d$-dimensional
  $\alpha'-$corrected black holes in string theory}},
  \href{https://doi.org/10.1007/JHEP09(2013)038}{\emph{JHEP} {\bfseries 09}
  (2013) 038} [\href{https://arxiv.org/abs/1105.5074}{{\ttfamily 1105.5074}}].

\bibitem{tHooft:1974toh}
G.~'t~Hooft and M.~J.~G. Veltman, \emph{{One loop divergencies in the theory of
  gravitation}}, {\emph{Ann. Inst. H. Poincare Phys. Theor. A} {\bfseries 20}
  (1974) 69}.

\bibitem{Goroff:1985th}
M.~H. Goroff and A.~Sagnotti, \emph{{The Ultraviolet Behavior of Einstein
  Gravity}}, \href{https://doi.org/10.1016/0550-3213(86)90193-8}{\emph{Nucl.
  Phys. B} {\bfseries 266} (1986) 709}.

\bibitem{Chen:2021qrz}
Y.~Chen, \emph{{Revisiting $R^4$ higher curvature corrections to black holes}},
   \href{https://arxiv.org/abs/2107.01533}{{\ttfamily 2107.01533}}.

\bibitem{Cvetic:1996kv}
M.~Cvetic and D.~Youm, \emph{{Entropy of nonextreme charged rotating black
  holes in string theory}},
  \href{https://doi.org/10.1103/PhysRevD.54.2612}{\emph{Phys. Rev. D}
  {\bfseries 54} (1996) 2612}
  [\href{https://arxiv.org/abs/hep-th/9603147}{{\ttfamily hep-th/9603147}}].

\bibitem{Cvetic:2021vxa}
M.~Cvetic, G.~W. Gibbons, C.~N. Pope and B.~F. Whiting, \emph{{Supergravity
  black holes, Love numbers, and harmonic coordinates}},
  \href{https://doi.org/10.1103/PhysRevD.105.084035}{\emph{Phys. Rev. D}
  {\bfseries 105} (2022) 084035}
  [\href{https://arxiv.org/abs/2109.03254}{{\ttfamily 2109.03254}}].

\bibitem{Emparan:2017qxd}
R.~Emparan, A.~Fernandez-Pique and R.~Luna, \emph{{Geometric polarization of
  plasmas and Love numbers of AdS black branes}},
  \href{https://doi.org/10.1007/JHEP09(2017)150}{\emph{JHEP} {\bfseries 09}
  (2017) 150} [\href{https://arxiv.org/abs/1707.02777}{{\ttfamily
  1707.02777}}].

\bibitem{Nair:2024mya}
S.~Nair, S.~Chakraborty and S.~Sarkar, \emph{{Asymptotically de-Sitter black
  holes have non-zero tidal Love numbers}},
  \href{https://arxiv.org/abs/2401.06467}{{\ttfamily 2401.06467}}.

\bibitem{DeLuca:2022tkm}
V.~De~Luca, J.~Khoury and S.~S.~C. Wong, \emph{{Implications of the weak
  gravity conjecture for tidal Love numbers of black holes}},
  \href{https://doi.org/10.1103/PhysRevD.108.044066}{\emph{Phys. Rev. D}
  {\bfseries 108} (2023) 044066}
  [\href{https://arxiv.org/abs/2211.14325}{{\ttfamily 2211.14325}}].

\bibitem{Cvetic:2011hp}
M.~Cvetic and F.~Larsen, \emph{{Conformal Symmetry for General Black Holes}},
  \href{https://doi.org/10.1007/JHEP02(2012)122}{\emph{JHEP} {\bfseries 02}
  (2012) 122} [\href{https://arxiv.org/abs/1106.3341}{{\ttfamily 1106.3341}}].

\bibitem{Cvetic:2011dn}
M.~Cvetic and F.~Larsen, \emph{{Conformal Symmetry for Black Holes in Four
  Dimensions}}, \href{https://doi.org/10.1007/JHEP09(2012)076}{\emph{JHEP}
  {\bfseries 09} (2012) 076} [\href{https://arxiv.org/abs/1112.4846}{{\ttfamily
  1112.4846}}].

\bibitem{Nair:2022xfm}
S.~Nair, S.~Chakraborty and S.~Sarkar, \emph{{Dynamical Love numbers for area
  quantized black holes}},
  \href{https://doi.org/10.1103/PhysRevD.107.124041}{\emph{Phys. Rev. D}
  {\bfseries 107} (2023) 124041}
  [\href{https://arxiv.org/abs/2208.06235}{{\ttfamily 2208.06235}}].

\bibitem{Chakraborty:2023zed}
S.~Chakraborty, E.~Maggio, M.~Silvestrini and P.~Pani, \emph{{Dynamical tidal
  Love numbers of Kerr-like compact objects}},
  \href{https://arxiv.org/abs/2310.06023}{{\ttfamily 2310.06023}}.

\bibitem{Porfyriadis:2021psx}
A.~P. Porfyriadis and G.~N. Remmen, \emph{{Large diffeomorphisms and accidental
  symmetry of the extremal horizon}},
  \href{https://doi.org/10.1007/JHEP03(2022)107}{\emph{JHEP} {\bfseries 03}
  (2022) 107} [\href{https://arxiv.org/abs/2112.13853}{{\ttfamily
  2112.13853}}].

\bibitem{Hadar:2020kry}
S.~Hadar, A.~Lupsasca and A.~P. Porfyriadis, \emph{{Extreme Black Hole
  Anabasis}}, \href{https://doi.org/10.1007/JHEP03(2021)223}{\emph{JHEP}
  {\bfseries 03} (2021) 223}
  [\href{https://arxiv.org/abs/2012.06562}{{\ttfamily 2012.06562}}].

\bibitem{Horowitz:2022leb}
G.~T. Horowitz, M.~Kolanowski and J.~E. Santos, \emph{{A deformed IR: a new IR
  fixed point for four-dimensional holographic theories}},
  \href{https://doi.org/10.1007/JHEP02(2023)152}{\emph{JHEP} {\bfseries 02}
  (2023) 152} [\href{https://arxiv.org/abs/2211.01385}{{\ttfamily
  2211.01385}}].

\bibitem{Horowitz:2023xyl}
G.~T. Horowitz, M.~Kolanowski, G.~N. Remmen and J.~E. Santos, \emph{{Extremal
  Kerr Black Holes as Amplifiers of New Physics}},
  \href{https://doi.org/10.1103/PhysRevLett.131.091402}{\emph{Phys. Rev. Lett.}
  {\bfseries 131} (2023) 091402}
  [\href{https://arxiv.org/abs/2303.07358}{{\ttfamily 2303.07358}}].

\bibitem{Bondi:1962px}
H.~Bondi, M.~G.~J. van~der Burg and A.~W.~K. Metzner, \emph{{Gravitational
  waves in general relativity. VII. Waves from axisymmetric isolated systems}},
  \href{https://doi.org/10.1098/rspa.1962.0161}{\emph{Proc. Roy. Soc. Lond. A}
  {\bfseries 269} (1962) 21}.

\bibitem{Sachs:1962zza}
R.~Sachs, \emph{{Asymptotic symmetries in gravitational theory}},
  \href{https://doi.org/10.1103/PhysRev.128.2851}{\emph{Phys. Rev.} {\bfseries
  128} (1962) 2851}.

\bibitem{Donnay:2015abr}
L.~Donnay, G.~Giribet, H.~A. Gonzalez and M.~Pino, \emph{{Supertranslations and
  Superrotations at the Black Hole Horizon}},
  \href{https://doi.org/10.1103/PhysRevLett.116.091101}{\emph{Phys. Rev. Lett.}
  {\bfseries 116} (2016) 091101}
  [\href{https://arxiv.org/abs/1511.08687}{{\ttfamily 1511.08687}}].

\bibitem{Donnay:2016ejv}
L.~Donnay, G.~Giribet, H.~A. Gonz\'alez and M.~Pino, \emph{{Extended Symmetries
  at the Black Hole Horizon}},
  \href{https://doi.org/10.1007/JHEP09(2016)100}{\emph{JHEP} {\bfseries 09}
  (2016) 100} [\href{https://arxiv.org/abs/1607.05703}{{\ttfamily
  1607.05703}}].

\bibitem{Penna:2015gza}
R.~F. Penna, \emph{{BMS invariance and the membrane paradigm}},
  \href{https://doi.org/10.1007/JHEP03(2016)023}{\emph{JHEP} {\bfseries 03}
  (2016) 023} [\href{https://arxiv.org/abs/1508.06577}{{\ttfamily
  1508.06577}}].

\bibitem{Grumiller:2019fmp}
D.~Grumiller, A.~P\'erez, M.~M. Sheikh-Jabbari, R.~Troncoso and C.~Zwikel,
  \emph{{Spacetime structure near generic horizons and soft hair}},
  \href{https://doi.org/10.1103/PhysRevLett.124.041601}{\emph{Phys. Rev. Lett.}
  {\bfseries 124} (2020) 041601}
  [\href{https://arxiv.org/abs/1908.09833}{{\ttfamily 1908.09833}}].

\bibitem{Donnay:2018ckb}
L.~Donnay, G.~Giribet, H.~A. Gonz\'alez and A.~Puhm, \emph{{Black hole memory
  effect}}, \href{https://doi.org/10.1103/PhysRevD.98.124016}{\emph{Phys. Rev.
  D} {\bfseries 98} (2018) 124016}
  [\href{https://arxiv.org/abs/1809.07266}{{\ttfamily 1809.07266}}].

\bibitem{Donnay:2020yxw}
L.~Donnay, G.~Giribet and J.~Oliva, \emph{{Horizon symmetries and hairy black
  holes in AdS}}, \href{https://doi.org/10.1007/JHEP09(2020)120}{\emph{JHEP}
  {\bfseries 09} (2020) 120}
  [\href{https://arxiv.org/abs/2007.08422}{{\ttfamily 2007.08422}}].

\bibitem{DeLuca:2022xlz}
V.~De~Luca, A.~Maselli and P.~Pani, \emph{{Modeling frequency-dependent tidal
  deformability for environmental black hole mergers}},
  \href{https://doi.org/10.1103/PhysRevD.107.044058}{\emph{Phys. Rev. D}
  {\bfseries 107} (2023) 044058}
  [\href{https://arxiv.org/abs/2212.03343}{{\ttfamily 2212.03343}}].

\bibitem{DeLuca:2023mio}
V.~De~Luca, J.~Khoury and S.~S.~C. Wong, \emph{{Nonlinearities in the tidal
  Love numbers of black holes}},
  \href{https://doi.org/10.1103/PhysRevD.108.024048}{\emph{Phys. Rev. D}
  {\bfseries 108} (2023) 024048}
  [\href{https://arxiv.org/abs/2305.14444}{{\ttfamily 2305.14444}}].

\bibitem{Riva:2023rcm}
M.~M. Riva, L.~Santoni, N.~Savi\'c and F.~Vernizzi, \emph{{Vanishing of
  Nonlinear Tidal Love Numbers of Schwarzschild Black Holes}},
  \href{https://arxiv.org/abs/2312.05065}{{\ttfamily 2312.05065}}.

\bibitem{NIST}
``{\it NIST Digital Library of Mathematical Functions}.''
  http://dlmf.nist.gov/, Release 1.1.0 of 2020-12-15.

\end{thebibliography}\endgroup

\end{document}